\documentclass[14pt]{article}
\pdfoutput=1
\usepackage{amsmath,amssymb,amsfonts,amsthm,bm,bbm,cancel,wasysym}
\usepackage{epsfig,graphics,graphicx,epstopdf,caption,subcaption}
\graphicspath{{Charts/}}
\usepackage{array,booktabs,colortbl,colordvi,multirow}
\usepackage{colordvi,color,xcolor}
\usepackage{hyperref}
\usepackage{rotating}
\usepackage{comment}
\usepackage{feynmf}

\usepackage{verbatim}
\usepackage{cite}
\usepackage{setspace}
\usepackage{url}
\usepackage[percent]{overpic}
\usepackage{slashed}
\usepackage{authblk}
\usepackage{xspace}
\usepackage{fullpage}
\usepackage[numbers,sort&compress]{natbib}
\def\ie{{\it i.e.}}
\def\eg{{\it e.g.}}

\newskip\zatskip \zatskip=0pt plus0pt minus0pt
\def\matth{\mathsurround=0pt}
\def\lsim{\mathrel{\mathpalette\atversim<}}
\def\gsim{\mathrel{\mathpalette\atversim>}}
\def\atversim#1#2{\lower0.7ex\vbox{\baselineskip\zatskip\lineskip\zatskip
  \lineskiplimit 0pt\ialign{$\matth#1\hfil##\hfil$\crcr#2\crcr\sim\crcr}}}

\newif\ifdiagrams
\diagramstrue
\ifdiagrams

\else
  \excludecomment{fmffile}
\fi

\begin{document}


\begin{flushright}
\today
\end{flushright}
\vspace*{5mm}

\renewcommand{\thefootnote}{\fnsymbol{footnote}}
\setcounter{footnote}{1}

\begin{center}

{\Large {\bf Kinetic Mixing from Kaluza-Klein Modes: A Simple Construction for Portal Matter}}\\

\vspace*{0.75cm}

{\bf George N. Wojcik$^{1}$}~\footnote{gwojcik@wisc.edu}

\vspace{0.5cm}

{$^1$Department of Physics, University of Wisconsin-Madison, Madison, WI 53703 USA}

\end{center}
\vspace{.5cm}

\begin{abstract}
 
\noindent The vector portal/kinetic mixing simplified model of dark matter, in which thermal dark matter of a mass ranging from a few MeV to a few GeV can be realized with a dark-sector $U(1)$, relies on a small kinetic mixing term between this dark $U(1)$ and the Standard Model (SM) hypercharge. It is well-known that kinetic mixing of the right magnitude can be generated at one loop by the inclusion of ``portal matter'' fields which are charged under both the dark $U(1)$ and the SM hypercharge, and it has been previously argued on phenomenological grounds that fermionic portal matter fields must exhibit a specific set of characteristics: They must be vector-like and have the same Standard Model group representations as existing SM fermions. A natural explanation for the presence of dark $U(1)$-charged copies of SM fermions would be to enlarge the dark gauge group and embed the SM and the portal matter into a single dark multiplet, however, previous works in this direction require significant ad hoc additions to ensure that the portal matter fields are vector-like while the SM fields are chiral, and rely on complicated dark Higgs sectors to realize the appropriate symmetry breaking. In this paper, we argue that a model with large extra dimensions can easily generate chiral SM fermions and vector-like portal matter from a single dark bulk multiplet, while allowing for a far simpler dark Higgs sector. We present a minimal construction of this ``Kaluza-Klein (KK) portal matter'' and explore its phenomenology at the LHC, noting that even our simple realization exhibits signatures unique from those of more conventional models with large extra dimensions and 4-D portal matter. Generically, the portal matter sector fields will be much lighter than the other KK fields in the model, so the portal matter sector has the potential to be the first observable experimental signature of an extra dimension.

\end{abstract}

\renewcommand{\thefootnote}{\arabic{footnote}}
\setcounter{footnote}{0}
\thispagestyle{empty}
\vfill
\newpage
\setcounter{page}{1}



\section{Introduction}
As the parameter space for popular dark matter (DM) candidates such as WIMPs \cite{Arcadi:2017kky} and axions \cite{Kawasaki:2013ae,Graham:2015ouw} continues to narrow under a growing body of null results in searches, efforts to consider somewhat more exotic DM candidates have borne a plethora of other DM models in a wide range of searchable parameter spaces \cite{Battaglieri:2017aum,Alexander:2016aln}. A popular and simple expansion of the parameter space of the WIMP paradigm is the so-called vector portal/kinetic mixing class of models  \cite{Holdom:1985ag,Holdom:1986eq,Pospelov:2007mp,Izaguirre:2015yja,Essig:2013lka,Curtin:2014cca}, in which the dark matter is an SM singlet charged under a hidden local $U(1)_D$ symmetry under which the SM field content is uncharged. Interaction between the SM and dark sectors is then achieved by small kinetic mixing between $U(1)_D$ and the SM hypercharge parameterized by a coefficient $\epsilon$, appearing in the action as a term
\begin{align}\label{eq:EpsilonDef}
    \frac{\epsilon}{2 c_w} B^{\mu \nu} X_{\mu \nu},
\end{align}
where $B^{\mu \nu}$ is the usual SM hypercharge field strength tensor, $X_{\mu \nu}$ is the field strength tensor for the dark photon field, and $c_w$ is the Weinberg angle. If the dark photon $A_D$ which carries the $U(1)_D$ force attains a mass, then SM fields gain an interaction strength with $A_D$ of $\sim \epsilon e Q$, where $e$ is the electron charge and $Q$ is the electromagnetic charge of a given SM particle. Assuming the coupling constant associated with $U(1)_D$ is roughly comparable in strength to those in the electroweak sector, the relic abundance measured by \emph{Planck} \cite{Planck:2018vyg} will in turn be recreated when both the dark matter and dark photon attain masses of approximately $0.1-1 \; \textrm{GeV}$, while $\epsilon \sim 10^{-(3-5)}$, circumventing the Lee-Weinberg bound on WIMP masses and therefore avoiding harsh direct detection constraints from DM-nucleon scattering searches \cite{PandaX-II:2020oim,XENON:2018voc,LUX:2016ggv,SuperCDMS:2017mbc,DarkSide:2018kuk}.

\begin{figure}
\centering
\includegraphics[width=3.4in]{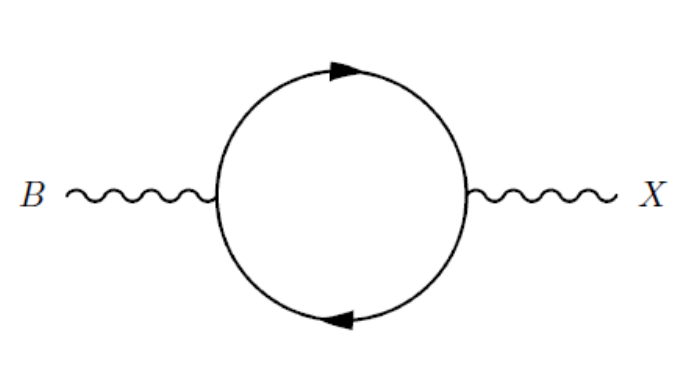}
\caption{The vacuum polarization-like Feynman graph which contributes to the kinetic mixing of the dark photon $X$ with the SM hypercharge gauge boson $B$.}
\label{fig1}
\end{figure}

It has long been known that the small kinetic mixing term $\epsilon$ can be generated at the one-loop level from vacuum polarization-like diagrams featuring particles that are charged under both $U(1)_D$ and $U(1)_Y$, so-called ``portal matter'' \cite{Holdom:1985ag,Holdom:1986eq}-- the diagram contributing to $\epsilon$ in the case of fermionic portal matter is depicted in Figure \ref{fig1}. Evaluating this diagram for a set of vector-like fermions with masses $m_i$, SM hypercharges $Q_{Y_i}$, and $U(1)_D$ charges $Q_{D_i}$ will lead to a kinetic mixing term,
\begin{align}\label{eq:4DKM}
    \epsilon = c_w \frac{g_D g_Y}{12 \pi^2} \sum_{i} Q_{Y_i} Q_{D_i} \log \frac{m_i^2}{\mu^2},
\end{align}
where $g_D$ is the $U(1)_D$ coupling, $g_Y$ is the SM hypercharge coupling, and $\mu$ is some renormalization scale. Inspection of Eq.(\ref{eq:4DKM}) indicates that the condition 
\begin{align}\label{eq:finiteMixingCondition}
    \sum_i Q_{D_i} Q_{Y_i} = 0
\end{align}
will ensure that the renormalization scale $\mu$ will drop out of the calculation and the resultant mixing will be finite and calculable, as long as the various portal matter fields don't have perfectly degenerate masses.

Because the portal matter itself might a priori appear at any high scale, relatively little work was done on probing its possible phenomenology until quite recently, when the question was broached in works such as \cite{Gherghetta:2019coi,Rizzo:2018vlb}. In particular, in \cite{Rizzo:2018vlb} the author argued that if the portal matter is fermionic, then the confluence of phenomenological constraints ranging from effects on BBN to precision electroweak measurements suggest that low-scale portal matter must take on a very specific form: It must transform as vector-like copies of SM fermion fields. More formally, a portal matter fermion must have the same quantum numbers as some SM fermion and be vector-like with respect to the SM gauge group and the dark $U(1)_D$. These portal matter fields in turn have a distinctive phenomenology from more commonly-considered vector-like fermions, decaying primarily via dark photon emission rather than via the emission of electroweak gauge bosons, and therefore have unique experimental signatures and constraints\footnote{We can also contrast these results from those for similar new fermions in models with a strict unbroken dark $U(1)$ symmetry \cite{Belotsky:2004ga,Khlopov:2006uv}, rather than the broken hidden $U(1)$ we consider here.} \cite{Kim:2019oyh}.

The requirement that portal matter fermions share SM quantum numbers with some SM fermion naturally suggests that in a more complete theory, $U(1)_D$ might be part of a larger gauge group $\mathcal{G}_D$ under which a portal matter fermion (or fermions) and the analogous SM fermion are part of a single multiplet. Then, the SM charge assignments of the portal matter fields are no longer \emph{ad hoc}, but a natural consequence of the UV construction. This possibility was explored, for example, in \cite{Rueter:2019wdf} and \cite{Wojcik:2020wgm}. In these constructions, however, it was clear that realizing non-minimal extensions to the framework of \cite{Rizzo:2018vlb} presented significant difficulties: In particular, to avoid phenomenological constraints the breaking of $\mathcal{G}_D$ down to $U(1)_D$ must be done at multi-TeV scales or higher, but the breaking of $U(1)_D$ itself must occur at the $O(\textrm{GeV})$ scale in order to reproduce the observed dark matter relic abundance-- this generally requires enormously complicated dark Higgs sectors with prodigious amounts of tree-level fine tuning. Furthermore, the fact that SM fermions are chiral while portal matter fermions must be vector-like (at least with respect to the SM) means that constructions with extended portal matter sectors require somewhat arbitrary particle content: An individual multiplet of $\mathcal{G}_D$ might contain both a left-handed SM fermion and the left-handed part of a vector-like portal matter field, but then the right-handed part of the portal matter field would have to be introduced separately. Both of these difficulties are substantially exaggerated as the dark group $\mathcal{G}_D$ becomes larger, as might be desirable if one wishes to construct a model in which the dark gauge group contains some other gauge symmetry beyond the standard model, such as a local flavor symmetry (as considered in \cite{Wojcik:2020wgm}) or lepton and baryon number.

In four dimensions, these model-building difficulties are generally unavoidable and present significant challenges. In this paper, however, we demonstrate how a model with an extended dark gauge group can elegantly be realized in models with compactified extra dimensions. In such a construction, portal matter and SM fields are part of the same $\mathcal{G}_D$ multiplet in the bulk, but boundary conditions in the higher-dimensional theory both break $\mathcal{G}_D$ down to the Abelian dark group $U(1)_D$ \emph{and} ensure that portal matter fermions only appear in the 4-dimensional theory as an infinite tower of $U(1)_D$-charged vector-like copies of an SM fermion. We construct a minimal 5D model in which such a setup is realized, and explicitly compute the kinetic mixing arising at low energies in this model.\footnote{We note that while computation of kinetic mixing from Kaluza-Klein modes is not to our knowledge well-explored in field theory, it has been the subject of inquiry in string theory, \eg \cite{Dienes:1996zr}.} From our computation of this kinetic mixing, we demonstrate that the \emph{same} conditions, in particular Eq.(\ref{eq:finiteMixingCondition}), that ensure finite and calculable kinetic mixing in a 4-dimensional theory are also valid in the low-energy limit of a five-dimensional theory. Our paper is organized as follows: In Section \ref{section:recipe}, we discuss the process of building a theory of portal matter in extra dimensions, keeping our discussion as generic as possible to explore the advantages attained and the difficulties encountered when constructing such a model. In Section \ref{section:modelSetup}, we then present a construction in this paradigm with minimal complexity, in order to demonstrate the feasibility of our program and explore the phenomenology of a specific model. In Section \ref{section:KineticMixing}, we then analytically compute the kinetic mixing arising in our model and demonstrate that the conditions for finite and calculable kinetic mixing in a 5-dimensional theory are analogous to those in a 4-dimensional one, arguing that our results are applicable to a far broader class of constructions than the minimal model in which the computation is performed. In Section \ref{section:Phenomenology}, we then explore the phenomenological signatures and constraints of our model, including searches for SM and dark sector Kaluza-Klein modes and precision constraints on extra dimensions. In Section \ref{section:Conclusion}, we conclude and discuss avenues for future work.

\section{Kaluza-Klein Portal Matter: A Generic Recipe}\label{section:recipe}

Before constructing a particular model, it is perhaps more useful to present a generic framework illustrating how extra dimensions may be leveraged to produce portal matter-like fields. In this Section, we shall endeavor to keep our treatment of the extra dimensions as generic as possible, to maximally illustrate the utility of this approach in comparison to analogous constructions in four dimensions.

We remind the reader that our goal is to create a model with the following characteristics:
\begin{itemize}
    \item The SM gauge group is augmented by some local symmetry given by the group $\mathcal{G}_D$ that contains an Abelian $U(1)_D$ symmetry
    \item At least one species of SM fermion is part of a multiplet of $\mathcal{G}_D$ containing both the SM particle, which is uncharged under $U(1)_D$ and remains chiral until electroweak symmetry breaking, and some number of portal matter particles, which are identical to the SM particle under the SM gauge group, but possess non-zero $U(1)_D$ charge and achieve a mass significantly in excess of the electroweak scale.
    \item At a high $\gg \textrm{GeV}$ scale, the symmetry $\mathcal{G}_D$ is broken down to $U(1)_D$, while $U(1)_D$ remains unbroken until $O(\textrm{GeV})$ scales.
    \item $U(1)_D$ undergoes kinetic mixing with the SM hypercharge, mediated at the one-loop level by the portal matter fields. The portal matter content of the model must be such that the naive 4-dimensional condition for this kinetic mixing to be finite and calculable, Eq.(\ref{eq:finiteMixingCondition}), is satisfied. In a theory with compactified extra dimensions, this is of course not a priori sufficient to ensure finite kinetic mixing because an infinite number of Kaluza-Klein modes for each portal matter field will appear in the loop diagram of Figure \ref{fig1}; that is to say, the sum in Eq.(\ref{eq:4DKM}) now has an infinite number of terms. Nevertheless, it remains a \emph{necessary} condition for finite and calculable mixing, since otherwise it is clear that the renormalization scale will not drop out of this sum. We shall demonstrate that an only slightly modified version of this condition is sufficient to ensure finite and calculable kinetic mixing at least in a 5-dimensional theory later on. Generalizing this condition to compactifications with 2 or more extra dimensions may be of interest, but is not explicitly considered in this paper.
\end{itemize}

The requirements for a heavy $\mathcal{G}_D \rightarrow U(1)_D$ breaking scale and a limited set of light chiral fermions accompanied by much heavier vector-like fermions are reminiscent of what emerges in a theory with compactified extra dimensions-- specifically, appropriately chosen boundary conditions can preserve the emergence of desirable light states (massless solutions to the bulk equations of motion, so-called ``zero-modes'') in the effective 4-dimensional theory for fields, while ensuring that other fields only have states with masses on the order of the compactification scale (and for fermions, are vector-like).  Using simple techniques outlined in, \eg, \cite{Hebecker:2003jt,Antoniadis:1990ew}, we can easily achieve a model of portal matter meeting all of the requirements we have listed above.

We begin our recipe by reminding the reader of the basic mechanics of the mechanism of orbifold symmetry breaking by a discrete Abelian $Z_N$ orbifold. We consider a theory compactified on some general extra-dimensional manifold $M$, orbifolded by $Z_N$.\footnote{The value of $N$ one requires will depend on the specifics of the construction. This value in turn informs our choice of a manifold $M$: For example, an $S_1$ manifold allows only $N=2$, while a $T_2$ manifold might accommodate $N=2,3,4,6$.} Reviewing the mechanism of orbifold symmetry breaking, the symmetry of the full extra-dimensional theory can be broken down to a subgroup by imposing an orbifold twist on the bulk gauge fields that acts non-trivially on different members of multiplets-- this will preserve only the subgroup as a gauge symmetry on the fixed point(s) of the orbifold twist and, in the effective 4-dimensional theory, correspond to breaking the other symmetries of the theory at the compactification scale. Systematically, one may take some $U(1)_X \subset \mathcal{G}_D$, and if we wish to break $\mathcal{G}_D \rightarrow \mathcal{G}'_D$ (where $\mathcal{G}'_D$ is a subgroup of $\mathcal{G}_D$ containing $U(1)_X$), we can impose an orbifold twist under which different members of the bulk gauge boson multiplet of $\mathcal{G}_D$ will acquire a phase proportional to their $U(1)_X$ charge. Formally, if under the decomposition $\mathcal{G}_D \rightarrow U(1)_X$, the adjoint representation $\mathcal{A}$ of $\mathcal{G_D}$ decomposes as
\begin{align}
    \mathcal{A} \rightarrow \sum_i \mathbf{1}_{Q^X_i},
\end{align}
then under the $Z_N$ orbifold twist, $\mathcal{A}$ will transform as
\begin{align}\label{eq:GaugeTransform}
    \sum_i \mathbf{1}_{Q^X_i} \rightarrow \sum_i \exp(2 \pi i \xi Q^X_i) \mathbf{1}_{Q^X_i},
\end{align}
where $Q^X_i$ is the $U(1)_X$ charge of some member of $\mathcal{A}$ (denoted by the index $i$), while $\xi$ is a constant such that invoking the $Z_N$ twist $N$ times will leave $\mathcal{A}$ invariant, that is, $exp(N \cdot 2 \pi i \xi Q^X_i) = 1$ for all indices $i$. At the fixed point(s) of the orbifold, the theory will only retain the gauge symmetry generated by those members of $\mathcal{A}$ which are left invariant under this twist, namely the subgroup $\mathcal{G}'_D$ of $\mathcal{G}_D$, since other gauge fields must vanish at these points. In the four-dimensional theory, this shall correspond to the symmetry $\mathcal{G}_D$ being broken down to $\mathcal{G}'_D$ at the scale of compactification: $\mathcal{G}'_D$ gauge bosons will have massless (zero) modes in their Kaluza-Klein towers corresponding to 4-dimensional gauge bosons, while the broken generators of $\mathcal{G}_D$ will lack zero modes and so all attain masses at the compactification scale.

The orbifold twist used to break $\mathcal{G}_D \rightarrow \mathcal{G}'_D$ will also affect other bulk fields with nontrivial representations under $\mathcal{G}_D$. In particular, if we assume that an SM fermion $\Psi$ is embedded in a representation $(\mathcal{R}_{SM}, \mathcal{R}_D)$ of the group $\mathcal{G}_{SM} \times \mathcal{G}_D$ (where $\mathcal{G}_{SM} \supset SU(3)_c \times SU(2)_L \times U(1)_Y$ is some group that contains the traditional SM gauge symmetry), then its decomposition under $\mathcal{G}_D \rightarrow U(1)_X$ will be
\begin{align}
    (\mathcal{R}_{SM}, \mathcal{R}_D) \rightarrow \sum_i (\mathcal{R}_{SM})_{Q^X_i},
\end{align}
and its transformation property under the orbifold twist will be
\begin{align}\label{eq:FermionTransform}
    \sum_i (\mathcal{R}_{SM})_{Q^X_i} \rightarrow \omega_{\Psi} \sum_i exp(2 \pi i \xi Q^X_i) (\mathcal{R}_{SM})_{Q^X_i},
\end{align}
where $\xi$ is the same $\xi$ previously selected for the gauge multiplet's orbifold transformation in Eq.(\ref{eq:GaugeTransform}), and $\omega_{\Psi}$ is any constant phase such that applying the $Z_N$ orbifold twist to $\Psi$ $N$ times will leave $\Psi$ invariant, that is, $\omega_{\Psi}^N exp(N \cdot 2 \pi i \xi Q^X_i) = 1$. As in the case of the gauge bosons, only members of the $\Psi$ multiplet which are left invariant under the orbifold twist will have zero modes; the other states will all be heavy and vector-like. This observation suggests that orbifold symmetry breaking will naturally generate the sort of fermion content in the 4-dimensional theory that a model of fermionic portal matter stipulates: If fermions are embedded in SM representations and multiplets of a dark gauge group $\mathcal{G}_D$, then under orbifold symmetry breaking of $\mathcal{G}_D$ we would expect zero-mode fermions to appear only with certain dark sector quantum numbers, while there would also emerge vector-like fermions with \emph{identical} SM charges but \emph{differing} dark sector quantum numbers, compared to these zero-modes. In order to realize a scenario consistent with our phenomenological expectations of fermionic portal matter, namely light SM fermions uncharged under a dark $U(1)_D$ together with heavy vector-like copies of these fields with non-zero $U(1)_D$ charge, we only need to ensure that for some $U(1)_D \subset \mathcal{G}'_D$, the only fermions which retain zero-modes have a $U(1)_D$-charge of 0, while all other fields vary under the orbifold twist. As an example of this construction, we might take $\mathcal{G}_D = SU(3)_D$. Then, we can select our fermion $\Psi$ to be in the $\mathbf{3}$ of $SU(3)_D$ and impose a $Z_2$ orbifold twist (so our parities are $+1$ and $-1$) that breaks $SU(3)_D \rightarrow SU(2)_D \times U(1)'_D$ (do not confuse $U(1)_D'$, an arbitrary Abelian subgroup of $SU(3)_D$, with $U(1)_D$, the gauge symmetry we identify with the dark photon), under which the gauge boson decomposes as
\begin{align}\label{eq:SU3ToyGauge}
    \mathcal{A} = \mathbf{8} \rightarrow \mathbf{3}^+_0 + \mathbf{1}^+_0 + \mathbf{2}^-_{+3} + \mathbf{2}^-_{-3}
\end{align}
and $\Psi$ decomposes as
\begin{align}\label{eq:SU3ToyMatter}
    (\mathcal{R}_{SM}, \mathbf{3}) \rightarrow (\mathcal{R}_{SM}, \mathbf{2})^-_{+1} +  (\mathcal{R}_{SM}, \mathbf{1})^+_{-2},
\end{align}
where the superscript $+(-)$ denotes $+1(-1)$ $Z_2$ parity. We can identify the zero mode of the $(\mathcal{R}_{SM}, \mathbf{1})^+_{-2}$ Kaluza-Klein tower with an SM fermion, and apply a second $Z_2$ twist with a different fixed point to ensure that this zero mode is chiral, in the usual manner employed in 5-dimensional orbifold theories \cite{Csaki:2004ay,Csaki:2005vy}.\footnote{While such an additional orbifolding to obtain chiral matter is necessary when the twist is a $Z_2$, it should be noted that there are a wider variety of strategies which might yield chiral zero-modes with $Z_N$ twists for $N \geq 3$, including scenarios in which the single $Z_N$ twist used in the orbifold symmetry breaking also preserves only one chirality of the zero-mode \cite{Kawamura:2013rj}.} With only the $Z_2$ orbifolding, the gauge symmetry of the 4-dimensional theory is $SU(2)_D \times U(1)_D'$, and because we want our zero-mode fermion to be uncharged under the dark photon gauge symmetry $U(1)_D$, we must identify $U(1)_D$ with an Abelian charge embedded in $SU(2)_D$.

We can see from the example in Eqs.(\ref{eq:SU3ToyGauge}) and (\ref{eq:SU3ToyMatter}) that following our recipe thus far will generally give us a theory with an unbroken dark gauge group $\mathcal{G}'_D$ that is larger than the dark photon gauge group $U(1)_D$.\footnote{We note that $Z_N$ orbifold breaking cannot by itself reduce the rank of the symmetry group $\mathcal{G}'_D$ \cite{Hebecker:2003jt}, so if $\mathcal{G}_D$ has rank greater than 1 some additional brane- or bulk-localized scalars with large vev's must be included in the theory to break $\mathcal{G}'_D \rightarrow U(1)_D$.} For the vector portal dark matter parameter space we are considering, we clearly need $U(1)_D$ to be broken at a small $O(\textrm{GeV})$ scale, while the remaining generators of $\mathcal{G}'_D$ must be broken at a much larger scale in order to ensure phenomenological viability of the model. In a construction with extra dimensions, we can encode at least some of these hierarchies in the geometry of model: The breaking of $U(1)_D$ might be achieved by small scalar vev's localized on the $Z_N$ orbifold fixed point(s), while the breaking $\mathcal{G}'_D$ can take place from much larger scalar vev's localized in the bulk or on a different brane. Furthermore, the orbifold construction can allow us considerably more freedom with the group representations of the $U(1)_D$-breaking scalars: Because only $\mathcal{G}'_D \subset \mathcal{G}_D$ is preserved at the $Z_N$ fixed point, any brane-localized scalar localized there can be a multiplet of $\mathcal{G}'_D$ and not the full dark gauge group $\mathcal{G}_D$. For example, in the model of Eqs.(\ref{eq:SU3ToyGauge}) and (\ref{eq:SU3ToyMatter}), $U(1)_D$ can be broken by a fixed-point-localized $SU(2)_D$ doublet. With appropriately chosen (and orbifold-consistent) boundary conditions for bulk scalars (or relying solely on scalars localized on different 3-branes from the small $U(1)_D$-breaking scalars), we might eliminate tree-level coupling between the large scalar vev's and the smaller ones. Such a decoupling between the large and small vev's can easily be realized, for example, in Randall-Sundrum-like models, which might naturally have Planckian vev's breaking $\mathcal{G}'_D \rightarrow U(1)_D$ on the UV brane \cite{Agashe:2003zs}, while the orbifold fixed point for the $Z_2$ breaking, at which the small $U(1)_D$-breaking vev is localized, is assumed to be the TeV-brane. Of course, if we limit our selection of $\mathcal{G}_D$ to rank-one groups, the problem of additional scalars to break $\mathcal{G}'_D$ becomes moot: We can assume that orbifold symmetry breaking only preserves the dark photon gauge symmetry $U(1)_D$ and that the entire dark Higgs sector simply consists of the scalar(s) which break $U(1)_D$ localized on the $Z_N$ fixed point(s).
Regardless of whether small $U(1)_D$-breaking vev's are decoupled from large vev's by some arrangement of the model geometry or if large vev's simply don't exist, we can sidestep the need for complicated tree-level fine-tuning in the Higgs sector, such as what appears in the prescribed symmetry breaking pattern in the models of \cite{Rueter:2019wdf,Wojcik:2020wgm}, and instead only confront a loop-level hierarchy problem not unlike that which is encountered for the SM Higgs. In fact, provided the compactification scale is much lower than the Planck scale, this fine-tuning should generically be far less severe than that which is encountered in the SM.

It is critical to note that we have assumed that the orbifold is a fundamental object in our construction (for example, if some string theory compactification leads to the $M/Z_N$ orbifold we require as a vacuum state) and not simply the limit of a large brane-localized scalar vev. As long as this is the case, we can rely on the orbifold to achieve all or part of our symmetry breaking \emph{without} introducing a complicated or finely-tuned Higgs sector. In this case, there are no additional dynamical scalars other than those we have discussed above, and we have achieved a large splitting between the scale of $\mathcal{G}_D$ and $U(1)_D$ breaking in whole or in part entirely due to geometry. If instead we were to assume that the boundary conditions consistent with this orbifold were actually generated by extremely large brane-localized scalar vev's (which in 5D will generate identical boundary conditions to our orbifold ones as their vev's are taken to infinity), much of the simplicity that we have gained by this technique will be lost-- we will need to introduce large-vev scalars at the $Z_N$ fixed points in order to attain the required boundary conditions, resulting in a return of the same splitting problems that we find in 4D portal matter constructions. The fact that the simplicity gains in this model are predicated on the idea that the orbifold being a physical object is of course not unique to our construction here-- the same assumption underlies the well-known attempts to address the doublet-triplet splitting problem in $SU(5)$ theories via orbifold boundary conditions \cite{Kawamura:2000ev,Csaki:2005vy}.

The final task left to be addressed in our general recipe for portal matter with Kaluza-Klein modes is simply to ensure that kinetic mixing remains finite and calculable, at least in our naive 4-dimensional intuition, by arranging our model such that Eq.(\ref{eq:finiteMixingCondition}) is satisfied in the effective 4-dimensional theory. A simple means to ensure this is to embed the SM fermions which have portal matter modes into sets of SM particles that form multiplets of semisimple unified groups, such as $SU(5)$ \cite{Georgi:1974sy} or the Pati-Salam group \cite{Pati:1974yy}, which contain the SM gauge group. Then, as was previously argued in \cite{Wojcik:2020wgm}, Eq.(\ref{eq:finiteMixingCondition}) will automatically be satisfied for kinetic mixing between any $U(1)$ in the SM gauge group (or its semisimple extension) and any Abelian subgroup $U(1) \subset \mathcal{G}_D$. More generally, we note that in the event that the SM fermions might be charged under some $U(1) \subset \mathcal{G}_D$, care must be taken to ensure that Eq.(\ref{eq:finiteMixingCondition}) is satisfied for kinetic mixing featuring this $U(1)$. For example, in the model of Eqs.(\ref{eq:SU3ToyGauge}) and (\ref{eq:SU3ToyMatter}), it is possible that if an intermediate stage of symmetry breaking exists in which $U(1)'_D$ remains unbroken, then the fermion we considered in Eq.(\ref{eq:SU3ToyMatter}), will have three infinite massive Kaluza-Klein towers of vector-like fermions which contribute to kinetic mixing between SM hypercharge and $U(1)'_D$: Two with $U(1)_D'$ charge of $+1$, and one with a $U(1)_D$ charge of $-2$. Clearly, the infinite Kaluza-Klein towers satisfy Eq.(\ref{eq:finiteMixingCondition}) themselves-- this cancellation is simply a consequence of the fact that $SU(3)_D$ is semisimple, and therefore the sum of the $U(1)$ charges of the members of \emph{any} $SU(3)_D$ multiplet will vanish. The zero-mode with a $U(1)_D'$ charge of $-2$, however, has no $+1$ charge fermions with which its contribution to kinetic mixing will cancel, indicating that this kinetic mixing is not finite and calculable in the theory, and therefore that a UV completion likely will require more portal matter states (although it should be noted here that $U(1)'_D$ is entirely orthogonal to the dark photon gauge group $U(1)_D$, so perhaps model building to keep the $U(1)'_D$ mixing finite and calculable is of limited phenomenological interest in this specific model). While this observation obviously doesn't modify the underlying condition of Eq.(\ref{eq:finiteMixingCondition}), it does demonstrate that the chiral zero-mode fermions' contribution to any kinetic mixing must cancel \emph{independently} of the cancellation that may be present among the Kaluza-Klein modes, a subtlety not present in the 4-dimensional theory.

The dark matter itself can be realized in several contexts in this class of constructions. The simplest realization is to embed the dark matter in a multiplet of the broken group $\mathcal{G}'_D$ on the fixed point where only $\mathcal{G}'_D$ is preserved-- in the model of Eqs.(\ref{eq:SU3ToyGauge}) and (\ref{eq:SU3ToyMatter}), dark matter can be readily realized as a brane-localized $SU(2)_D$ doublet scalar or fermion. A perhaps more intriguing possibility would be to embed the dark matter as an SM singlet and $\mathcal{G}_D$ multiplet in the bulk, in the $SU(3)_D$ toy model an $SU(3)_D$ triplet, for example, and set an intrinsic phase to the field's orbifold parity such that the $U(1)_D$-charged components receive a zero-mode and the $U(1)_D$-neutral component is massive. Such a more complex setup may have intriguing dark matter phenomenology, but it will be highly model dependent, and as we are focusing on the nature of kinetic mixing in these extra-dimensional theories, further discussion on the dark matter itself will simply be considered beyond the scope of this work.

There exist significant advantages from a model-building perspective in using a Kaluza-Klein realization of portal matter to extend the minimal model in \cite{Rizzo:2018vlb}. First, the multiplet structure of any model where Kaluza-Klein modes play the role of portal matter will likely be substantially simpler than an analogous 4-dimensional construction. There is no need to include additional ad hoc vector-like partners for each portal matter field, which can become particularly taxing if one chirality of the portal matter fields is assumed to be in a dark multiplet with an SM field; rather, both chiralities of the portal matter fields arise naturally as part of the same bulk field. 
Meanwhile, we have seen that the large hierarchy between the scale of breaking $\mathcal{G}_D \rightarrow U(1)_D$ and the $U(1)_D$ breaking scale does not need to be the consequence of a complicated and highly-tuned Higgs sector, but can instead be encoded in the geometry of the theory; it is in fact entirely feasible that any fine-tuning of the scalar sector of the model can be relegated to the loop level if it is present at all.
Finally, while in the  minimal model of \cite{Rizzo:2018vlb} the mass scale of the portal matter was arbitrary (as long as it was high enough to escape existing experimental constraints), the connection of the $\mathcal{G}_D$ breaking scale to the compactification scale in the extra dimensional theory allows us to easily relate the scale of the portal matter sector to new physics at other phenomenologically interesting scales: As a simple example, we note that Randall Sundrum models \cite{Randall:1999ee} address the gauge-gravity hierarchy problem with a single TeV-scale warped extra dimension and can be naturally stabilized at this size \cite{Goldberger:1999uk}. Following our model building recipe in this construction, then, the scale of the fermionic portal matter masses in such a model is no longer arbitrary: The new physics associated with portal matter must emerge at the TeV-scale as well.\footnote{There are many other models, dating back to \cite{Antoniadis:1990ew}, which posit TeV-scale extra dimensions for a variety of phenomenological concerns, such as supersymmetry breaking. As with the RS construction, many of these models and the phenomenological problems they address might be adapted to incorporate our TeV-scale Kaluza-Klein portal matter paradigm.}

In presenting our recipe in this Section, we have been quite agnostic about the specifics of how a construction of a model with Kaluza-Klein modes functioning as fermionic portal matter might look, specifying a simple example only insofar as it was useful to discuss our general arguments about this class of models. In an effort to take speculation to reality, in the following Sections we shall present a more fully realized semi-realistic toy model of such a construction, where we can hope to address a number of the questions left unanswered in our general discussion: In particular, how we might guarantee that kinetic mixing remains finite and calculable when individual portal matter modes are replaced by Kaluza-Klein towers, what sort of phenomenology might be associated with the new portal matter sector, and how the existence of this sector might affect other phenomenological constraints on theories of extra dimensions. We shall therefore move on from the vague generalities we have addressed up to this point and into more concrete model building.

\section{Model Setup}\label{section:modelSetup}

Our setup begins by embedding the entire SM in a flat 5D theory compactified on an interval of length $\pi R$, where the 5th dimension is parameterized by the coordinate $\phi$ ranging from $\phi=0$ to $\phi=\pi$. At this point we have no need to specify a scale for $R$, but in the interest of exploring the phenomenology of TeV-scale portal matter we shall assume a compactification scale of $R^{-1} \sim \textrm{TeV}$ here. Assuming a TeV-scale compactification of a flat extra dimension does introduce some potentially troubling pitfalls-- in particular, we would require fine tuning to stabilize $R$, and the non-renormalizable effective 5D theory will break down at a comparatively low scale of $\sim 10-20 \; \textrm{TeV}$, requiring a UV completion. A somewhat more complicated treatment with a warped extra dimension avoids these pitfalls, allowing the compactification scale to be stabilized at $O(\textrm{TeV})$ via a Goldberger-Wise field \cite{Goldberger:1999uk} and deferring the need for a UV completion to an exponentially higher energy scale, but we shall restrict all of our computations and most of our discussions in this work to the much simpler flat case, which will demonstrate the crucial points of the construction's phenomenology in either geometry, and only briefly touch on the potentially more realistic warped scenario.

Our construction is equivalent to an $S_1/{Z_2 \times Z_2'}$ orbifold, with $\phi=0$ being the $Z_2$ fixed point and $\phi=\pi$ being the $Z_2'$ one, and we shall use orbifold arguments to motivate our boundary conditions and symmetry breaking before transitioning to the somewhat more convenient interval framework for performing actual calculations. Following our recipe from the previous Section, we shall assume that $Z_2'$ is the orbifold twist that will break the dark group $\mathcal{G}_D$ down to $U(1)_D$. In order to allow the orbifold twist to perform this breaking without the need for additional scalars in the bulk or on either of the branes, we shall take our dark group $\mathcal{G}_D = SU(2)_D$, a rank-one group. Our recipe then calls for us to find a representation $\mathbf{R}_D$ of $SU(2)_D$ with at least one $U(1)_D$-neutral member, into which we can put fermions that shall become both SM fields and portal matter. The smallest such non-trivial representation is the adjoint $\mathbf{3}$ of $SU(2)_D$, which breaks down to $U(1)_D$ as
\begin{align}
    \mathbf{3} \rightarrow \mathbf{1}_{-1} + \mathbf{1}_0 + \mathbf{1}_{+1}.
\end{align}
Following Eq.(\ref{eq:FermionTransform}), we see that a fermion in the representation $\mathcal{R}_D$ of $SU(2)_D$ will \emph{only} have zero-modes for its $U(1)_D$-uncharged state-- the $Q^D=\pm 1$ states have negative $Z_2'$ parity and are therefore not invariant at the $Z_2'$ fixed point. Conveniently, because the bulk $SU(2)_D$ gauge field must also transform according to Eq.(\ref{eq:FermionTransform}) and is also in the adjoint representation of $SU(2)_D$, we see that the same orbifolding also only leaves the $U(1)_D$ gauge boson with a zero-mode, breaking $SU(2)_D$ down to $U(1)_D$ at the $Z_2'$ fixed point $\phi=\pi$. Since only the $U(1)_D$ dark gauge symmetry remains at $\phi=\pi$, we can then break $U(1)_D$ with a $U(1)_D$-charged complex scalar $\Phi_D$ localized at the $\phi=\pi$ brane; if $\langle \Phi_D \rangle \ll R^{-1}$, then the scale of this breaking will be much lower than that of the compactification-scale breaking of the $SU(2)_D \rightarrow U(1)_D$ from the orbifold twist. While certainly somewhat fine-tuned, the large hierarchy between the scale of $SU(2)_D \rightarrow U(1)_D$ breaking and the breaking of $U(1)_D$ is not unreasonable: Breaking $SU(2)_D$ to $U(1)_D$ and then to nothing in a 4-dimensional theory would require substantial tree-level fine tuning of scalar potential terms to effect this hierarchy, but here, the vev of a brane-localized dark Higgs and the compactification scale have no relationship at tree-level.\footnote{As noted in our general discussions in Section \ref{section:recipe}, a hierarchy still persists between the $U(1)_D$ breaking scale and the compactification scale still persists at the loop level, but given that such a difficulty exists even in the Standard Model we do not propose to address it here.}

We can then use the $Z_2$ parity to eliminate several phenomenologically undesirable light states, such as scalars associated with the 5th component of gauge fields and opposite-chirality fermions (in the familiar manner for 5-dimensional theories \cite{Csaki:2004ay,Csaki:2005vy,Agashe:2003zs}). We have therefore realized our primary model building goal: The SM fermions remain light, but there are now a number of heavy fermions of mass $\sim O(R^{-1})$ with identical SM quantum numbers, but are charged under $U(1)_D$-- these conveniently fit the description of portal matter fields. Furthermore, because $SU(2)_D$ is semisimple (and therefore the trace of the $U(1)_D$ charges of any representation of $SU(2)_D$ must vanish) and the chiral zero modes of our fermion Kaluza-Klein towers have no $U(1)_D$ charge, our naive 4-dimensional condition for finite and calculable mixing in Eq.(\ref{eq:finiteMixingCondition}) is automatically satisfied, regardless of what SM fermion(s) we embed in the bulk in this manner. Of course, because there is now an \emph{infinite} tower of pairs of portal matter particles, this naive condition is not sufficient to ensure that the kinetic mixing remains finite and calculable, however later on we shall demonstrate that that the mixing remains finite even when the contributions of the entire Kaluza-Klein towers are included.

Having established the general outlines of our model (and how it relates to our general strategy for constructing a model with Kaluza-Klein portal matter fields), we can move on to specifics, determining the bulk profiles of the various fermions and gauge bosons appearing in the construction.

\subsection{Gauge Bosons}\label{section:gauges}
We shall begin our discussion with an overview of the gauge sector of the theory, focusing on the new gauge bosons associated with our dark gauge group $SU(2)_D$, before moving on to the fermionic sector containing the portal matter fields. For convenience, we shall write the $SU(2)_D$ gauge bosons in a basis of $U(1)_D$-charge eigenstates-- under this decomposition the three new gauge bosons in our model are the $U(1)_D$ gauge boson $X$ and the two $U(1)_D$-charged states $I^+$ and $I^-$, with $U(1)_D$ charges of $+1$ and $-1$ respectively. We must translate our orbifold-inspired discussions above into boundary conditions on the interval which realize our desired gauge mass spectrum: Namely, that the gauge bosons $I^\pm$ will lack zero modes and hence acquire masses at the compactification scale $R^{-1}$, while $X$ will have a light state which is massless up to contributions from the small vev of a brane-localized dark Higgs; we shall identify this light state of $X$ with the dark photon. 
In the language of the interval, the orbifold parities we have identified at the beginning of Section \ref{section:modelSetup} correspond to differing boundary conditions at the $\phi=0$ and $\phi=\pi$ branes-- a positive $Z_2$ ($Z_2'$) orbifold parity corresponds to a von Neumann boundary condition $\partial_\phi X,I^{\pm} =0$ at the $\phi=0 (\pi)$ brane, while a negative parity corresponds to a Dirichlet boundary condition $X,I^{\pm} = 0$ at these branes.\footnote{Brane-localized terms, such as the mass terms arising from the vev of a brane-localized dark Higgs or brane-localized kinetic terms, shall modify these boundary conditions; the modified boundary conditions can be found as usual \cite{Casagrande:2008hr} by displacing the $\delta$-functions that appear in the action from these terms infinitesimally into the bulk and taking the limit as this displacement goes to 0.}.
To have a zero-mode in the Kaluza-Klein spectrum, a field must have von Neumann boundary conditions (up to modifications from brane-localized terms) at both branes.
In Table \ref{table:gaugeBosons}, the boundary conditions of the various gauge fields (both their 4-dimensional vector components $X_\mu$, $I^{\pm}_\mu$ and their fifth component scalars $X_\phi$, $I^{\pm}_\phi$) are depicted, where we can see that our orbifold-inspired choices preserve the full $SU(2)_D$ gauge symmetry in the bulk and on the $\phi=0$ brane, but on the $\phi=\pi$ brane, only $U(1)_D$ is preserved. The fact that only $U(1)_D$, rather than the full $SU(2)_D$, remains unbroken on the $\phi=\pi$ brane is critical: This allows us to assume that the brane-localized dark Higgs which breaks $U(1)_D$ is simply a complex scalar with charge $Q_H$ under $U(1)_D$, rather than an $SU(2)$ multiplet, and furthermore will allow us to write $SU(2)_D$-violating terms on this brane in our fermionic action, which shall prove essential to generating non-zero kinetic mixing. 

\begin{table}[h!]
\centering
\begin{tabular}{| c | c | c |}
\hline
Gauge boson & $\phi=0$ BC & $\phi=\pi$ BC \\
\hline
$I^{\pm}_\mu$ & + & - \\
\hline
$I^{\pm}_\phi$ & - & + \\
\hline
$X_\mu$ & + & + \\
\hline
$X_\phi$ & - & - \\
\hline
\end{tabular}
\caption{The bulk $SU(2)_D$ gauge bosons and their boundary conditions (BCs) on the branes, excluding the effect of the dark Higgs and any brane-localized kinetic terms. A $+$ denotes a von Neumann BC, while a $-$ denotes a Dirichlet BC. Notably, only fields which possess a von Neumann ($+$) BC on both the $\phi=0$ and $\phi=\pi$ branes will have a zero mode. The gauge bosons are arranged in a basis of charge eigenstates of $U(1)_D$, the gauge symmetry associated with the the gauge boson $X_\mu$ that remains unbroken by the BCs.}
\label{table:gaugeBosons}
\end{table}

In the basis of $U(1)_D$ eigenstates, the $SU(2)_D$ gauge boson action becomes (omitting interactions among the gauge bosons that we can address specifically later)
\begin{align}\label{eq:oldGaugeAction}
    S=\int d^4 x \int_0^\pi d\phi \, R \bigg\{ &-\frac{1}{4} X_{\mu \nu}^2 \bigg( 1+\omega_X \frac{\delta(\phi)}{R}+ \tau_X \frac{\delta(\phi)}{R} \bigg)-\frac{1}{2} I^+_{\mu \nu} I^-_{\mu \nu} \bigg( 1+\omega_X \frac{\delta(\phi)}{R} \bigg) \nonumber\\
    &+\frac{1}{R^2} I^-_{\phi \mu} I^+_{\phi \mu}+\frac{1}{2 R^2} X_{\phi \mu}^2 + \frac{\delta(\phi-\pi)}{R} \bigg( |(\partial_\mu +i Q_H g_{D 5} X_\mu)\Phi_D|^2 -V(\Phi_D) \bigg)\bigg\},
\end{align}
where
\begin{align}
    X_{\mu \nu} \equiv (\partial_\mu X_\nu -\partial_\nu X_\mu),\;\;\; X_{\mu \phi} \equiv (\partial_\mu X_\phi -\partial_\phi X_\mu), \;\;\; I^\pm_{\mu \nu} \equiv (\partial_\mu I^\pm_\nu - \partial_\nu I^\pm_\mu), \;\;\; I^\pm_{\mu \phi} \equiv (\partial_\mu I^\pm_\phi - \partial_\phi I^\pm_\mu),
\end{align}
and $\omega_X$ and $\tau_X$ are brane-localized kinetic terms \cite{Georgi:2000ks,Dvali:2000hr}, which we retain because of their nontrivial influence on the model phenomenology. Note that both $X$ and $I^{\pm}$ must have the same brane-localized kinetic term $\omega_X$ on the $\phi=0$ brane because $SU(2)_D$ symmetry is preserved there\footnote{For simplicity, we omit the possibility of $\phi=\pi$ brane-localized kinetic terms for the $I^\pm$ fields. In any case, because these fields have odd parity at the $\phi=\pi$ brane, if such terms are absent at tree level they will not be generated radiatively \cite{delAguila:2003kd}.}. Furthermore, to avoid the emergence of tachyonic or ghost-like states in the Kaluza-Klein towers, we shall assume that all brane-localized kinetic terms are positive, that is $\omega_X, \tau_X \geq 0$. For convenience, we have separated the 4-dimensional Lorentz indices (indicated by Greek characters) from the Lorentz index associated with the 5-dimensional coordinate $\phi$. Finally, $\Phi_D$ denotes the dark Higgs localized on the $\phi=\pi$ brane which shall break the  $U(1)_D$ gauge symmetry that survives after our boundary conditions are imposed. For clarity, we have written $\Phi_D$'s covariant derivative explicitly here, with $g_{D5}$ being the 5-D dimensionful gauge coupling constant for $SU(2)_D$. We shall find that $g_{D5}$ will be related to $g_D$, the effective 4-dimensional gauge coupling for the dark photon in our model, by the expression
\begin{align}
    g_{D5} = \sqrt{R(\pi + \tau_X + \omega_X)} g_D,
\end{align}
up to small $O(m_D^2 R^2)$ corrections, where $m_D$ is the dark photon mass. Since $g_D$ makes far better direct contact with the measurable physical parameters of our model, we shall use the dimensionless parameter $g_D$ rather than the dimensionful $g_{D5}$ throughout the remainder of this work.
We also note that in Eq.(\ref{eq:oldGaugeAction}), while we have left the $U(1)_D$ charge of $\Phi_D$, $Q_H$, as a free parameter, later phenomenological considerations in the fermion sector actually constrain its choice: We shall find that in order for the portal matter to mix with SM matter at the tree level and therefore be short-lived, the dark Higgs $\Phi_D$ must have a $U(1)_D$ charge equal in magnitude to the $U(1)_D$ charge of the gauge bosons $I^\pm$. In our charge normalization convention, we therefore have $Q_D=\pm 1$.

As is usual in theories of extra dimensions, our selection of $Z_2$ parities for the fifth components of the $X$ and $I^{\pm}$ vector fields allow us to gauge them away \cite{Muck:2001yv,Flacke:2008ne,Casagrande:2008hr}, setting $I^\pm_\phi=X_\phi=0$. Our remaining action is then
\begin{align}\label{eq:fullGaugeAction}
    S= S_{X_\mu} + S_{I^\pm_\mu},
\end{align}
where
\begin{align}\label{eq:gaugeActionExpressions}
    S_{X_\mu} &= \int d^4 x \int_0^\pi d\phi \, R \bigg\{ -\frac{1}{4}X^{\mu \nu} X_{\mu \nu} -\frac{1}{2 R^2} X^\mu \partial_\phi^2 X_\mu + \delta(\phi-\pi) (\pi + \tau_X + \omega_X)\frac{Q_H^2 g_{D}^2 v_D^2}{2} X^\mu X_\mu \bigg\},\\
    S_{I^\pm_\mu} &= \int d^4 x \int_0^\pi d\phi \, R \bigg\{ -\frac{1}{2} (I^-)^{\mu \nu} I^+_{\mu \nu} -\frac{1}{R^2} I^-_\mu \partial_\phi^2 I^+_\mu - \frac{1}{\xi} (\partial^\mu I^-_\mu)(\partial^\nu I^+_\nu) \bigg\}, \nonumber
\end{align}
where for brevity we have written $\langle \Phi_D \rangle = v_D/\sqrt{2}$.
We can now perform a Kaluza-Klein decomposition of the gauge fields, arriving at
\begin{align}\label{eq:gaugeKKs}
    &X_\mu = \frac{1}{\sqrt{R}} \sum_n \chi^{X}_n(\phi) X_\mu^{(n)}(x), &I^\pm_\mu = \frac{1}{\sqrt{R}} \sum_n \chi^+_n (\phi) (I^\pm_\mu)^{(n)}(x)
\end{align}
together with the normalization condition
\begin{align}\label{eq:gaugeNorm}
    &\int_0^\pi d\phi \, \chi^{X,+}_n(\phi) \chi^{X,+}_n(\phi) = \delta_{nm}.
\end{align}
We can now solve for the $\chi^{X,+}_n(\phi)$ functions that produce mass eigenstates in the 4-dimensional theory after integrating over $\phi$. Starting with the $X$ gauge boson, we find that for $v_D \ll R^{-1}$, the lightest Kaluza-Klein mode of this field's Kaluza-Klein tower (which we shall identify with the dark photon), has a bulk wave function
\begin{align}\label{eq:DarkPhotonZeroMode}
    \chi^X_0 (\phi) \approx \sqrt{\frac{1}{\pi + \tau_X + \omega_X}} \bigg[ 1 + \frac{m_D^2 R^2}{6} \bigg(3(\pi^2-\phi^2) - 6 \phi \omega_X +\frac{2 \pi (-\pi^2 + 3 \tau_X \omega_X)}{(\pi + \tau_X + \omega_X)}\bigg) + O(m_D^4 R^4)\bigg],
\end{align}
where $m_{D}$ is the dark photon mass, given by
\begin{align}
    m_D^2 = g_D^2 v_D^2 (1 + O(m_D^2 R^2)).
\end{align}
While their large mass and suppressed coupling to SM states renders them irrelevant to our analysis, for completeness we note that the more massive Kaluza-Klein tower modes of the $X$ boson have bulk wave functions given by
\begin{align}
    \chi^X_n (\phi) = N^X_n \big( \cos{(m^X_n R \phi)} - m^X_n \omega_X \sin{(m^X_n R \phi)} \big) + O(m_D^2 R^2),\\
    N^X_n \equiv \sqrt{\frac{2(1+(m^X_n)^2 R^2 \tau_X^2)}{\pi(1+(m^X_n)^2 R^2 \tau_X^2)(1+(m^X_n)^2 R^2 \omega_X^2)+(\tau_X+\omega_X)(1+(m^X_n)^2 R^2 \tau_X \omega_X)}} \nonumber,
\end{align}
where $m^X_n$ is the mass of that mode. The allowed values of $m^X_n$ are given as the solutions of the equation
\begin{align}
    ((m^X_n)^2 R^2 (\tau_X + \omega_X) \cos{(m^X_n R \pi)} + m^n_X R (1-(m^X_n)^2 R^2 \tau_X \omega_X ) \sin{(m^X_n R \pi)}) \\
    = (\pi + \tau_X + \omega_X) m_D^2 R^2 (\cos{(m^X_n R \pi)} - m^X_n R \omega_X \sin{(m^X_n R \pi)}). \nonumber
\end{align}
Because of the absence of any brane-localized mass terms, the bulk wave functions and masses for the members of the $I^{\pm}$ Kaluza-Klein tower are somewhat simpler. There is no light mode for these gauge bosons, and the tower mode of mass $m^I_n$ has the bulk wave function
\begin{align}
    \chi^I_{n} (\phi) = \sqrt{\frac{2}{\pi(1+ (m^I_n)^2 R^2 \omega_X^2) + \omega_X}} \big( \cos (m^I_n R \phi) - m^I_n R \omega_X \sin(m^I_n R \phi) \big),
\end{align}
where the allowed values of $m^I_n$ are given as the solutions to the equation
\begin{align}
    \cos{(m^I_n R \pi) = m^I_n R \omega_X \sin{(m^I_n R \pi)}}.
\end{align}
Finally, before moving on to the discussing the gauge sector, it is useful to briefly discuss the analogous bulk wave functions and mass spectrum for the SM gauge bosons, in particular the electroweak sector. For simplicity, we have assumed that the SM gauge fields do not possess brane-localized kinetic terms. In practice this assumption shall have a limited effect on our phenomenological explorations of the model, since we shall be focusing on the dark/portal sector, however we shall comment on certain cases in which non-trivial brane-localized kinetic terms for the SM gauge fields may become relevant.

To avoid phenomenologically undesirable mixing between the dark scalar $\Phi_D$ and the SM Higgs, we decide to localize the SM Higgs field (and hence the $W$ and $Z$ boson mass terms) on the $\phi=0$ brane.\footnote{A reader may be concerned that although mixing terms may be absent at tree-level, they may still be generated from bulk loops. Because current constraints, such as those from \cite{Gorbunov:2021ccu,BNL-E949:2009dza}, place limits on the mixing angle of $\sim O(10^{-4})$ (or $\sim O(10^{-3})$ for a dark Higgs mass $\gsim 300 \; \textrm{MeV}$ \cite{Rizzo:2018vlb}), we have assumed that radiative suppression will be sufficient to keep the mixing between the scalars within constraints.} Furthermore, we shall assume that the the SM Higgs is a singlet under the dark gauge group $SU(2)_D$, to avoid phenomenological pitfalls associated with additional SM Higgs multiplets. Then the $W$ and $Z$ boson Kaluza-Klein towers will have light modes that we shall identify with the corresponding SM electroweak gauge bosons; these shall have bulk wave functions given by
\begin{align}\label{eq:WZZeroMode}
    \chi^{W/Z}_0 (\phi) = \frac{1}{\sqrt{\pi}} \bigg( 1 + m_{Z/W}^2 R^2 \bigg(\pi \phi- \frac{\phi^2}{2} - \frac{\pi^2}{3}\bigg) + O(m_{Z/W}^4 R^4) \bigg),
\end{align}
where $m_{W/Z}$ is the mass of the SM $W$ and $Z$ bosons. Notably, mixing between the zero mode and the heavier Kaluza-Klein modes of the $W$ and $Z$ gauge boson towers slightly shift the $W$ and $Z$ boson masses from their SM predictions. The masses are now given by
\begin{align}\label{eq:WZMasses}
    m_{W}^2 = \frac{g_{L5}^2 R v^2}{4 \pi} \bigg( 1 - \frac{g_{L5}^2 R v^2 \pi}{12} +O(m_{W}^4 R^4)\bigg),\\
    m_{Z}^2 = \frac{(g_{L5}^2+ g_{Y5}^{2}) R v^2}{4 \pi} \bigg( 1 - \frac{(g_{L5}^2 + g_{Y5}^{2}) R v^2 \pi}{12}  + O(m_{Z}^4 R^4)\bigg),
\end{align}
where $v$ is the SM Higgs vev, $g_{L5}$ is the 5-dimensional $SU(2)_L$ coupling constant, and $g_{Y5}$ is the 5-dimensional $U(1)_Y$ coupling constant. As is typical in theories with TeV-scale extra dimensions, this slight shift will result in a small correction to the electroweak $\rho$ parameter; this shall provide a significant phenomenological constraint on our model.

Meanwhile, the heavier Kaluza-Klein modes of the $W$ and $Z$ bosons will have bulk wave functions given by
\begin{align}
    \chi^{W/Z}_n(\phi) = \sqrt{\frac{2}{\pi}} \cos \big( m^{W/Z}_n R (\pi-\phi) \big) + O(m_{W/Z}^2 R^2),
\end{align}
with the Kaluza-Klein mode masses $m^{W/Z}_n$ given by solutions to the equation
\begin{align}\label{eq:WZKKMasses}
    \sin(m^{W/Z}_n R \pi) = \frac{ \pi m_{W/Z}^2 R}{ m^{W/Z}_n} \sin (m^{W/Z}_n R \pi).
\end{align}
Since $m_{W}$ and $m_{Z}$ are both much smaller than the compactification scale, we note that to excellent approximation we can simply write that the mass of the $n^{th}$ Kaluza-Klein mode of the $W$ or $Z$ boson tower is simply $m^{W/Z}_n = n R^{-1}$. From the expressions in Eqs.(\ref{eq:WZMasses}-\ref{eq:WZKKMasses}), we can in turn straightforwardly derive the analogous results for the gluon and photon zero modes and Kaluza-Klein towers, simply by setting $m_{W/Z} = 0$.

Before moving on, we also note that there will be occasions in which sums of the form
\begin{align}
    \sum_n \frac{\chi^{W/Z}_n(\phi_1) \chi^{W/Z}_n(\phi_2)}{p^2-(m^{W/Z}_n)^2}
\end{align}
will be useful in calculations-- such as computing the contribution of the exchange of the entire Kaluza-Klein tower of $W$ gauge bosons to the Fermi constant. As is well-known \cite{Randall:2001gb,Rizzo:2020ybl}, these sums can be computed in closed form by simply solving for the bulk profile of the gauge field propagator in the 5-dimensional theory. For completeness, we note that the relevant sum in our case is
\begin{align}\label{eq:ZWBosonSum}
    \sum_n \frac{\chi^{W}_n(\phi_1) \chi^{W}_n(\phi_2)}{p^2-(m^{W/Z}_n)^2} = \frac{R^2 \big[ p R \cos(p R \phi_<) + \frac{g_{L5}^2 v^2 R}{4} \sin(p R \phi_<) \big] \big[ \cos(p R (\pi - \phi_>))\big]}{p R \big[p R \sin(p \pi R) - \frac{g_{L5}^2 v^2 R}{4} \cos(p \pi R)\big]},\\
    p \equiv \sqrt{p^2}, \;\;\; \phi_< \equiv \textrm{min}(\phi_1,\phi_2), \;\;\; \phi_> \equiv \textrm{max}(\phi_1,\phi_2),
\end{align}
with the equivalent sum for the $Z$ boson being given by substituting $g_{L5}^2 \rightarrow g_{L5}^2+g_{Y5}^2$.

\begin{figure}
    \centering
    \includegraphics[width=3.5in]{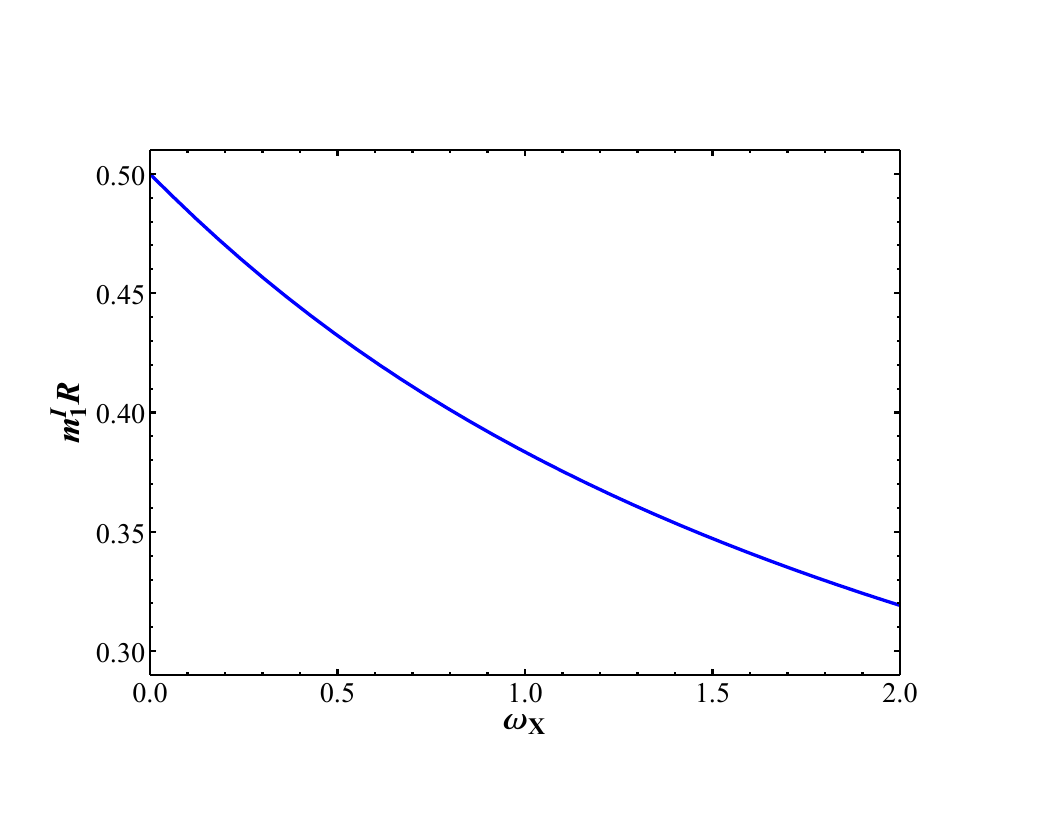}
    \caption{The ratio of the mass of the lightest $I^\pm$ boson Kaluza-Klein state to the compactification scale $R^{-1}$, as a function of the brane-localized kinetic term $\omega_X$. Note that in our construction, the lowest-lying heavy Kaluza-Klein modes of the SM gauge bosons will all have a mass of $R^{-1}$, up to small corrections from the SM Higgs.}
    \label{fig3}
\end{figure}

An important phenomenological point to emphasize here is that because of their boundary conditions and their brane-localized kinetic terms, the gauge bosons $I^\pm$ will invariably be significantly lighter than the lightest SM gauge boson Kaluza-Klein modes (other than the zero modes). Recalling that the lightest SM gauge boson Kaluza-Klein tower modes will have a mass given by $R^{-1}$, in Figure \ref{fig3} we depict the ratio of these masses, $m^I_1 R$, as a function of the brane-localized kinetic term $\omega_X$, and we see that for $\omega_X > 0$ the lightest $I^\pm$ Kaluza-Klein state is less than half of the mass of the lightest Kaluza-Klein state of the SM gauge bosons. While this ratio can be somewhat ameliorated by the introduction of brane-localized kinetic terms for the SM gauge bosons, unless these terms are very large it is unlikely to make the SM Kaluza-Klein modes lighter than the portal matter fields. Nor is this mass difference unique to flat extra dimensions: Even larger mass splittings between the lightest portal matter modes and the lightest massive SM gauge bosons can attained in warped extra dimensions, depending on which brane the symmetry-breaking orbifold boundary conditions are applied. This larger mass splitting will likely increase the branching fraction of these gauge bosons to portal matter fields and further weaken constraints from direct searches for Kaluza-Klein gauge boson searches in these models, even compared to the robust suppression we observe in our flat case. We shall find that this mass ordering will also generally hold for the portal matter fermions, which will have analogous orbifold boundary conditions to those observed for $I^\pm$.\footnote{Intuitively, this mass splitting is obvious for our flat construction: In the absence of brane-localized kinetic terms, a bulk wavefunction with mixed Dirichlet and von Neumann boundary conditions, like $I^\pm$ or our later fermionic portal matter fields, will have a sinusoidal solution which contains only half of its total wavelength on the interval, while a wavefunction with the same boundary conditions at both branes must have only sinusoidal solutions for which the wavelength is entirely contained in the interval. Since smaller-wavelength solutions correspond to greater momentum in the 5th dimension and hence greater Kaluza-Klein mass, it is clear that the mixed boundary conditions will lead to at least some state that is significantly lighter than any massive modes emerging from an otherwise equivalent state with the same boundary conditions on both branes.} This leads us to an important point: \emph{New physics from the portal matter-associated sector will generally enter our scenario at a significantly lower scale than new physics from SM Kaluza-Klein modes}.
In turn, we shall see that this has the potential to allow portal matter sectors from TeV-scale extra dimensions to probe larger compactification scales than searches which focus solely on the SM sector of a theory of large extra dimensions.

\subsection{Fermions}\label{section:fermions}

With our gauge sector fully accounted for, we can move on to discuss our treatment of fermions in the model, yielding both SM fermions and portal matter. Given our choice in the previous Section to make the SM Higgs an $SU(2)_D$ singlet localized on the $\phi=0$ brane, we can consider the minimal structure of bulk fermions (with portal matter) necessary to produce a light fermion which acquires a mass from the SM Higgs mechanism in this framework -- in this Section, we shall consider the scenario in which this minimal set of fields are the \emph{only} fermionic fields that propagate in the bulk; the extension to the case with additional bulk fermions, or to bulk fermions which aren't part of an $SU(2)_D$ multiplet, is trivial. We find that in order to arrive at a single massive SM fermion, we require two bulk fermion fields in the adjoint representation of $SU(2)_D$: An electroweak doublet $\mathbf{F}$ and an electroweak singlet $\mathbf{f}$. Under the SM gauge group, $\mathbf{F}$ will be in the representation of some electroweak doublet SM fermion, while $\mathbf{f}$ will be a corresponding electroweak singlet fermion with which $\mathbf{F}$ forms a Dirac fermion after spontaneous symmetry breaking: For leptonic bulk fermions, for example, we could assume that $\mathbf{F}$ is in the representation $(\mathbf{1},\mathbf{2})_{-1/2}$ of $SU(3)_c \times SU(2)_L \times U(1)_Y$, while $\mathbf{f}$ is in the representation $(\mathbf{1}, \mathbf{1})_1$ of the gauge group. Notably, because the SM Higgs is a singlet under $SU(2)_D$ and is localized at the $\phi=0$ brane, where the full $SU(2)_D$ symmetry is preserved, it is essential that\emph{both} $\mathbf{F}$ and $\mathbf{f}$ propagate in the bulk: If $\mathbf{F}$ were a bulk fermion and $\mathbf{f}$ were a chiral fermion localized on the $\phi=0$ brane, then for the SM Higgs to provide the appropriate mass term to the component of $\mathbf{F}$ with 0 $U(1)_D$ charge (that is, the SM fermion), either the SM Higgs or the brane-localized chiral fermion $\mathbf{f}$ would have to be in the adjoint representation of $SU(2)_D$. If the former were the case, we would introduce additional SM Higgs multiplets into our theory, while if the latter were the case, the theory would feature extra massless chiral fermions with non-zero $U(1)_D$ and SM charges. Neither of these outcomes is phenomenologically desirable, so we insist that both $\mathbf{F}$ and $\mathbf{f}$ are bulk fermions in the adjoint representation of $SU(2)_D$.

For practical computation, it is the most convenient to work in the basis of $U(1)_D$ charge eigenstates of both $\mathbf{F}$ and $\mathbf{f}$, that is, the three members of each $SU(2)_D$ triplet can be written
\begin{align}
    &\mathbf{F} = (F^0 ,F^+, F^-), &\mathbf{f} = (f^0, f^+, f^-)
\end{align}
The boundary conditions for $\mathbf{F}$ and $\mathbf{f}$ are then chosen to be consistent with the orbifold twist that breaks $SU(2)_D \rightarrow U(1)_D$ on the $\phi=\pi$ brane, and preserve light Kaluza-Klein modes only for those fields which are part of the SM-- that is, a left-handed component of $\mathbf{F}$ and a right-handed component of $\mathbf{f}$, both with 0 $U(1)_D$ charge. The appropriate boundary conditions are given in Table \ref{table:fermions}

\begin{table}[h!]
\centering
\begin{tabular}{| c | c | c |}
\hline
Fermion & $\phi=0$ BC & $\phi=\pi$ BC \\
\hline
$F^0_L$ & + & + \\
\hline
$F^0_R$ & - & - \\
\hline
$F^\pm_L$ & + & - \\
\hline
$F^\pm_R$ & - & + \\
\hline
$f^0_L$ & - & - \\
\hline
$f^0_R$ & + & + \\
\hline
$f^\pm_L$ & - & + \\
\hline
$f^\pm_R$ & + & - \\
\hline
\end{tabular}
\caption{The bulk fermions $F$ and $f$ and their boundary conditions (BCs) on the branes. A $+$ denotes a von Neumann BC, while a $-$ denotes a Dirichlet BC. Notably, only fields which possess a von Neumann ($+$) BC on both the $\phi=0$ and $\phi=\pi$ branes will have a zero mode. The gauge bosons are arranged in a basis of charge eigenstates of $U(1)_D$, the gauge symmetry that remains unbroken by the BCs.}
\label{table:fermions}
\end{table}

We can write the fermionic action of $F$ and $f$ as
\begin{align}\label{eq:fermionAction}
    S_{\textrm{ferm}} &= \int d^4 x \int_{0}^{\pi} d \phi \, R \bigg\{ \nonumber \\
    &\bigg[ i \overline{\mathbf{F}} \cdot \slashed{\partial} (\mathbf{1} + \omega_F P_L \mathbf{1} \delta(\phi) + \delta(\phi-\pi) \boldsymbol{\tau}^F) \mathbf{F} + i \overline{\mathbf{f}} \cdot \slashed{\partial} (\mathbf{1} + \omega_f P_R \mathbf{1} \delta(\phi) + \delta(\phi-\pi) \boldsymbol{\tau}^f) \mathbf{f}\bigg] \\
    &-\frac{1}{R} \bigg[ \overline{\mathbf{F}}_L \cdot \partial_\phi \mathbf{F}_R - \overline{\mathbf{F}}_R \cdot \partial_\phi \mathbf{F}_L + \overline{\mathbf{f}}_L \cdot \partial_\phi \mathbf{f}_R - \overline{\mathbf{f}}_R \cdot \partial_\phi \mathbf{f}_L\bigg] \nonumber \\
    &- \delta(\phi-\pi) \frac{v_D}{\sqrt{2}}\bigg( \overline{\mathbf{F}}_L \cdot \boldsymbol{\mathcal{Y}}^F_D \cdot \mathbf{F}_R + \overline{\mathbf{f}}_L \cdot \boldsymbol{\mathcal{Y}}^f_D \cdot \mathbf{f}_R + h.c.\bigg) -\delta(\phi) \bigg( \frac{y_h v}{\sqrt{2}} \overline{\mathbf{F}}_L \cdot \mathbf{f}_R + h.c. \bigg) \bigg\}, \nonumber
\end{align}
where $P_L(R)$ is the left(right)-handed chiral projection operator, and
\begin{align}\label{eq:yFDefs}
    \begin{matrix}
    \boldsymbol{\mathcal{Y}}^F_D \equiv \begin{pmatrix}
    0 & y^F_+ & y^F_-\\
    0 & 0 & 0\\
    0 & 0 & 0
    \end{pmatrix}, &\boldsymbol{\mathcal{Y}}^f_D \equiv \begin{pmatrix}
    0 & 0 & 0\\
    y^f_+ & 0 & 0\\
    y^f_- & 0 & 0
    \end{pmatrix},\\
    \boldsymbol{\tau}^F \equiv diag(\tau_{F0} P_L, \tau_{F+} P_R, \tau_{F-} P_R), & \boldsymbol{\tau}^f \equiv diag(\tau_{f0} P_R, \tau_{f+} P_L, \tau_{f-} p_L).
    \end{matrix}
\end{align}
%
Here, 
$\omega_{F,f}$, $\tau_{F0,F+,F-}$, and $\tau_{f0,f+,f-}$ are brane-localized kinetic terms at the $\phi=0$ and $\phi=\pi$ branes, respectively. Note that we have only included brane-localized kinetic terms on a given brane for fields which have von Neumann boundary conditions on that brane, as we did in Section \ref{section:gauges}. Also following the analogous logic in Section \ref{section:gauges}, we shall assume that $\omega_{F,f},\tau_{F0,F+,F-},\tau_{f0,f+,f-} \geq 0$ to avoid the emergence of tachyonic or ghost-like states in the Kaluza-Klein towers. Meanwhile,
$v_D$ is the dark Higgs vev, $v$ is the SM Higgs vev, and $y_{+,-}^{F,f}$ and $y_h$ are both brane-localized Yukawa couplings arising from Yukawa couplings to the dark Higgs $\Phi_D$ and to the SM Higgs, respectively-- we have absorbed a factor of $R$ into the dimensionful Yukawa couplings of the five-dimensional theory in order to render these couplings dimensionless. We recall that $v_D \sim O(1 \; \textrm{GeV})$ in order to explore the region of dark photon parameter space that motivates our model, and as we have no motivation not to do so, we shall assume that $y_{+,-}^{F,f}$ are of $O(1)$. Since $y_h v/\sqrt{2}$ corresponds to the SM fermion's mass, meanwhile, we can estimate its magnitude to be approximately equal to the mass of whichever SM fermion we are embedding in the bulk. Note that the $y_h$ term must respect the $SU(2)_D$ gauge symmetry of the theory, since it remains unbroken on the $\phi=0$ brane, but the $y_{+,-}^{F,f}$ terms, which emerge from the dark Higgs which breaks $U(1)_D$ on the $\phi=\pi$ brane, clearly do not. Since the $y_{+,-}^{F,f}$ terms will result in the mixing between the portal matter states and the SM states, these mass terms are also phenomenologically necessary in order to permit the portal matter states to decay; as such, it is essential that we select the $U(1)_D$ charge of the dark Higgs $\Phi_D$ such that these terms will appear in the action. As noted in Section \ref{section:gauges}, this is easily accomplished by assuming that $\Phi_D$ has a $U(1)_D$ charge equal in magnitude to that of the portal matter fermions $F^\pm$ and $f^\pm$, or equivalently, to that of the $U(1)_D$-charged gauge bosons $I^\pm$.
Finally, we note that in the interest of simplicity, the action in Eq.(\ref{eq:fermionAction}) omits fermion bulk mass terms, which are \emph{a priori} permitted in this interval theory. Because they are forbidden in the equivalent orbifold theory (and for our model to simplify the Higgs sector, we remind the reader that the orbifold is assumed to be a fundamental object here), we do not find this assumption unreasonable, and it significantly simplifies our later computation.\footnote{Fermion bulk mass terms can be reintroduced by mass terms with a non-trivial parity under the $Z_2 \times Z_2'$ orbifold, which can be introduced by parity-odd bulk scalars, as in, \eg \cite{Ahmed:2019zxm}. However, because introducing such terms ultimately relies on introducing these bulk scalars to the model, there is no fine-tuning issue associated with simply omitting the bulk masses entirely.}

Before performing a Kaluza-Klein decomposition of the fields in Eq.(\ref{eq:fermionAction}), we note that the brane mass coupling $y_h$ leads to mixing between the $F$ and $f$ fields, while the $y^{F,f}_{+,-}$ couplings mix the $U(1)_D$ charge eigenstates. Therefore in order to produce a diagonalized Kaluza-Klein tower, all fermion fields must be considered together. When performing a Kaluza-Klein decomposition, it is most convenient to borrow from the treatment of bulk fermions in \cite{Casagrande:2008hr,Casagrande:2010si}, which handled generic quark flavor mixing in a 5-dimensional model by solving for the bulk profiles as vectors in flavor space, and write the bulk profiles of these fermions as vectors in the space of $U(1)_D$ charge eigenstates. So, we write our Kaluza-Klein decomposition as
\begin{align}\label{eq:fermionKK}
    &\mathbf{F}_L = \frac{1}{\sqrt{R}} \sum_{k} \mathbf{C}^{F}_k (\phi) \cdot f^{(k)}_L (x), &\mathbf{F}_R = \frac{1}{\sqrt{R}} \sum_{k} \mathbf{S}^F_k(\phi) \cdot f^{(k)}_R (x),\\
    &\mathbf{f}_L = \frac{1}{\sqrt{R}} \sum_{k} \mathbf{S}^f_k(\phi) f^{(k)}_L (x), &\mathbf{f}_R = \frac{1}{\sqrt{R}} \sum_{k} \mathbf{C}^f_k(\phi) f^{(k)}_R (x), \nonumber
\end{align}
where $\mathbf{C}^{F,f}_k$ and $\mathbf{S}^{F,f}_k$ are 3-component vectors in the space of $U(1)_D$ charge eigenstates, given as
\begin{align}\label{eq:fermionKKVectors}
\begin{matrix}
    \mathbf{C}^{F}_k (\phi) \equiv \begin{pmatrix}
    C^{F^0}_{k}(\phi)\\
    c^{F^+}_{k}(\phi)\\
    c^{F^-}_{k}(\phi)
    \end{pmatrix}, &\mathbf{S}^F_k(\phi) \equiv \begin{pmatrix}
    s^{F^0}_{k}(\phi)\\
    S^{F^+}_{k}(\phi)\\
    S^{F^-}_{k}(\phi)
    \end{pmatrix}, &\mathbf{S}^f_k(\phi) \equiv \begin{pmatrix}
    s^{f^0}_{k}(\phi)\\
    S^{f^+}_{k}(\phi)\\
    S^{f^-}_{k}(\phi)
    \end{pmatrix}, &\mathbf{C}^f_k(\phi) \equiv \begin{pmatrix}
    C^{f^0}_{k}(\phi)\\ 
    c^{f^+}_{k}(\phi)\\
    c^{f^-}_{k}(\phi)
    \end{pmatrix}.
\end{matrix}
\end{align}
The notation we have developed here, in particular the bulk profile symbols in Eq.(\ref{eq:fermionKKVectors}), merits some explanation. First, as stated before, each fermion is expressed as a three-dimensional vector of bulk profiles with different $U(1)_D$ charges. A bulk profile subject to von Neumann (+) boundary conditions on the $\phi=0$ brane is denoted by a $C$ or a $c$, while a profile subject to Dirichlet boundary conditions on this brane is denoted by an $S$ or an $s$. The boundary conditions on the $\phi=\pi$ brane are indicated by the case of the letter: A $C$ or an $S$ indicates that the bulk profile has von Neumann (+) boundary conditions on the $\phi=\pi$ brane, while a $c$ or an $s$ indicates that the bulk profile has Dirichlet (-) boundary conditions on this brane. Furthermore, we note that the Kaluza-Klein index $k$ treats the Kaluza-Klein modes arising from the fields $F^{0,+,-}$ and $f^{0,+,-}$ as a \emph{single} tower of states-- to emphasize this fact here we have used the letter $k$ to denote the Kaluza-Klein index rather than our usual choice of $n$. Later on it will benefit us to consider this single tower as a perturbation of 6 separate towers (one for each of the three $F$ and $f$ $U(1)_D$ charge eigenstates); when we do so we shall note this minor abuse of notation explicitly.
\begin{table}[h!]
\centering
\begin{tabular}{| c | c | c |}
\hline
Profile & $\phi=0$ BC & $\phi=\pi$ BC \\
\hline
$C$ & + & + \\
\hline
$S$ & - & + \\
\hline
$c$ & + & - \\
\hline
$s$ & - & - \\
\hline
\end{tabular}
\caption{The $\phi=0$ and $\phi=\pi$ boundary conditions for the different classes of fermion bulk profile symbols used in Eqs.(\ref{eq:fermionKK}) and (\ref{eq:fermionKKVectors}), as described in the text. As in Tables \ref{table:gaugeBosons} and \ref{table:fermions}, a +(-) symbol denotes von Neumann (Dirichlet) boundary conditions.}
\label{table:fermionFuncs}
\end{table}

Under this Kaluza-Klein decomposition, a canonically normalized effective 4-dimensional action will be achieved if
\begin{align}\label{eq:fermionNorm}
    \delta_{k l} = &\int_0^\pi d\phi \, \bigg\{ \mathbf{C}^{F,f}_k \cdot \mathbf{C}^{F,f}_l + \mathbf{S}^{f,F}_k \cdot \mathbf{S}^{f,F}_l + \delta(\phi) \omega^{F,f} \mathbf{C}^{F,f}_k \cdot \mathbf{C}^{F,f}_l \\
    &+ \delta(\phi-\pi) \mathbf{C}^{F,f}_k \cdot \boldsymbol{\tau}^{F,f}_0 \cdot \mathbf{C}^{F,f}_l + \delta(\phi-\pi) \mathbf{S}^{f,F}_k \cdot \boldsymbol{\tau}^{f,F}_\pm \cdot \mathbf{S}^{f,F}_k \bigg\}, \nonumber
\end{align}
where
\begin{align}\label{eq:tauDefs}
    \boldsymbol{\tau}_{F0,f0} \equiv \begin{pmatrix}
    \tau_{F0,f0} & 0 & 0\\
    0 & 0 & 0\\
    0 & 0 & 0
    \end{pmatrix}, \;\;\; \boldsymbol{\tau}_{F\pm,f\pm} \equiv \begin{pmatrix}
    0 & 0 & 0\\
    0 & \tau_{F+,f+} & 0\\
    0 & 0 & \tau_{F-,f-}
    \end{pmatrix}.
\end{align}
Meanwhile, the Kaluza-Klein tower modes will be mass eigenstates if the equations of motion
\begin{align}\label{eq:fermionEOMs}
    (\partial_\phi + \delta(\phi-\pi) \frac{v_D R}{\sqrt{2}} \boldsymbol{\mathcal{Y}}^F_D)\cdot \mathbf{S}^F_k + \delta(\phi) \frac{y_h v R}{\sqrt{2}} \mathbf{C}^f_k = R m^f_k (1 + \omega_F \delta(\phi) + \delta(\phi-\pi) \boldsymbol{\tau}^F_0) \cdot \mathbf{C}^F_k \nonumber\\
    (\partial_\phi - \delta(\phi-\pi) \frac{v_D R}{\sqrt{2}}\boldsymbol{\mathcal{Y}}^{F \dagger}_D)\cdot \mathbf{C}^F_k = - R m^f_k (1 + \delta(\phi-\pi) \boldsymbol{\tau}^F_\pm) \cdot \mathbf{S}^F_k\\
    (-\partial_\phi - \delta(\phi-\pi) \frac{v_D R}{\sqrt{2}} \boldsymbol{\mathcal{Y}}^f_D) \cdot \mathbf{C}^f_k = - R m^f_k (1 + \delta(\phi-\pi) \boldsymbol{\tau}^f_\pm) \cdot \mathbf{S}^f_k \nonumber\\
    (\partial_\phi - \delta(\phi-\pi) \frac{v_D R}{\sqrt{2}} \boldsymbol{\mathcal{Y}}^{f \dagger}_D)\cdot S^f_k - \delta(\phi) \frac{y_h v R}{\sqrt{2}} \mathbf{C}^F_k = - R m^f_k (1 + \omega_f \delta(\phi)+ \delta(\phi-\pi) \boldsymbol{\tau}^f_0) \cdot \mathbf{C}^f_k \nonumber
\end{align}
are satisfied, where $m^f_k$ is the mass of the $k^{\textrm{th}}$ Kaluza-Klein tower mode. At this point, we shall find it convenient to define a series of dimensionless quantities to replace the dimensionful mass scales $m^f_k$, $y_h v /\sqrt{2}$, and $y^{F,f}_{+,-} v_D / \sqrt{2}$ that arise in Eq.(\ref{eq:fermionEOMs}). We therefore define
\begin{align}
    \begin{matrix}
    x_k \equiv m^q_k R, &\mu_h \equiv \frac{y_h v R}{\sqrt{2}}, &\mu^{F,f}_{\pm} \equiv \frac{y^{F,f}_\pm v_D R}{\sqrt{2}}.
    \end{matrix}
\end{align}
Integrating the expressions in Eq.(\ref{eq:fermionEOMs}) near $\phi=0$ now gives the boundary conditions
\begin{align}\label{eq:fermion0BCs}
    \mathbf{S}^F_k (0) - x_k \omega_F \mathbf{C}^F_k(0) &= -\mu_h \mathbf{C}^f_k(0)\\
    \mathbf{S}^f_k(0) + x_k \omega_f \mathbf{C}^f_k(0) &= \mu_h \mathbf{C}^F_k (0) \nonumber
\end{align}
at the $\phi=0$ brane and
\begin{align}\label{eq:fermionPiBCsQ}
    s^{F^0}_k(\pi) + x_k \tau_{F0} C^{F^0}_k(\pi) &= \mu^F_+ S^{F^+}_k(\pi) + \mu^F_- S^{F^-}_k(\pi), \nonumber\\
    c^{F^+}_k (\pi) - x_k \tau_{F+} S^{F^+}_k(\pi) &= -\mu^F_+ C^{F^0}_k (\pi),\\
    c^{F^-}_k (\pi) - x_k \tau_{F-} S^{F^-}_k(\pi) &= -\mu^F_- C^{F^0}_k (\pi) \nonumber,
\end{align}
and
\begin{align}\label{eq:fermionPiBCsq}
    s^{f^0}_k (\pi) - x_k \tau_{f0} C^{f^0}_k (\pi) &= -\mu^f_+ S^{f^+}_k (\pi) -\mu^f_- S^{f^-}_k (\pi), \nonumber\\
    c^{f^+}_k (\pi) + x_k \tau_{f+} S^{f^+}_k (\pi) &= \mu^f_+ C^{f^0}_k(\pi),\\
    c^{f^-}_k (\pi) + x_k \tau_{f-} S^{f^-}_k (\pi) &= \mu^f_- C^{f^0}_k(\pi), \nonumber
\end{align}
at the $\phi=\pi$ brane.
%
We may solve the differential equations in Eq.(\ref{eq:fermionEOMs}) in the bulk, identify mass eigenstates using the coupled boundary conditions of Eqs.(\ref{eq:fermion0BCs}), (\ref{eq:fermionPiBCsQ}), and (\ref{eq:fermionPiBCsq}), and normalize the result using the conditions of Eq.(\ref{eq:fermionNorm}). Of course, in practice solving this system exactly will be cumbersome. However, because the brane mass terms $M_h$ and $M^{F,f}_{+,-}$ are much smaller than the compactification scale $R^{-1}$ (or in other words, $\mu_h, \mu^{F,f}_{+,-} \ll 1$), finding the exact solution will be entirely unnecessary-- instead, we can work perturbatively to $O(\mu_h, \mu^{F,f}_{+,-})$. Because the $\mu_h$ terms mix electroweak doublet and singlet fields, and the $\mu^{F,f}_{+,-}$ terms mix fields of different $U(1)_D$ charges, we see that in the perturbative limit, the massive modes of the fermion Kaluza-Klein tower will consist of six \emph{separate} towers of vector-like $U(1)_D$ and electroweak eigenstates, which are mixed at $O(\mu_h, \mu^{F,f}_{+,-})$. For the sake of completeness, we shall present the results for all six towers of heavy Kaluza-Klein states, plus the light state that we shall identify with the SM fermion, up to and including $O(\mu_h, \mu^{F,f}_{+,-})$ corrections. For convenience, we shall define the functions
\begin{align}\label{eq:FGdef}
\begin{matrix}
    \alpha_x (\omega, \phi) \equiv x \omega \cos (x \phi) + \sin (x \phi),\\
    \beta_x (\omega,\phi) \equiv \bigg( \cos (x \phi) - x \omega \sin ( x \phi)\bigg).
\end{matrix}
\end{align}
With these definitions, we can write the approximate mass eigenstates forthe heavy modes of all six of the the fermion Kaluza-Klein towers, which we shall label based on what electroweak ($F$ for weak isospin doublet fermions and $f$ for weak isospin singlets) and $U(1)_D$ eigenstate (a superscript of $0$, $+$, or $-$ for a $U(1)_D$ charge of $0$, $+1$, or $-1$, respectively) serves as the dominant component of the mixed state. For the sake of brevity, here we shall present only the results for the weak isospin doublet tower with $U(1)_D$ equal to $+1$ (in our notation, $F^+$),\footnote{Much like the $SU(2)_D$ gauge bosons, we find that the SM fermion Kaluza-Klein towers that lack a zero-mode will have a lowest-lying Kaluza-Klein mode that is significantly lighter than those which have zero-modes, and so the lightest modes of the $F^\pm$ and $f^\pm$ towers will generally be much more phenomenologically relevant than those of the $F^0$ and $f^0$ towers.} with the results for the full list of towers included in Appendix \ref{appendix:FermionTowers}.

\begin{align}\label{eq:Q+Tower}
    &\mathbf{F^+:} \nonumber\\
    &\begin{matrix}
        \mathbf{C}^F_n = \begin{pmatrix}\frac{\mu^F_+}{1 + x_n^2 \tau_{F0} \tau_{F+}}\\ 1\\ 0 \end{pmatrix} N^{F_+}_n \beta_{x_n} (\omega_F, \phi), & \mathbf{S}^F_n (\phi) = \begin{pmatrix} \frac{\mu^F_+}{1 + x_n^2 \tau_{F0} \tau_{F+}}\\ 1\\ 0 \end{pmatrix} N^{F_+}_n \alpha_{x_n} (\omega_F, \phi),\\
        \mathbf{C}^f_n (\phi) = -\begin{pmatrix} 0\\ 1\\ 0\end{pmatrix} \frac{\mu_h N^{F_+}_n \alpha_{x_n} (\tau_{f+}, \pi-\phi)}{\beta_{x_n} (\omega_f,\pi)-xn \tau_{f+} \alpha_{x_n} (\omega_f,\pi)}, & \mathbf{S}^f_n (\phi) = \begin{pmatrix} 0\\ 1\\ 0\end{pmatrix} \frac{\mu_h N^{F_+}_n \beta_{x_n} (\tau_{f+}, \pi-\phi)}{\beta_{x_n}(\omega_f,\pi) - x_n \tau_{f+} \alpha_{x_n} (\omega_f, \pi)}
    \end{matrix},\\
    &N^{F_+}_n = \bigg[ \frac{2 (1 + x_n^2 \tau_{F+}^2)}{\pi (1 + x_n^2 \omega_F^2) (1 + x_n^2 \tau_{F+}^2)+ (\tau_{F+} + \omega_F)(1 + x_n^2 \tau_{F+} \omega_F)} \bigg]^{1/2},  \nonumber\\
    &x_n \; \textrm{such that} \;\; \beta_{x_n} (\omega_F,\pi) = x_n \tau_{F+} \alpha_{x_n} (\omega_F,\pi), \nonumber
\end{align}

It is important to note that the expressions in Eqs.(\ref{eq:Q+Tower}) (and the others appearing in Appendix \ref{appendix:FermionTowers}) are predicated on the assumption that the theory does \emph{not} possess very near degeneracy between the weak isospin singlet brane localized kinetic terms $\omega_f, \, \tau^f_{+,0,-}$ and the weak isospin doublet terms $\omega_F, \, \tau^F_{+,0,-}$-- otherwise the mixing between electroweak singlet and doublet towers will become large, since, for example, the expression $(\alpha_{x_n}(\omega_f,\pi) + x_n \tau_{f0} \beta_{x_n} (\omega_f, \pi))^{-1}$ in Eq.(\ref{eq:Q0Tower}) would diverge. In the limit of complete singlet-doublet brane term degeneracy, the mass eigenstates of the Kaluza-Klein towers would in fact become equal combinations of $SU(2)_L$ singlet and doublet fields, with dramatic implications for the chirality structure of couplings between Kaluza-Klein tower fermions and the light SM fermions. While this scenario might have some intriguing phenomenological consequences, we do not consider it any further here, both because the degeneracy must be numerically very close in order to render our expressions invalid, and because we shall find that for our particle content, such an exact (or near-exact) symmetry would also severely reduce the kinetic mixing generated by the portal matter, thus defeating the entire purpose of our original model building.

Having found the bulk wave functions for the \emph{massive} fermions, we have not yet fully determined the bulk wave functions for all allowable Kaluza-Klein modes of the fermion towers-- the expressions above only apply for the Kaluza-Klein modes which have mass in the limit of $\mu_h, \mu^{F,f}_{+,-} \rightarrow 0$, but for the both the $F^0$ and $f^0$ there exists a single massless mode in the unperturbed system. With the introduction of the brane mass terms, these two massless chiral fermions will form a single Dirac fermion with mass of $O(\mu_h)$, which serves as our SM fermion. To find the bulk profiles and the mass of the SM fermion, we can expand solutions to Eqs.(\ref{eq:fermionEOMs}) and (\ref{eq:fermion0BCs}-\ref{eq:fermionPiBCsq}) in the limit of small $x_n$ and brane mass terms. Up to linear order in $\mu_h$ and $\mu^{F,f}_{+,-}$, we have
\begin{align}\label{eq:qZeroModeTower}
&\begin{matrix}
    \mathbf{C}^{F}_{0} (\phi) = \begin{pmatrix} 1\\ -\mu^F_+\\ -\mu^F_- \end{pmatrix}\frac{1}{\sqrt{\pi + \omega_F + \tau_{F0}}}, &\mathbf{S}^{F}_{0} (\phi) = \begin{pmatrix} 1\\ 0\\ 0 \end{pmatrix}\frac{x_0 (\tau_{F0} + \pi- \phi)}{\sqrt{\pi + \omega_F + \tau_{F0}}},\\
    \mathbf{C}^{f}_0 (\phi) = \begin{pmatrix} 1\\ \mu^f_+\\ \mu^f_- \end{pmatrix} \frac{1}{\sqrt{\pi + \tau_{f0} + \omega_f}}, &\mathbf{S}^{f}_0 (\phi) = \begin{pmatrix} 1\\ 0\\ 0\end{pmatrix} \frac{x_0 (\pi + \tau_{f0} - \phi)}{\sqrt{\pi + \tau_{f0} + \omega_f}},
\end{matrix}\\
&x_0 = \frac{\mu_h}{\sqrt{\pi + \tau_{F0} + \omega_F}\sqrt{\pi + \tau_{f0} + \omega_f}} \nonumber.
\end{align}

We end our list of useful fermion-related formulas in our model by briefly noting that much as in the case of the SM gauge bosons, there will arise computations in our phenomenological explorations that will require summing over entire Kaluza-Klein towers of fermions. In particular, we shall need to compute the sums over an exchange of a complete tower of portal matter fields, such as $F^+$. In the case of the $F^+$ tower, we find the sum identities
\begin{align}\label{eq:FermionSums}
    \sum_n \frac{(N^{F_+}_n)^2 \beta_{x_n}(\omega_F,\phi_1)\beta_{x_n}(\omega_F,\phi_2)}{p^2 - x_n^2 R^{-2}} = \frac{R^2 \beta_{p R} (\omega_F, \phi_<) \alpha_{p R}(\tau_{F+},\pi-\phi_>)}{p R \big[- \beta_{p R} (\omega_F,\pi) + p R \tau_{F+} \alpha_{p R} (\omega_F,\pi) \big]}, \nonumber\\
    \sum_n \frac{(N^{F_+}_n)^2 \alpha_{x_n}(\omega_F,\phi_1)\alpha_{x_n}(\omega_F,\phi_2)}{p^2 - x_n^2 R^{-2}} = \frac{R^2 \alpha_{p R}(\omega_F, \phi_<) \beta_{p R} (\tau^F,\pi-\phi_>)}{p R \big[ -\beta_{p R}(\omega_F,\pi) + p R \tau_{F+} \alpha_{p R}(\omega_F, \pi) \big]},\\
    \sum_n \frac{x_n R^{-1} (N^{F_+}_n)^2 \beta_{x_n}(\omega_F,\phi_1)\alpha_{x_n}(\omega_F,\phi_2)}{p^2 - x_n^2 R^{-2}} = \frac{R\left\{ \begin{array}{lr}
        \alpha_{p R}(\tau_{F+},\pi-\phi_1) \alpha_{p R}(\omega_F, \phi_2), & \phi_1>\phi_2\\
        \beta_{p R}(\omega_F,\phi_1) \beta_{p R} (\tau_{F+}, \pi-\phi_2), & \phi_1 \leq \phi_2
    \end{array} \right\}}{\big[ -\beta_{p R}(\omega_F,\pi) + p R \tau_{F+} \alpha_{p R}(\omega_F, \pi) \big]}, \nonumber
\end{align}
with $p$, $\phi_<$, and $\phi_>$ defined as in Eq.(\ref{eq:ZWBosonSum}).

Finally, we can conclude with a brief note on the relative masses of the various fermion Kaluza-Klein tower modes arising in the model. Just as was the case for the gauge bosons in Section \ref{section:gauges}, we note that the fermions which lack a zero mode (the portal matter) will have significantly lighter lowest-lying massive Kaluza-Klein modes than those with zero modes (the SM) due to differing boundary conditions. The situation is now somewhat complicated here because both the SM and portal matter modes are permitted to have brane-localized kinetic terms, but the principle observed in Section \ref{section:gauges} remains: The portal matter modes will appear at a significantly lower scale than the Kaluza-Klein modes associated with the SM.

\section{Kinetic Mixing from Kaluza-Klein Fermions}\label{section:KineticMixing}

\begin{figure}
\centering
\includegraphics[width=3.5in]{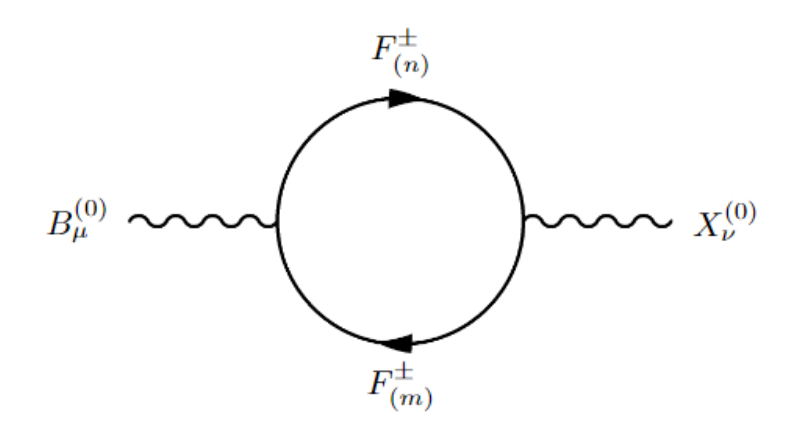}
\caption{The vacuum polarization-like Feynman graph which contributes to the kinetic mixing of the dark photon $X$ with the SM hypercharge gauge boson $B$, from exchange of the $n^{\textrm{th}}$ and $m^{\textrm{th}}$ Kaluza-Klein modes of a portal matter fermion, $F^+$ or $F^-$}
\label{fig2}
\end{figure}

With the information in Section \ref{section:modelSetup}, we are now poised to compute how the Kaluza-Klein fermions contribute to kinetic mixing. The computation of radiative corrections in a 5D theory is, of course, highly non-trivial, since the full theory itself is not renormalizable. Fortunately, for our purposes we are concerned with the behavior of the theory at \emph{low} energies, namely those which are well below the compactification scale $R^{-1}$. In this case, we can work in the effective four-dimensional theory, and are only concerned with the mixing of the massless (or in the case of the dark photon, light) modes of the SM hypercharge and the dark photon Kaluza-Klein towers-- the heavier modes are inaccessible.\footnote{Formally speaking, we're working in the limit that $p^2 \ll R^{-2}$, where $p$ is the 4-momentum of the external gauge fields in Figure \ref{fig2}. In this case, the full 5D theory's radiative correction to its propagator will be equivalent to our massless-mode only approximation-- up to $O(p^2 R^{2})$ corrections.} Assuming that kinetic mixing vanishes at some high, $O(R^{-1})$ energy (which it should in our model, because at higher scales the dark gauge symmetry is restored to $SU(2)_D$ and therefore can't mix with the SM hypercharge), our only task remains to compute the kinetic mixing between the massless dark photon $X^{(0)}$ and SM hypercharge boson $B^{(0)}$ emerging from the exchange of the infinite Kaluza-Klein tower of portal matter fermions, for example the $F^+$ or $F^-$ fields in our model. In doing this computation, we should then be able to determine if our naive condition for ensuring that kinetic mixing remains finite and calculable, $\sum_i Q_{D_i} Q_{Y_i} = 0$, remains sufficient if there are an infinite number of portal matter fermions.


Considering only the kinetic mixing between the hypercharge and $U(1)_D$ massless modes simplifies our work substantially-- in particular, before spontaneous symmetry breaking, fermion states with different $U(1)_D$ charges don't mix, and both $B^{(0)}$ and $X^{(0)}$ have perfectly flat profiles in the bulk. As a result, normalization of the fermion bulk profiles means that for a tower of portal matter fields with SM hypercharge $Q_Y$ and $U(1)_D$ charge $Q_D$, the diagram in Figure \ref{fig2} will give a vacuum polarization contribution of
\begin{align}\label{eq:ZeroModeVacuumPolarization}
    i \Pi_{\mu \nu}(p^2) = -4 g_D g_Y Q_Y Q_D \sum_n \int \frac{d^4 k}{(2 \pi)^4} \frac{(k + p)_\mu k_\nu + (k + p)_\nu k_\mu - g_{\mu \nu} (k^2 + k \cdot p - m_n^2 )}{((k + p)^2 - m_n^2)(k^2 - m_n^2)} + O(p^2 R^2),
\end{align}
where $m_n$ is the mass of the $n^{\textrm{th}}$ portal matter Kaluza-Klein mode, $k$ is a loop 4-momentum, $p$ is the external momentum of the gauge bosons. For clarity, we have included an $O(p^2 R^2)$ term to remind the reader that the mixing of only zero modes will only remain accurate up to corrections from the Kaluza-Klein tower modes of higher-energy gauge bosons. The coupling constants are $g_D$ for the dark photon and $g_Y$ for the SM hypercharge. We note that as long as the $O(p^2 R^2)$ terms are neglected, the results here are entirely independent of the specific properties of the bulk gauge field-- in particular, we might have introduced brane-localized kinetic terms to the gauge fields $B$ or $X$ without altering Eq.(\ref{eq:ZeroModeVacuumPolarization}). Using Feynman parameterization and a universal shift of the integration variable, this becomes
\begin{align}\label{eq:originalKMIntegral}
    i \Pi_{\mu \nu}(p^2) = &-4 i g_D g_Y Q_Y Q_D \int_0^1 dx \int \frac{d^{D_4} l_E}{(2 \pi)^{D_4}} \sum_n \bigg[ p_\mu p_\nu u(m_n) + g_{\mu \nu} w(m_n)  \bigg] + O(p^2 R^2),\\
    &u(m_n) \equiv \frac{-2 x(1-x)}{(l_E^2+m_n^2 - x(1-x)p^2)^2}, \;\;\; w(m_n) \equiv \frac{-\big(\frac{2}{D_4}-1\big) l_E^2 +m_n^2 + x(1-x)p^2}{(l_E^2+m_n^2 - x(1-x)p^2)^2}, \nonumber
\end{align}
%
where $l_E$ is a loop momentum (after Wick rotation, so that the integral is in 4-dimensional Euclidean space), and we have taken the liberty of writing the dimensionality of the noncompact spacetime as a variable, $D_4$, in anticipation of our later use of dimensional regularization. We have also defined the shorthand functions $u(m_n)$ and $w(m_n)$ to represent this integrand. Overall, the expression in Eq.(\ref{eq:originalKMIntegral}) is simply equivalent to a textbook computation of vacuum polarization in, for example, QED, albeit in a theory with a compactified extra dimension. In Eq.(\ref{eq:originalKMIntegral}), there are two forms of divergence: First, each integral in the sum diverges, just as in a4-dimensional  theory. Second, the fact that there are an \emph{infinite} number of fermion Kaluza-Klein modes over which we are summing adds new divergences. To regulate these divergences appropriately, we can turn to a dimensional regularization strategy such as that outlined in \cite{GrootNibbelink:2001bx}. In this setup, the infinite sum in Eq.(\ref{eq:originalKMIntegral}) can be replaced by a contour integral in the complex plane of $p_5$ over an infinitesimally thin clockwise contour surrounding the entire real line, using the method of residues. This yields, for example,
\begin{align}\label{eq:sumToResidue}
    \sum_n \int \frac{d^{D_4} l_E}{(2 \pi)^D_4} u(m_n) = -\frac{1}{2}\frac{1}{2 \pi i} \int_{\rightleftharpoons} d p_5 \int \frac{d^{D_4} l_E}{(2 \pi)^{D_4}} \mathcal{P}(p_5) u(p_5),
\end{align}
where $\mathcal{P}(p_5)$ is a ``pole function''. The pole function $\mathcal{P}(p_5)$ is constructed such that it only has poles at each $p_5=m_n$, it has the residue 1 at all of its poles, and for $p_5 \rightarrow \infty$ with a positive (negative) imaginary part of $p_5$, $\mathcal{P}(p_5) \rightarrow -(+) i r$, where $r$ is some constant. Because the initial sum only includes positive masses, we have introduced an additional factor of $\frac{1}{2}$ when moving to the complex integral over $p_5$ in Eq.(\ref{eq:sumToResidue})-- because the pole function and the functions $u$ and $w$ that we wish to integrate via this method are symmetric under $p_5 \rightarrow -p_5$, this factor is equivalent to ignoring the negative-$p_5$ poles. We should also note that by invoking the method of residues here, we have implicitly assumed that the function over which we are summing (in Eq.(\ref{eq:sumToResidue}) $u(m_n)$) has no poles in on the real line-- for some values of $l_E$ (namely $l_E^2<x(1-x)p^2$), this is not the case for either $u(m_n)$ or $w(m_n)$. In practice, we can introduce a small parameter $\delta$ to displace these poles by $\pm i \delta$ into the complex plane, and then take the limit as $\delta \rightarrow 0$ in the end. We shall not show this displacement explicitly here, because the specifics of our prescription have no bearing on our final results.

We can deform the contour above to $\ominus$, which we define as two counterclockwise semicircles, one in the upper half-plane (and infinitesimally above the real axis) and the other in the lower half-plane (infinitesimally below the real axis)-- as long as the integrand vanishes at $|p_5| \rightarrow \infty$, this contour integral will be equivalent to our cigar-shaped contour around the real line. By introducing dimensional regulators for 4-dimensional momentum \emph{and} $p_5$, we can make this deformation valid for divergent sum-integrals such as we encounter here, by taking the dimensionality of the compact and noncompact dimensions to be small enough. In the end, our computation of the integral in Eq.(\ref{eq:originalKMIntegral}) will have the form,
\begin{align}\label{eq:KMIntegral}
    i \Pi_{\mu \nu}(p^2) = -\frac{-4 i g_D g_Y Q_Y Q_D}{4 \pi i} \int_0^1 dx \int_\ominus d^{D_5} p_5 \int \frac{d^{D_4} l_E}{ (2 \pi)^{D_4}} \mathcal{P}(p_5) (u(p_5)p_\mu p_\nu + w(p_5) g_{\mu \nu}) + O(p^2 R^2),
\end{align}
where $D_4$ and $D_5$ are the dimensionalities of the noncompact and compact parts of the spacetime, respectively. When writing the integral over $p_5$, we note that we have defined
\begin{align}
    \int_\ominus d^{D_5} p_5 \mathcal{P}_(p_5) f(p_5) \equiv \frac{\pi^{D_5/2}}{\Gamma(D_5/2)}\int_{-\infty}^{\infty} \big( f(p_5) \mathcal{P} (p_5) + e^{i D_5} f(e^{i \pi} p_5) \mathcal{P} (e^{i \pi} p_5) \big)\bigg( \frac{p_5}{\mu_5} \bigg)^{D_5-1} d p_5,
\end{align}
for any function $f(p_5)$. That is to say, when we use the dimensionality of the compact space as a regulator, we tacitly assume that it is small enough that the contributions from the upper arcs of the semicircles in the contour $\ominus$ vanish.

We can then evaluate the sum-integral of Eq.(\ref{eq:KMIntegral}) as long as we appropriately specify our pole function $\mathcal{P}(p_5)$ to reflect the mass spectrum of our portal matter states. Consulting the boundary conditions which govern Kaluza-Klein tower masses in, \eg, Eq.(\ref{eq:Q+Tower}) allows us to arrive at
\begin{align}
    \mathcal{P}(p_5) = R \bigg( \frac{- p_5 R (\pi(\tau+\omega) + 2 \tau \omega) \cot(\pi p_5 R) - (\pi(1-p_5^2 R^2 \tau \omega) + \tau + \omega)}{(1 - p_5^2 R^2 \tau \omega) \cot (\pi p_5 R) - p_5 R (\tau + \omega)} \bigg)
\end{align}
for a fermion with the boundary conditions of the portal matter states in our model, namely $F^\pm$ and $f^\pm$, with $\omega$ and $\tau$ specifying the $\phi=0$ and and $\phi=\pi$ brane-localized kinetic terms for the fermion field.
Using the identities
\begin{align}
    \cot (x) = -i \bigg( 1 + \frac{2}{e^{-2 i x}-1} \bigg) = i \bigg( 1 + \frac{2}{e^{2 i x}-1} \bigg),
\end{align}
is straightforward to verify that as $|p_5| \rightarrow \infty$ and $Im(p_5) = \pm \varepsilon$, for some $\varepsilon>0$, then $\mathcal{P}(p_5) \rightarrow \mp i \pi R$.

Having computed $\mathcal{P}(p_5)$, we are now equipped to compute the vacuum polarization integral in Eq.(\ref{eq:KMIntegral}). It is convenient to split $\mathcal{P}(p_5)$ into pieces which contain different divergences. We define
\begin{align}
    &\mathcal{P}_\pm (p_5) = \mp i \pi R + \sigma_\pm (p_5) + \rho_\pm (p_5), \nonumber\\
    &\sigma_\pm (p_5) \equiv \frac{R (-(\omega + \tau) \pm 2 i p_5 R \omega \tau)}{-p_5 R (\tau + \omega) \mp i (1 - p_5^2 R^2 \tau \omega)},\\
    &\rho_\pm (p_5) \equiv 2R \bigg( \frac{(\pi(1 + p_5^2 R^2 \tau^2)(1 + p_5^2 R^2 \omega^2)+\tau + \omega + p_5^2 R^2 \tau \omega (\tau + \omega))}{(\pm i + p_5 R \tau)(\pm i + p_5 R \omega)\big( p_5 R (\tau+\omega)(e^{\mp 2 i \pi p_5 R}-1) \mp i (-1 + p_5^2 R^2 \tau \omega)(e^{\mp 2 i \pi R} + 1)\big)} \bigg), \nonumber
\end{align}
where $\mathcal{P}_+(p_5)$ is the pole function in the upper half-plane, while $\mathcal{P}_- (p_5)$ is the pole function in the lower half-plane. We can now consider the contributions of these pieces of the pole function separately:
\begin{align}\label{eq:IntegralDefs}
    i \Pi_{\mu \nu}(p^2) = g_D g_Y Q_Y Q_D \big( I^{\pi R}_{\mu \nu} + I^{\sigma}_{\mu \nu} + I^{\rho}_{\mu \nu} \big) + O(p^2 R^2), \nonumber\\
    I^{\pi R}_{\mu \nu} \equiv -\frac{-4 i}{4 \pi i} \int_0^1 dx \int_\ominus d^{D_5} p_5 \int \frac{d^{D_4} l_E}{ (2 \pi)^{D_4}} (\mp i \pi R) (u (p_5) p_\mu p_\nu + w (p_5) g_{\mu \nu} ),\\
    I^{\sigma}_{\mu \nu} = -\frac{-4 i}{4 \pi i} \int_0^1 dx \int_\ominus d^{D_5} p_5 \int \frac{d^{D_4} l_E}{(2 \pi)^{D_4}} \sigma_\pm (p_5) (u (p_5) p_\mu p_\nu + w (p_5) g_{\mu \nu} ), \nonumber\\
    I^{\rho}_{\mu \nu} = -\frac{-4 i}{4 \pi i} \int_0^1 dx \int_\ominus d^{D_5} p_5 \int \frac{d^{D_4} l_E}{(2 \pi)^{D_4}} \rho_\pm (p_5) (u (p_5) p_\mu p_\nu + w (p_5) g_{\mu \nu} ), \nonumber
\end{align}
where we have slightly abused notation to imply that, for example, $\sigma_{\pm} (p_5)$ denotes a sum over the integral with $\sigma_+ (p_5)$ over the upper half-plane's contour and $\sigma_- (p_5)$ over the lower half-plane's contour. We shall find that separating our kinetic mixing into these pieces helps us identify different divergences: $I^{\pi R}_{\mu \nu}$ corresponds to the divergence we would find with an \emph{uncompactified} fifth dimension: This is a purely 5-D divergence that can be regulated by either $D_4$ or $D_5$. $I^{\sigma}_{\mu \nu}$ meanwhile corresponds to an additional divergence stemming from the brane-localized kinetic terms. Finally, $I^{\rho}_{\mu \nu}$ is convergent and will give only a finite contribution to kinetic mixing.

We now begin our kinetic mixing computation by considering the contribution of $I^{\pi R}_{\mu \nu}$ to the vacuum polarization. Including the integral along both the upper and lower half-planes, this contribution becomes
\begin{align}\label{eq:PiRDivergence1}
    I^{\pi R}_{\mu \nu} \equiv -\frac{4 i R \pi^{(D_4+D_5)/2}}{\Gamma \big( \frac{D_4}{2} \big) \Gamma \big( \frac{D_5}{2} \big) \mu_4^{D_4-4} \mu_5^{D_5-1}} \int_0^1 dx \int_0^\infty p_5^{D_5-1} d p_5 \int_0^\infty \frac{ l_E^{D_4-1} d l_E}{(2 \pi)^{D_4}} \big( 1 - e^{i \pi D_5}\big) (u(p_5)p_\mu p_\nu + w(p_5) g_{\mu \nu}),
\end{align}
where we have exploited the fact that the entire integrand is symmetric under $p_5 \rightarrow -p_5$ to limit our integral over $p_5$ to the interval $[0, \infty]$. Integrating over the loop momenta here eventually yields
\begin{align}\label{eq:PiRDivergence3}
    I^{\pi R}_{\mu \nu} &= (g_{\mu \nu} p^2 - p_\mu p_\nu ) I^{\pi R},\\
    I^{\pi R} &\equiv \int_0^1 dx \, \frac{2 i (1- e^{i \pi D_5}) \pi^{(D_4+D_5)/2}}{\mu_4^{D_4-4} \mu_5^{D_5-1}} \frac{(x(1-x)(g_{\mu \nu} p^2 - p_\mu p_\nu)) \Gamma(2-(D_4+D_5)/2)}{(-x(1-x)p^2)^{(4-D_4-D_5)}R^{D_4 + D_5 - 5}}. \nonumber
\end{align}
We note that our result for this correction respects 4-dimensional  gauge and Lorentz invariance, as it should given our choice of regulators. As an artifact of dimensional regularization, this expression is actually finite as $D_4 \rightarrow 4$ and $D_5 \rightarrow 1$. this occurs because this divergence is inherently five-dimensional in character: In odd numbers of total dimensions, dimensional regularization will always yield finite regulated integrals.
Next, we consider the contribution of the $\sigma_\pm$ piece of the pole function, given by
\begin{align}\label{eq:ResidueDivergence}
    I^{\sigma}_{\mu \nu} -\frac{-4 i}{4 \pi i} \int_0^1 dx \int_\ominus d p_5 \int \sigma_\pm (p_5) \frac{-2 x(1-x) p_\mu p_\nu + \big( \big(1-\frac{2}{D_4}\big) l_E^2 +p_5^2 + x(1-x)p^2\big) g_{\mu \nu}}{(l_E^2 + p_5^2 -x(1-x)p^2 )^2},
\end{align}
where $\sigma_+ (p_5)$ is used for the upper half-plane contour and $\sigma_- (p_5)$ is used for the lower half-plane contour. It is in principle possible to use the same dimensionally regularized treatment of the $p_5$ integral that we employed when finding $I^{\pi R}_{\mu \nu}$, but it's far simpler here to take the integral over the contour $\ominus$ when $D_5=1$ and regulate the divergence purely 4-dimensionally.\footnote{We could have actually done this for $I^{\pi R}_{\mu \nu}$ as well, since the expression we found only actually depends on the \emph{sum} $D_4 + D_5$ and therefore can be regulated by $D_4$ alone. Intuitively, this would just have corresponded to letting $D_4$ be small enough that after integration over the 4-dimensional momenta, the $p_5$ integrand vanishes at $|p_5| \rightarrow \infty$ and the arcs of the semicircles don't contribute to the integral over the contour $\ominus$. However, it was instructive to consider $I^{\pi R}_{\mu \nu}$ as a purely 5D divergence.} We can then perform the contour integral over $p_5$ using the method of residues: Examining the integrand, the upper and lower half-plane contours only enclose a single pole each: $p_5 = i\sqrt{ l_E^2 - x(1-x)p^2}$ in the upper half-plane and $p_5 = -i\sqrt{ l_E^2 -x(1-x)p^2}$ in the lower half-plane (as discussed earlier, these poles as written are \emph{real} when $l_E^2< x(1-x) p^2$, but by displacing the first pole by $+i \delta$ and the second by $-i \delta$, both poles remain inside their respective contours, and we can take $\delta \rightarrow 0$ at the end of our computation). Note that the functions $\sigma_+ (p_5)$ and $\sigma_- (p_5)$ only have poles in their respective contours if the brane-localized kinetic terms $\tau$ and/or $\omega$ are allowed to be negative, however as mentioned previously, we always assume that these terms are positive to avoid ghost-like or tachyonic instabilities.

Using the method of residues to compute the $p_5$ integral, we can then evaluate the resulting 4-dimensional momentum integral using conventional techniques in loop integration, arriving at
\begin{align}\label{eq:ISigmaFiniteD4}
    &I^\sigma_{\mu \nu} = (g_{\mu \nu} p^2 - p_\mu p_\nu ) I^{\sigma}\\
    &I^\sigma \equiv -\int_0^1 dx \, \frac{2^{2-d} i x (1-x) R^{4-D_4}}{\pi^{D_4/2} \mu_4^{D_4-4} (-x(1-x)p^2 R^2)^{-\frac{D_4}{2}}} \bigg\{ \frac{\tau^4 \Gamma \big( \frac{2-D_4}{2} \big)}{(-1-x(1-x)p^2 R^2 \tau^2)^2} \bigg( 1 + \frac{1}{x (1-x)p^2 R^2 \tau^2}\bigg) + \nonumber\\
    &\frac{\Gamma \big(\frac{3-D_4}{2} \big)}{\tau \sqrt{\pi} (-x(1-x)p^2 R^2)^{5/2}} \bigg[ (D_4-1) {}_2 F_1 \big( 1, \frac{3-D_4}{2}; \frac{3}{2}; -\frac{1}{x(1-x)p^2 R^2} \big)-2 {}_2 F_1 \big( 2, \frac{3-D_4}{2}; \frac{3}{2}; -\frac{1}{x(1-x)p^2 R^2} \big) \bigg]  + \tau \rightarrow \omega \bigg\}, \nonumber
\end{align}
where ${}_2 F_1 (a_1,a_2; a_3; z)$ is the Gauss hypergeometric function, and analytically continuing to $D_4$ values where this integral no longer converges. Letting $D_4 = 4-2 \varepsilon$ and expanding about $\varepsilon = 0$ then gives
\begin{align}\label{eq:ISigmaResult}
    I^{\sigma} = \int_0^1 dx \, \frac{i x (1-x) }{2 \pi^2} \bigg\{ \frac{1}{\varepsilon} + \bigg[ \gamma_E - 2 \log(4 \pi R^2 \mu_4^2) + \log \bigg( \frac{(1 + R\sqrt{-x(1-x)p^2} \tau)}{\tau}\bigg) + \log \bigg( \frac{ (1 + R\sqrt{-x(1-x)p^2} \omega)}{\omega} \bigg) \bigg] \bigg\},
\end{align}
where $\gamma_E$ is the Euler gamma constant. Finally, we can  complete our calculation of the kinetic mixing by finding $I^{\rho}_{\mu \nu}$. Unlike the previous components of the pole function, the $I^{\rho}_{\mu \nu}$ is convergent and therefore does not require any sort of ultraviolet regulator. Using the method of residues for the $p_5$ integral again, and setting $D_4=4$, we arrive at
\begin{align}\label{eq:IRhoResult}
    &I^{\rho}_{\mu \nu} = (g_{\mu \nu} p^2 - p_\mu p_\nu ) I^\rho,\\
    &I^\rho \equiv \int_0^1 dx \, \frac{i (g_{\mu \nu}p^2 - p_\mu p_\nu) x(1-x)}{2 \pi^2} \bigg\{ 2 \pi R\sqrt{-x(1-x)p^2} \nonumber\\
    &- \log \bigg( e^{2 \pi R\sqrt{-x(1-x)p^2} } + \frac{(-1+ R\sqrt{-x(1-x)p^2} \tau)(-1 + R\sqrt{-x(1-x)p^2} \omega)}{(1+ R\sqrt{-x(1-x)p^2} \tau)(1 + R\sqrt{-x(1-x)p^2} \omega)} \bigg) \bigg\}. \nonumber
\end{align}
We can combine these results to get the total contribution of a single portal matter Kaluza-Klein tower to the kinetic mixing of the dark photon and the hypercharge gauge boson. It is straightforward to then find the contribution to kinetic mixing by an ensemble of different portal matter fermions with dark charges $Q_{D_i}$, SM hypercharges $Q_{Y_i}$, $\phi=0$ brane-localized kinetic terms $\omega_i$, and $\phi=\pi$ brane-localized kinetic terms $\tau_i$. We find that the total contribution of such an ensemble of towers to the vacuum polarization diagram will be
\begin{align}\label{eq:PiMuNuFinal}
    i \Pi_{\mu \nu} (p^2) &= \frac{i g_D g_Y}{2 \pi^2} (g_{\mu \nu} p^2 - p_\mu p_\nu) \sum_i Q_{D_i} Q_{Y_i} \bigg( \mathcal{I}^{const} + \mathcal{I}^{var}(\tau_i, \omega_i) \bigg) + O(p^2 R^2),\\
    \mathcal{I}^{var} &\equiv \int_0^1 dx \, \bigg\{\mathcal\log \bigg( \frac{(1 + R\sqrt{-x(1-x)p^2} \tau_i)}{\tau_i}\bigg) + \log \bigg ( \frac{(1 + R\sqrt{-x(1-x)p^2} \omega_i)}{\omega_i} \bigg) \nonumber\\
    &- \log \bigg( e^{2 \pi R\sqrt{-x(1-x)p^2}} + \frac{(-1 + R\sqrt{-x(1-x)p^2} \tau_i)(-1 + R\sqrt{-x(1-x)p^2} \omega_i)}{(1 + R\sqrt{-x(1-x)p^2} \tau_i) (1 + R\sqrt{-x(1-x)p^2} \omega_i)}\bigg) \bigg\}, \nonumber
\end{align}
where $\mathcal{I}^{const}$ is simply a constant independent of the brane-localized kinetic terms. Then, we see that the usual 4-dimensional condition for finite and calculable kinetic mixing from portal matter, namely that $\sum_i Q_{D_i} Q_{Y_i} = 0$, continues to hold in the theory with a compact fifth dimension and an infinite Kaluza-Klein tower of portal matter fermions. There is, however, an additional condition that arises because of our inclusion of brane-localized kinetic terms. Specifically, we note that $\mathcal{I}^{const}$ contains terms which arise specifically from the divergence computed from $I^\sigma$, and we recall that this divergence arises from the brane-localized kinetic terms themselves. If $\tau_i = 0$ and/or $\omega_i = 0$ for some field in the portal matter ensemble, all or part of the contribution of this field to $\mathcal{I}^{const}$ will also vanish, and therefore not be cancelled in the summation over all portal matter fields. Careful inspection of the results of Eqs.(\ref{eq:ISigmaFiniteD4}) and (\ref{eq:ISigmaResult}) allows us to determine that both the $\phi=0$ brane term $\omega$ and the $\phi=\pi$ brane term $\tau$ have an equal divergent contribution to $\mathcal{I}^{const}$. As a result, if one portal matter field has $\tau_i = 0$ or $\omega_i = 0$, \emph{all} portal matter fields in an ensemble such that $\sum_i Q_{D_i} Q_{Y_i}$ must also have $\tau_i = 0$ or $\omega_i = 0$, respectively, in order to ensure a finite and calculable kinetic mixing.

Having established that our naive condition of Eq.(\ref{eq:finiteMixingCondition}) for finite and calculable kinetic mixing in a 4-dimensional theory only needs to be slightly modified in our 5-dimensional setting, we can now use our vacuum polarization result in Eq.(\ref{eq:PiMuNuFinal}) to actually compute the contribution to the kinetic mixing $\epsilon$ defined in Eq.(\ref{eq:EpsilonDef}) Working in the limit of $p^2 \ll R^{-2}$ in turn allows us to compute the kinetic mixing at low energies, arriving at
\begin{align}\label{eq:SimpleKM}
    \epsilon = -\frac{g_D g_Y c_w}{12 \pi^2} \sum_i Q_{D_i} Q_{Y_i} \big( \log (\tau_i) + \log(\omega_i) \big),
\end{align}
where we remind the reader that $c_w$ is the usual electroweak Weinberg angle. Our treatment in arriving at Eq.(\ref{eq:PiMuNuFinal}), and therefore the validity of the condition of Eq.(\ref{eq:finiteMixingCondition}) in a 5-dimensional theory, is of course heavily dependent on the specifics of our model construction. We have only calculated the kinetic mixing arising from bulk fermions embedded in a flat extra dimension without any bulk mass terms (and only negligibly small brane mass terms), and with specific orbifold parities. Therefore, our calculation may seem at first blush to be a highly specialized one, with little broader applicability to more general models in 5 dimensions. However, at the risk of losing some rigor, we can easily argue that while the analytical results for a computation done with such complicating factors as bulk and brane masses and a more general 5-dimensional geometry may differ from the result of Eq.(\ref{eq:SimpleKM}), our results suggest that such computations will remain \emph{finite and calculable}. Specifically, we note that our naive condition for finite and calculable kinetic mixing, Eq.(\ref{eq:finiteMixingCondition}), will only be insufficient to guarantee finite and calculable mixing in the 5-dimensional theory if the divergence associated with the infinite sum over Kaluza Klein modes is not cancelled in a given construction. Therefore, when determining whether Eq.(\ref{eq:finiteMixingCondition}) remains valid, we only need to reference the contribution of the Kaluza-Klein modes with masses greater than some high ultraviolet cutoff-- the contribution of whatever (finite) tower of states lie below that cutoff is of course finite if Eq.(\ref{eq:finiteMixingCondition}) is satisfied. However, we know that at sufficiently high energies, the effect of a bulk mass or a curvature of the metric shall vanish from the Kaluza-Klein spectrum, as the mass (or rather the 5-momentum) of the Kaluza-Klein modes greatly exceeds the scales of curvature and the bulk mass. So, since our toy model in flat space has precisely the same Kaluza-Klein spectrum in the high UV as a warped model and/or a model with fermion bulk and brane masses, we argue that the same cancellation of divergences occur for these constructions. A reader may also be concerned that our results are dependent on the fact that all fields mediating the kinetic mixing have mixed boundary conditions-- that is, von Neumann at one brane and Dirichlet at the other, with the different chiralities of the fermion having both boundary conditions flipped. It is unclear if the divergences from a Kaluza-Klein tower with the same boundary conditions at both branes will cancel with the divergences from a tower with mixed boundary conditions. However, we note that in realistic constructions with chiral fermions, the Kaluza-Klein tower for a fermion with the same boundary conditions at both branes will produce a chiral zero-mode, which is missing from a tower with mixed boundary conditions. So, as noted briefly in Section \ref{section:recipe}, if the same-boundary-condition fields don't cancel their contribution among themselves, the theory clearly won't satisfy Eq.(\ref{eq:finiteMixingCondition}) anyway. Meanwhile, sufficiently high up the tower of Kaluza-Klein states, the Kaluza-Klein spectrum of towers of fields with the same boundary conditions at both branes will simply match that of fields with mixed boundary conditions, albeit displaced by some finite amount. Since we know that the contribution to kinetic mixing for mixed-boundary-condition fields is finite and calculable if Eq.(\ref{eq:finiteMixingCondition}) is satisfied, we know that this is also true for fields with the same boundary conditions at both branes.

Based on these discussions, therefore, we can posit the following well-motivated condition for finite and calculable kinetic mixing in 5-dimensions: Kinetic mixing mediated by bulk fermions will remain finite and calculable at one loop if Eq.(\ref{eq:finiteMixingCondition}) is satisfied for all each set of fermion fields with the same boundary conditions, and sources of additional divergences (such as brane-localized kinetic terms) that are present in one field contributing to the kinetic mixing are present in all fields with the same boundary conditions. These conditions should hold for \emph{any} construction in five dimensions.

\section{Phenomenology}\label{section:Phenomenology}
To get a sense of the experimental signatures for Kaluza-Klein portal matter models, it is useful for us to do a brief survey of the collider phenomenology of our toy model. In order to proceed, we must of course specify how we shall embed the SM into the general model construction of Section \ref{section:modelSetup}. For simplicity, we assume that \emph{all} SM fermions are embedded in the 5-dimensional bulk (and therefore, gauge symmetry demands that we embed all SM gauge bosons in the bulk as well)-- in practical terms, this helps us evade constraints arising from the physics of the SM in an extra dimension: Any fermions localized on a brane would couple to Kaluza-Klein modes of SM gauge bosons a factor of $\sqrt{2}$ more strongly than they would couple to these gauge bosons' zero modes; a bulk embedding of the fermion can avoid this enhancement. Instead, the zero-modes of bulk fermions couple to the SM gauge boson Kaluza-Klein modes with a strength scaled by $\sqrt{2}(\tau_{F0} +(-1)^n \omega_F)/(\pi+\tau_{F0}+\omega_F)$ relative to the 4-dimensional gauge coupling (where $n$ is the Kaluza-Klein tower level of the gauge boson, or equivalently, $n R^{-1}$ is the Kaluza-Klein mode's mass): It is straightforward to demonstrate that this scaling factor will always be less than $\sqrt{2}$, and therefore the bulk fermions will always be less strongly coupled to the SM gauge boson Kaluza-Klein modes than their brane-localized counterparts. In fact, we can also see that the bulk coupling scaling factor here \emph{vanishes} when $\tau_{F0} = \omega_F$ (up to small corrections due to the effect of the brane-localized SM Higgs field on the gauge boson bulk profile); in this limit the theory has attained the ``KK parity'' observed in theories of Universal Extra Dimensions (UED) \cite{Appelquist:2000nn,Datta:2010us}.

In spite of embedding the entire SM fermion content in the bulk, we do \emph{not} assume that all SM fermions are part of $SU(2)_D$ multiplets that contain portal matter-- we can easily modify the bulk wavefunction results of Eqs.(\ref{eq:Q0Tower} \ref{eq:q-Tower}) in Appendix \ref{appendix:FermionTowers} and Eq.(\ref{eq:qZeroModeTower}) in Section \ref{section:fermions} for $SU(2)_D$ singlet fields by removing the $U(1)_D$-charged Kaluza-Klein towers from the expressions and setting the brane-localized mass terms associated with the dark Higgs to 0. To generate our kinetic mixing, we shall assume subsets of the SM fermions will be embedded in these portal multiplets such that the relevant brane mass terms and brane-localized kinetic terms featured in Section \ref{section:fermions} can be realized, \eg, a weak isospin doublet quark, a singlet up-like quark, and a singlet down-like quark might all be promoted to portal matter multiplets. Finally, in the name of simplicity and in order to avoid potentially catastrophic flavor constraints, we shall assume that brane-localized kinetic terms are flavor-universal, so that, for example, the electroweak singlet up, charm, and top quarks all share the same brane-localized kinetic terms. The fermion content of the model, along with the brane-localized kinetic terms for each field, is summarized in Table \ref{table:fermionModelContent}

\begin{table}[h!]
\centering
\begin{tabular}{| c | c | c | c | c | c |}
\hline
$SU(3)_c \times SU(2)_L \times U(1)_Y $ & $SU(2)_D$ & Copies & $Q_D$ & $\phi=0$ BLKT & $\phi=\pi$ BLKT \\
\hline
\multirow{3}{*}{ $(\mathbf{3},\mathbf{2})_{\frac{1}{6}}$} & \multirow{3}{*}{$\mathbf{3}$} & \multirow{3}{*}{$n_Q$} & $+1$ & \multirow{3}{*}{$\omega_Q$} & $\tau_{Q+}$ \\
& & & $0$ & & $\tau_{Q0}$\\
& & & $-1$ & & $\tau_{Q-}$\\
\hline
\multirow{3}{*}{ $(\mathbf{\overline{3}},\mathbf{1})_{-\frac{2}{3}}$} & \multirow{3}{*}{$\mathbf{3}$} & \multirow{3}{*}{$n_Q$} & $+1$ & \multirow{3}{*}{$\omega_u$} & $\tau_{u+}$ \\
& & & $0$ & & $\tau_{u0}$\\
& & & $-1$ & & $\tau_{u-}$\\
\hline
\multirow{3}{*}{ $(\mathbf{\overline{3}},\mathbf{1})_{\frac{1}{3}}$} & \multirow{3}{*}{$\mathbf{3}$} & \multirow{3}{*}{$n_Q$} & $+1$ & \multirow{3}{*}{$\omega_d$} & $\tau_{d+}$ \\
& & & $0$ & & $\tau_{d0}$\\
& & & $-1$ & & $\tau_{d-}$\\
\hline
\multirow{3}{*}{ $(\mathbf{1},\mathbf{2})_{-\frac{1}{2}}$} & \multirow{3}{*}{$\mathbf{3}$} & \multirow{3}{*}{$n_L$} & $+1$ & \multirow{3}{*}{$\omega_L$} & $\tau_{L+}$ \\
& & & $0$ & & $\tau_{L0}$\\
& & & $-1$ & & $\tau_{L-}$\\
\hline
\multirow{3}{*}{ $(\mathbf{1},\mathbf{1})_{1}$} & \multirow{3}{*}{$\mathbf{3}$} & \multirow{3}{*}{$n_L$} & $+1$ & \multirow{3}{*}{$\omega_e$} & $\tau_{e+}$ \\
& & & $0$ & & $\tau_{e0}$\\
& & & $-1$ & & $\tau_{e-}$\\
\hline
$(\mathbf{3},\mathbf{2})_{\frac{1}{6}}$ & $\mathbf{1}$ & $3-n_Q$ & 0 & $\omega_Q$ & $\tau_{Q0}$\\
\hline
$(\mathbf{3},\mathbf{1})_{-\frac{2}{3}}$ & $\mathbf{1}$ & $3-n_Q$ & 0 & $\omega_u$ & $\tau_{u0}$\\
\hline
$(\mathbf{3},\mathbf{1})_{\frac{1}{3}}$ & $\mathbf{1}$ & $3-n_Q$ & 0 & $\omega_d$ & $\tau_{d0}$\\
\hline
$(\mathbf{1},\mathbf{2})_{-\frac{1}{2}}$ & $\mathbf{1}$ & $3-n_L$ & 0 & $\omega_L$ & $\tau_{L0}$\\
\hline
$(\mathbf{1},\mathbf{1})_{1}$ & $\mathbf{1}$ & $3-n_L$ & 0 & $\omega_e$ & $\tau_{e0}$\\
\hline
\end{tabular}
\caption{The bulk fermion content of the model, together with the brane-localized kinetic terms (BLKT's) for each fermion field. Note that $n_Q$ and $n_L$ denote the number of copies of SM quarks and leptons that are embedded in $SU(2)_D$ multiplets, respectively. In total, there must be exactly three generations of chiral SM fermions in the effective 4-dimensional theory, so if $n_{Q(L)} < 3$ then there must be an additional number of bulk $SU(2)_D$ singlet fields to arrive at three generations of quarks (leptons). To avoid flavor constraints, brane-localized kinetic terms are assumed to be flavor-universal; note that this includes that $SU(2)_D$ singlet fields must have the same BLKT's as the $U(1)_D$-neutral component of the $SU(2)_D$ triplet field with the same SM representation. This is of course artificial, but might be enforced via some form of flavor symmetry, especially if all fermions of a given species are embedded in $SU(2)_D$ triplets and hence in the same representation under $\mathcal{G}_{SM} \times SU(2)_D$.}
\label{table:fermionModelContent}
\end{table}

Referencing Eq.(\ref{eq:SimpleKM}) and Table \ref{table:fermionModelContent}, we can see that the kinetic mixing realized in this setup will be
\begin{align}\label{eq:ModelKM}
    \epsilon = -\frac{g_D g_Y c_W}{12 \pi^2} \bigg\{ 3 n_Q \bigg[ \frac{1}{3}\log \bigg( \frac{\tau_{Q+}}{\tau_{Q-}}\bigg) -\frac{2}{3} \log \bigg( \frac{\tau_{u+}}{\tau_{u-}} \bigg) + \frac{1}{3} \log \bigg( \frac{\tau_{d+}}{\tau_{d-}} \bigg) \bigg] + n_L \bigg[ \log \bigg( \frac{\tau_{e+}}{\tau_{e-}} \bigg) -\log \bigg( \frac{\tau_{L+}}{\tau_{L-}} \bigg) \bigg] \bigg\},
\end{align}
where $n_{Q(L)}$ is the number of $SU(2)_D$ triplet quark(lepton) generations embedded in the bulk, while the various $\tau$ symbols denote the $\phi=\pi$ brane-localized kinetic term for different portal matter fields, as indicated in Table \ref{table:fermionModelContent}.

Our setup can be subject to a number of possible experimental constraints. Some of these are of secondary interest to us here: Probes of the underlying sub-GeV dark matter model are of limited utility here, since kinetic mixing is only logarithmically sensitive to the brane-localized kinetic terms, and a sub-GeV dark matter model model with virtually any realistic value of the dark coupling $g_D \sim O(1)$ might be devised to realize the observed relic abundance, provided that the sub-GeV dark matter parameter space is not disallowed completely. As mentioned briefly in Section \ref{section:gauges}, Higgs invisible decays provide significant constraints on mixing between the SM Higgs and the brane-localized dark Higgs that breaks the $U(1)_D$ subgroup of $SU(2)_D$, however by construction mixing between these operators is suppressed at tree level. Other Higgs observables associated with the extra dimension, such as tree-level modifications to SM Yukawa couplings and loop-level modifications to the Higgs production rate from gluon fusion, only enter at second order in $v R \ll 1$ and are hence well within current constraints \cite{ATLAS:2022ooq,CMS:2020djy}.\footnote{It should be noted that this situation may not hold in some warped constructions of the paradigm, where sizable corrections to these Higgs observables have been realized in Randall-Sundrum constructions \cite{Azatov:2010pf,Casagrande:2010si}.} Flavor-changing currents are highly suppressed as long as we assume flavor-universality among the brane-localized kinetic terms and no flavor-dependent fermion bulk masses, as mentioned earlier in this Section.
Our setup then has three major remaining classes of constraints: Collider searches for Kaluza-Klein modes of SM particles, precision electroweak constraints from embedding the SM electroweak gauge group in an extra dimension, and collider searches for portal matter fields and $SU(2)_D$ gauge bosons. the first two are present in any model with a large extra dimension, while the third is of course purely reliant on the physics of the new portal matter sector. We shall explore each of these constraints in turn, and examine how the phenomenology of the extra dimension is influenced by the existence of Kaluza-Klein portal matter.

\subsection{Conventional Extra Dimensions Searches}\label{section:SMExtraDimensions}
Because the SM-only constraints are well-studied and frequently harsh \cite{Csaki:2010az,Cheung:2001mq}, we shall begin our discussion of extra dimensions with a brief exploration of ``conventional'' signals of extra dimensions featuring only SM states or their Kaluza-Klein modes. In spite of the absence of portal matter or dark gauge bosons in the final states, however, we shall see that some of these signals will be significantly modified by the presence of the portal sector.

To start, we consider collider searches for SM Kaluza-Klein particles. In many models of extra dimensions, the Drell-Yan production of the heavy Kaluza-Klein modes of the $Z$ boson that then decay into SM leptons can provide a meaningful constraint on the masses of these modes; it is straightforward to estimate this constraint here. We note, however, that even though such searches involve only SM fields and their Kaluza-Klein modes, the existence of the portal matter fields will dramatically alter the expected signals from these results. Specifically, we note that in our construction, the lightest massive Kaluza-Klein mode of the $Z$ boson, which we shall refer to as $Z^{(1)}$, will have a mass of $m^Z_1 = R^{-1}$. However, we find that for any portal matter fermion with positive $\phi=0$ and $\phi=\pi$ brane-localized kinetic terms, the lightest Kaluza-Klein modes of its tower will have mass \emph{less} than $R^{-1}/2$-- the same mass difference pattern we observed for the $I^\pm$ gauge bosons relative to the Kaluza-Klein modes of the SM gauge bosons in Section \ref{section:gauges}. As a result, the Kaluza-Klein $Z$ boson always has new, non-SM decay channels that don't appear in a construction of extra dimensions without portal matter; for a portal matter mode with $U(1)_D$ charge of $\pm 1$, a $\phi=\pi$ brane-localized kinetic term of $\tau_{F\pm}$, a $\phi=0$ brane-localized kinetic term of $\omega_F$, and a mass of $m_{F\pm}$, the partial width of the lightest Kaluza-Klein $Z$ boson decaying into a pair of such fermions will be
\begin{align}\label{eq:ZKKPortalDecay}
    \Gamma_{Z^{(1)} \rightarrow F^{\pm} \overline{F}^{\pm}}  = (g_L^2+g_Y^2) (T^F_3 - s_w^2 Q^F)^2 \frac{R^{-1}}{12 \pi} \gamma_{Z^{(1)} \rightarrow F^{\pm} \overline{F}^{\pm}}, \nonumber\\
    \gamma_{Z^{(1)} \rightarrow F^{\pm} \overline{F}^{\pm}} \equiv \frac{2(4 (1+4 x_F^2) \omega_F \tau_{F \pm} x_F^2 A_{\tau_{F \pm}} A_{\omega_F} + 2 (\omega_F^2 A^2_{\tau_{F \pm}} + \tau_{F \pm}^2 A^2_{\tau_{F \pm}})(1 - x_F^2 - 4 x_F^4))}{(1- 4 x_F^2)^{\frac{1}{2}} [(\tau_{F \pm} + \omega_F)(1 + x_F^2 \tau_{F \pm} \omega_F)+ \pi A_{\tau_{F \pm}} A_{\omega_F}]^2},\\
    A_\tau \equiv 1 + x_F^2 \tau^2, \;\;\; x_F \equiv m_{F\pm} R, \nonumber
\end{align}
where $T_3^F$ and $Q^F$ represent the weak isospin and the electromagnetic charge of the portal matter fermion. Note that the above expression appears to diverge when $x_F \rightarrow 1/2$, \ie, when the decay width should vanish as the phase space vanishes, however, this is simply an artifact of the fact that $x_F$, $\tau_{F \pm}$, and $\omega_F$ are \emph{not} independent variables: $x_F$ is the root of a transcendental equation governed by the brane-localized kinetic terms. If the limit $\tau_{F \pm}, \omega_F \rightarrow 0$ (in which case $x_F \rightarrow 1/2$) is taken carefully, the decay width in Eq.(\ref{eq:ZKKPortalDecay}) will vanish.

\begin{figure}
    \centerline{\includegraphics[width=3.5in]{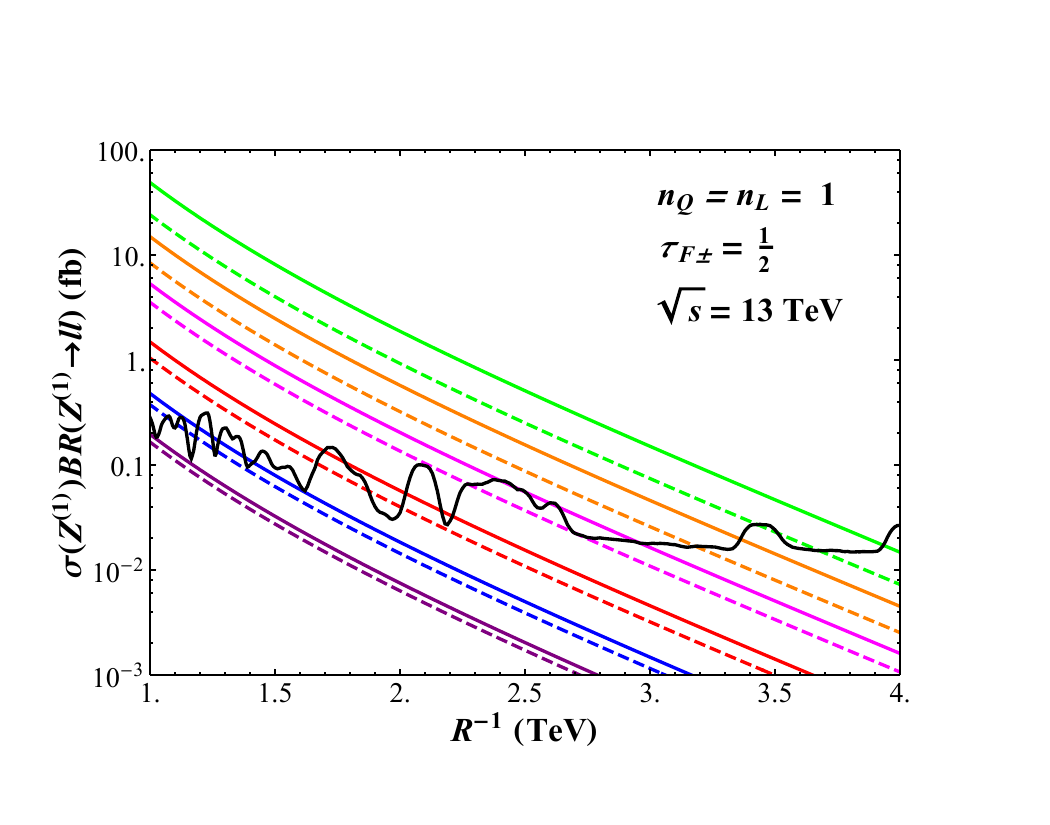}
    \hspace{-0.75cm}
    \includegraphics[width=3.5in]{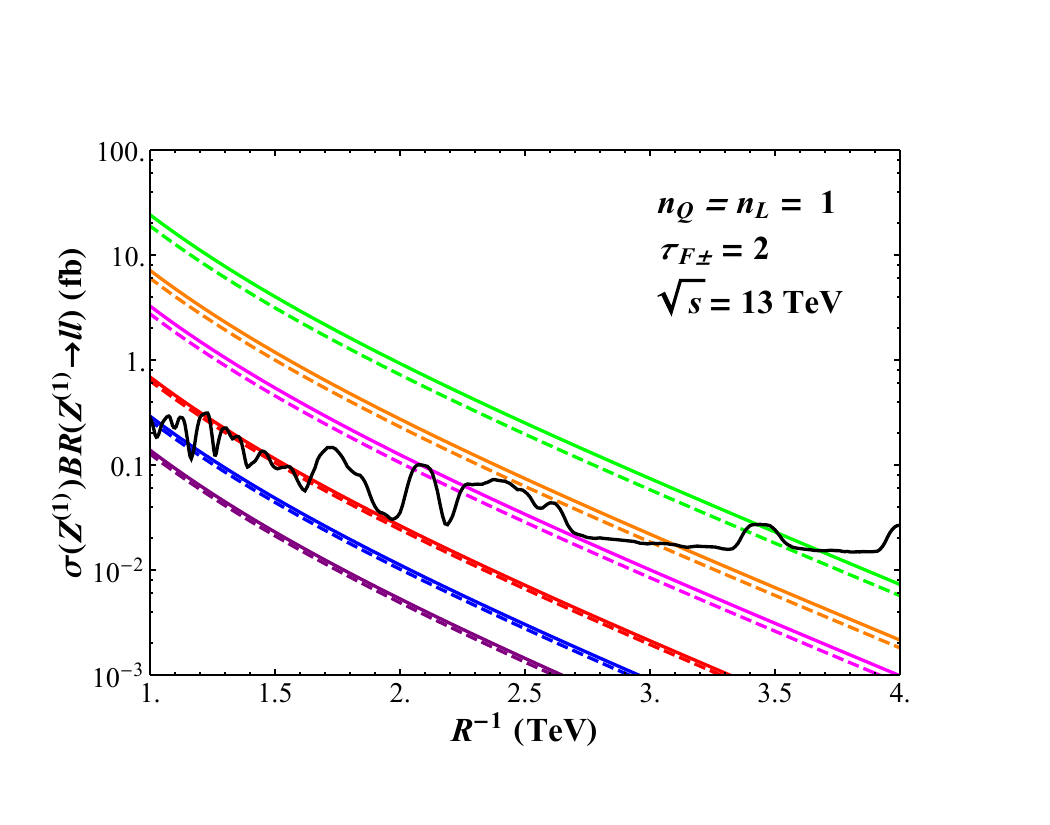}}
    \vspace*{-1.5cm}
    \centerline{\includegraphics[width=3.5in]{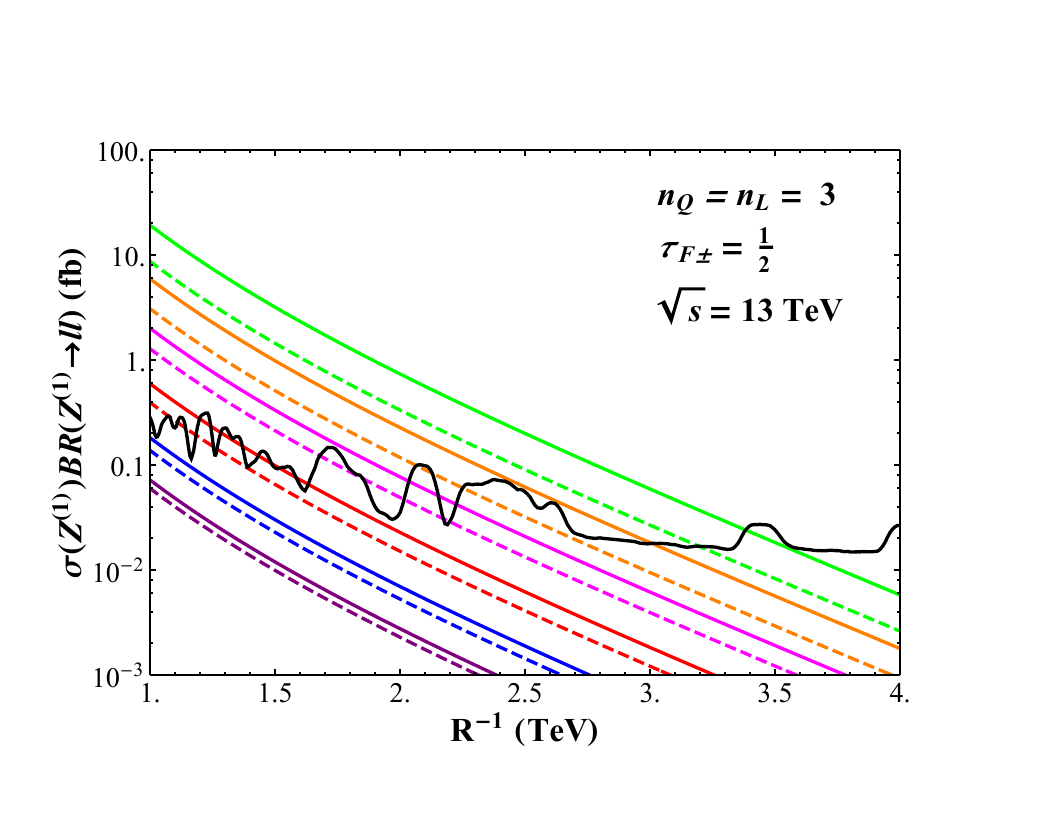}
    \hspace{-0.75cm}
    \includegraphics[width=3.5in]{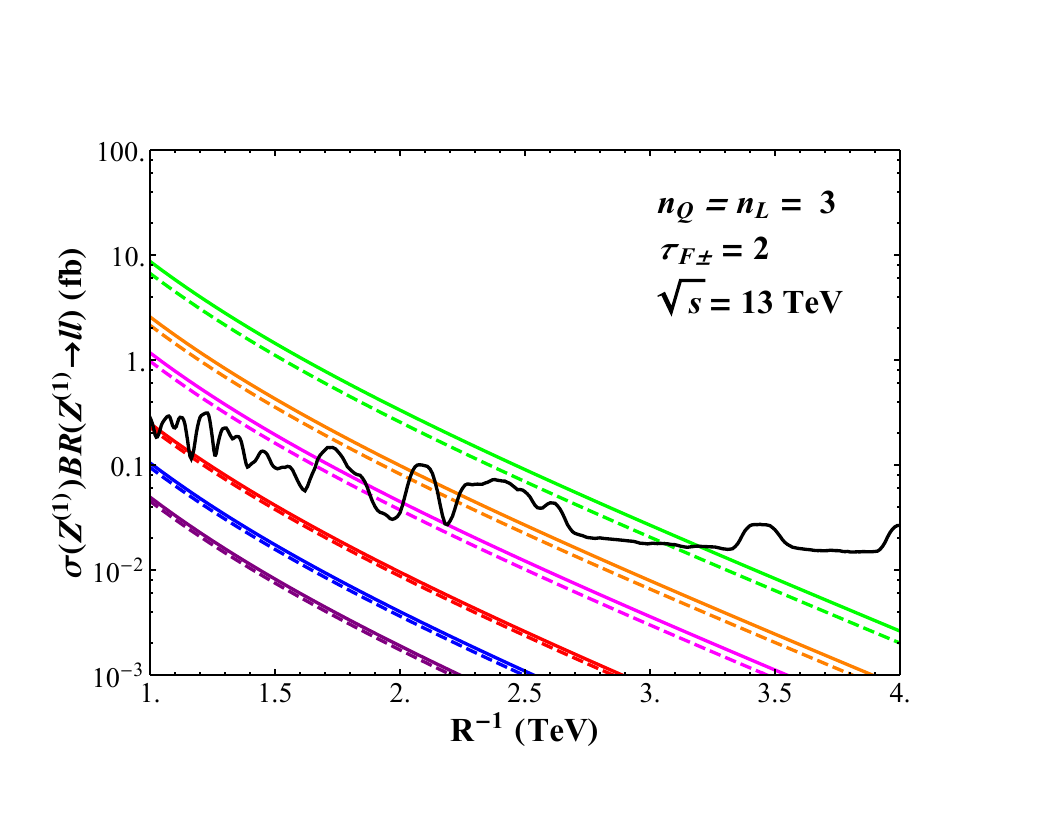}}
    \vspace*{-1.5cm}
    \centering
    \includegraphics[width=3.5in]{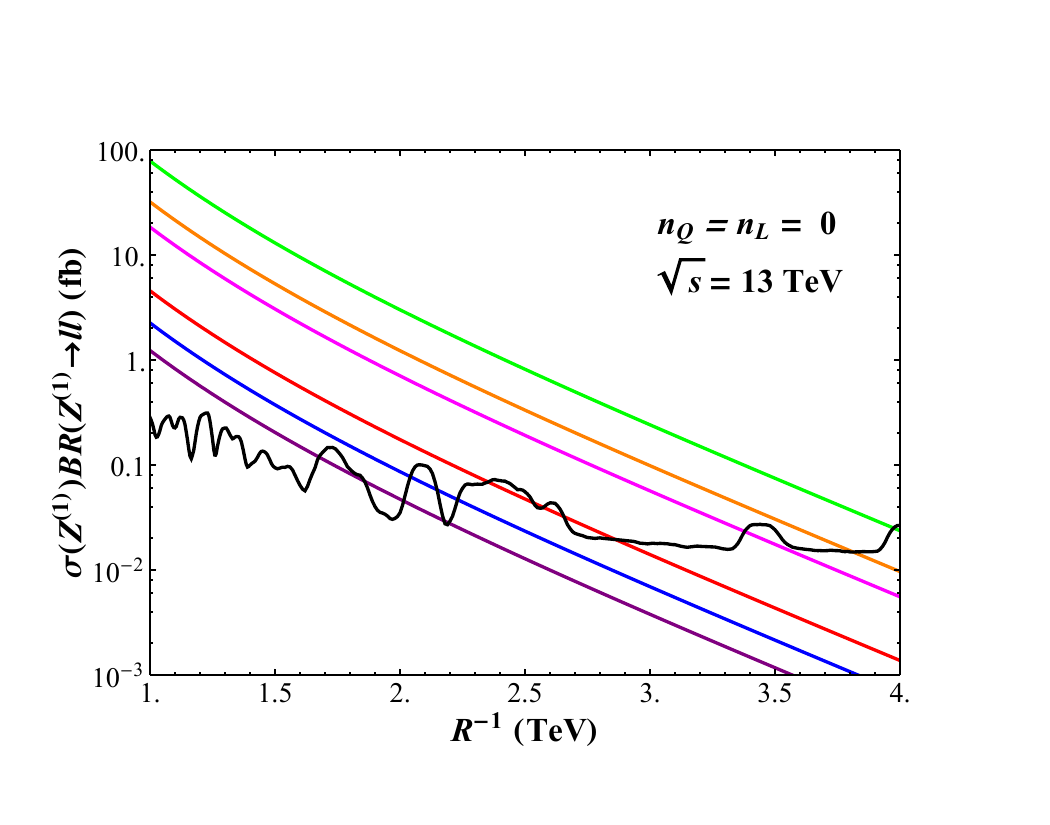}
    \caption{Predicted LHC cross sections with a center-of-mass energy of $\sqrt{s} = 13 \; \textrm{TeV}$ for Drell-Yan production of $Z^{(1)}$, with the choice of the number of quark and lepton portal matter generations ($n_Q$ and $n_L$, respectively, as defined in Table \ref{table:fermionModelContent}) and the $\phi=\pi$ brane-localized kinetic term for the portal matter fields denoted by a plot label. For the solid lines, SM fermion brane-localized kinetic terms are assumed to be $(\omega_F,\tau_{F0}) \, = \, (1/2,1)$ (Red), $(1/2,3/2)$ (Orange), $(1/2,2)$ (Green), $(1,3/2)$ (Blue), $(1,2)$ (Magenta), and $(3/2,2)$ (Purple) for all SM fields. For simplicity, a high degree of universality is assumed among the brane-localized kinetic terms: In reference to the brane-localized kinetic terms in Table \ref{table:fermionModelContent}, we have assumed $\tau_{F\pm}=\tau_{Q,u,d,L,e+}=\tau_{Q,u,d,L,e-}$, $\tau_{F0} = \tau_{Q,u,d,L,e0}$, and $\omega_F = \omega_{Q,u,d,L,e}$. Dashed lines denote the same SM brane term choices, but with $\omega_F$ and $\tau_{F0}$ exchanged-- these lines are absent in the $n_Q=n_L=0$ case (that is, the case without any portal matter fields), since the cross section and branching fraction is symmetric under this interchange here. The black line denotes the 95\% CL exclusion limit from \cite{ATLAS:2019erb}.}
    \label{fig4}
\end{figure}

\begin{figure}
    \centerline{\includegraphics[width=3.5in]{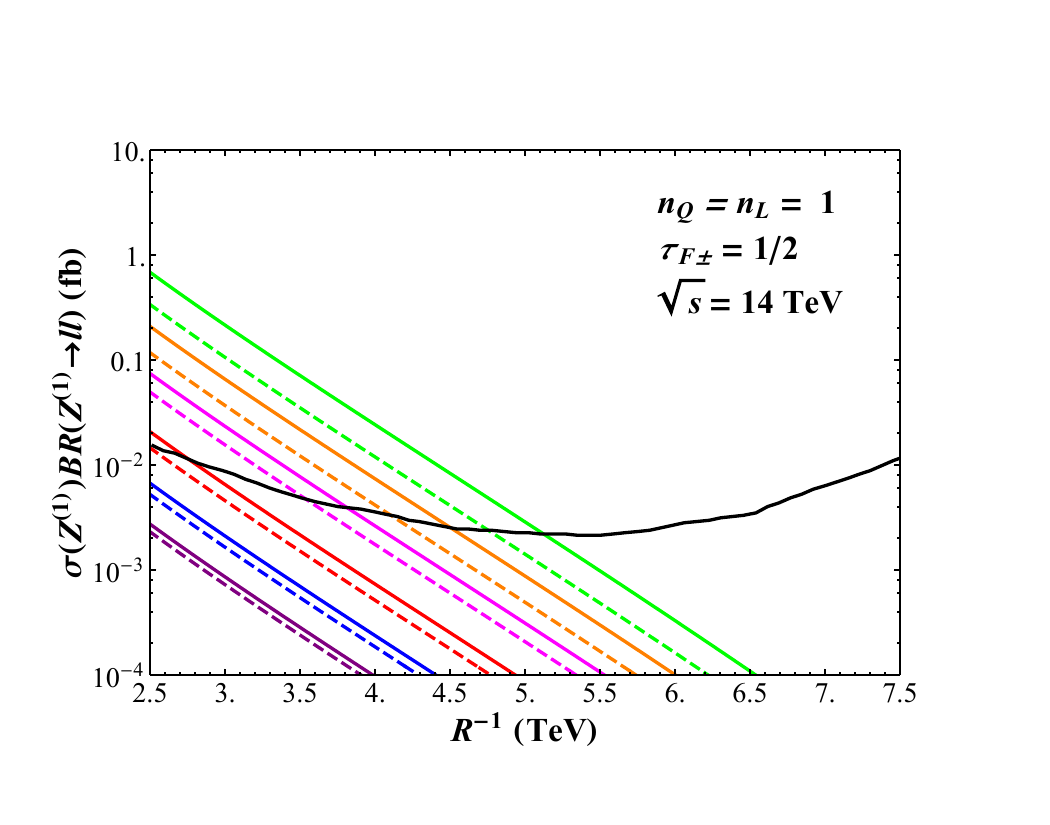}
    \hspace{-0.75cm}
    \includegraphics[width=3.5in]{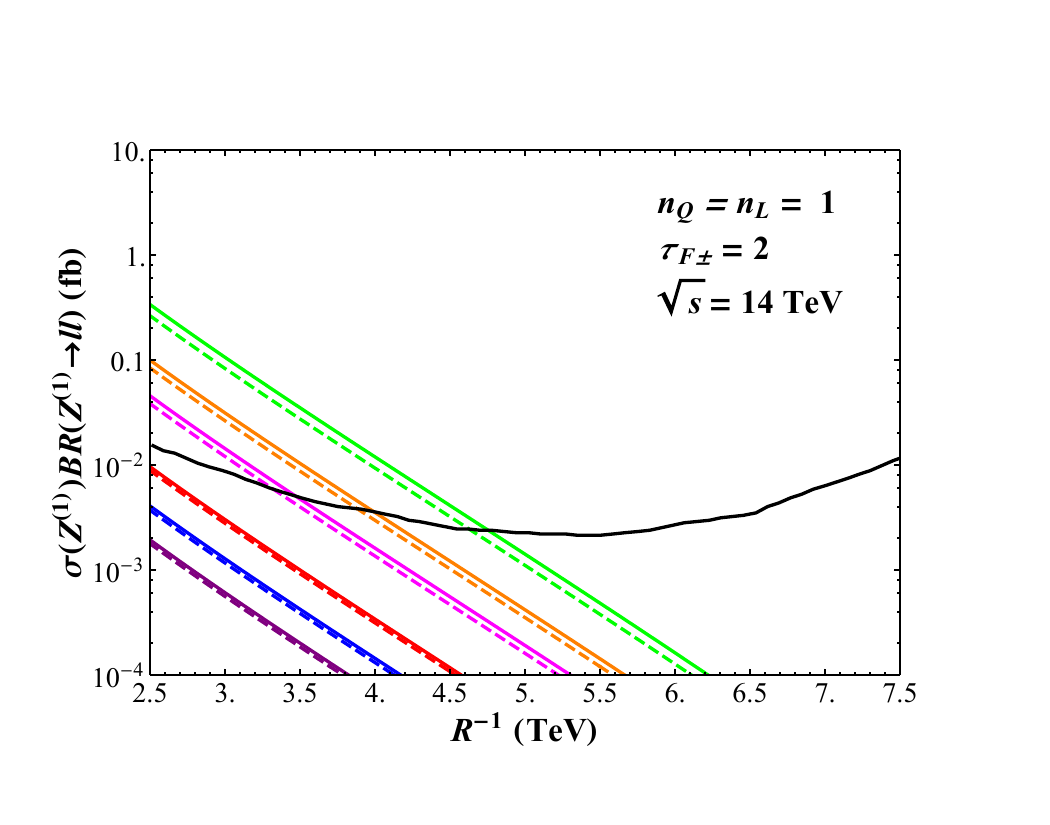}}
    \vspace*{-1.5cm}
    \centerline{\includegraphics[width=3.5in]{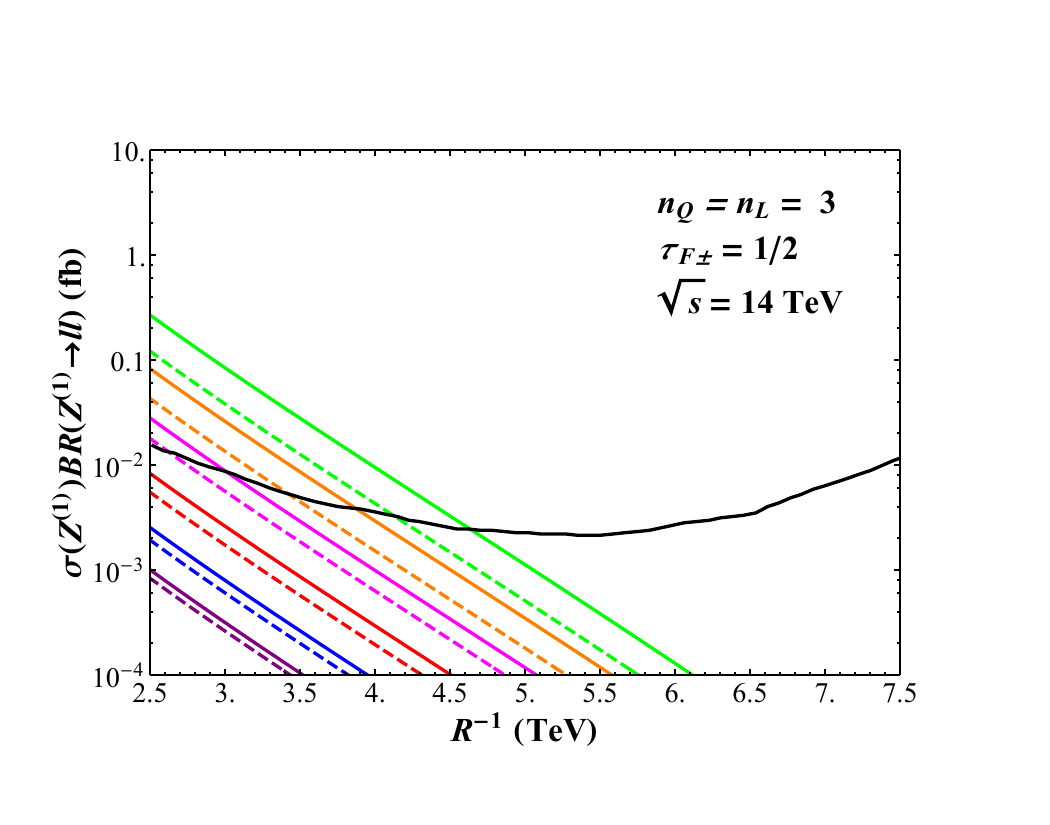}
    \hspace{-0.75cm}
    \includegraphics[width=3.5in]{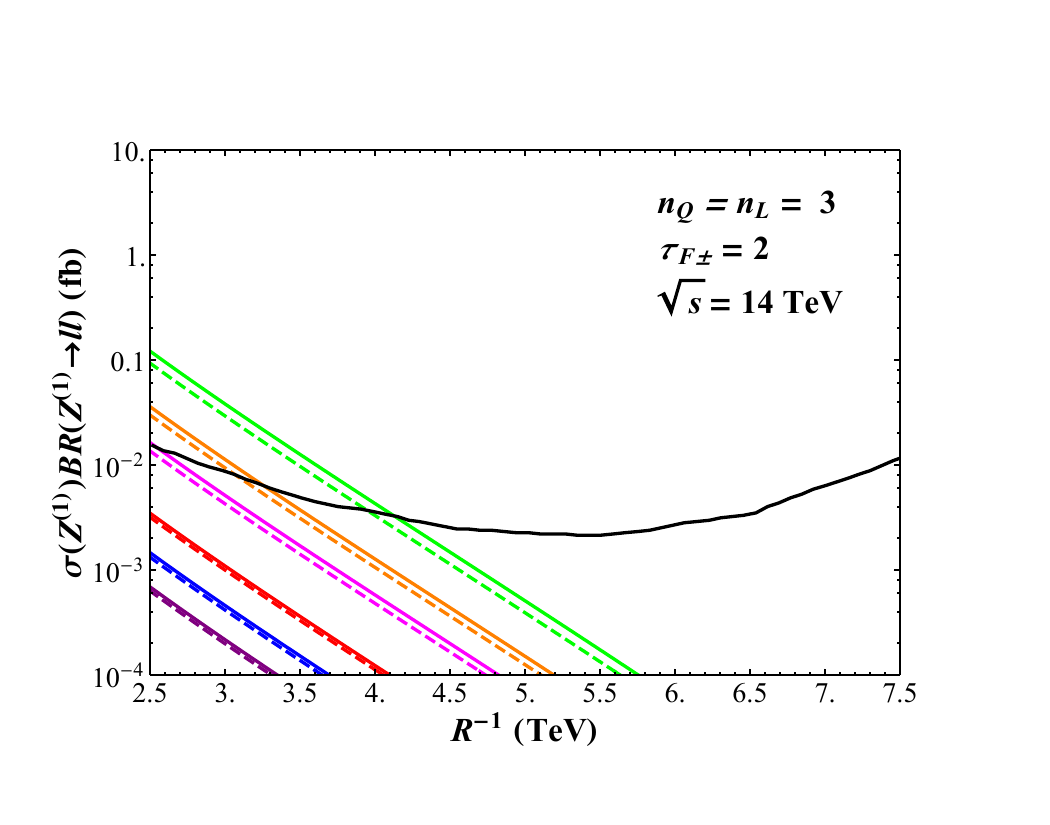}}
    \vspace*{-1.5cm}
    \centering
    \includegraphics[width=3.5in]{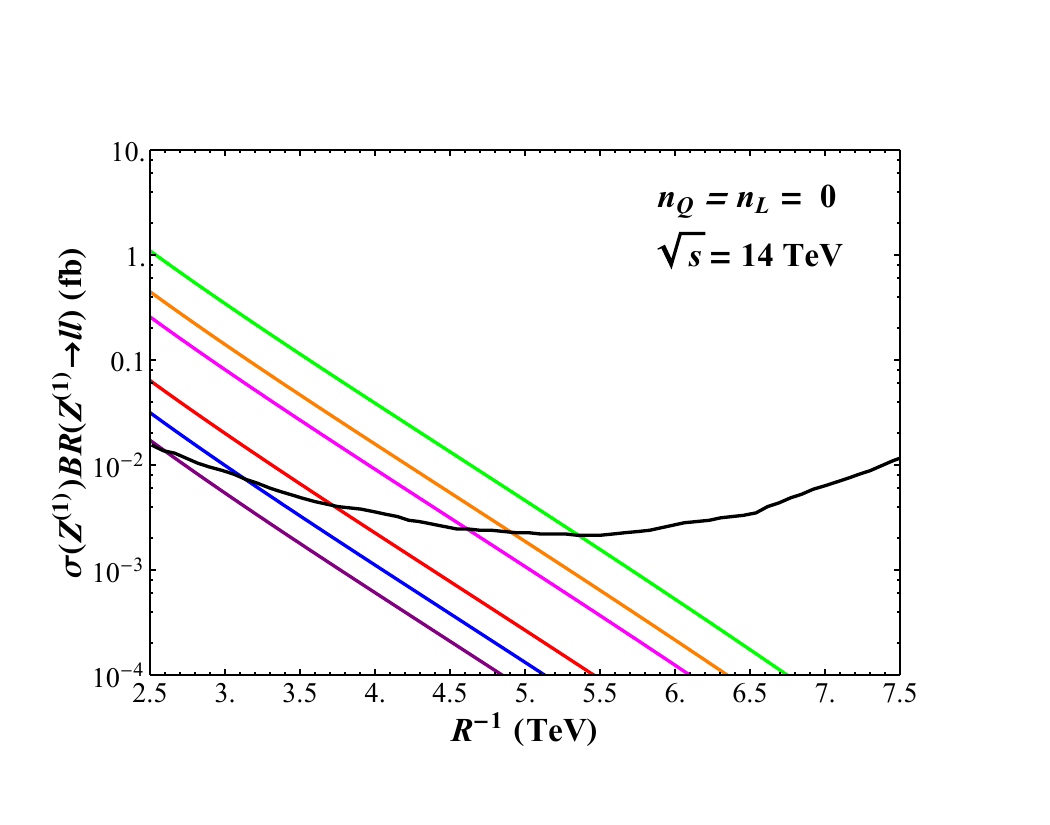}
    \caption{As in Figure \ref{fig4}, but depicted for a center-of-mass energy of $\sqrt{s}=14 \; \textrm{TeV}$ and with the black line depicting the expected 95\% CL exclusion limit from a null result for Drell-Yan production of a heavy spin-1 resonance given in \cite{ATLAS:2018tvr}.}
    \label{fig5}
\end{figure}

The other significant decay channels for $Z^{(1)}$ are all into SM particles. For the decay into an SM fermion-antifermion pair, we arrive at a partial width
\begin{align}
    \Gamma_{Z^{(1)} \rightarrow f \overline{f}} = \bigg( (T_3^F - s_w^2 Q^F)^2 \bigg(\frac{\tau_{F0}-\omega_F}{\pi + \tau_{F0} + \omega_F}\bigg)^2 + s_w^4 (Q^F)^2 \bigg(\frac{\tau_{f0}-\omega_f}{\pi + \tau_{f0} + \omega_f}\bigg)^2 \bigg) \frac{(g_L^2 + g_Y^2) R^{-1}}{12 \pi},
\end{align}
where $\tau_{F(f)0}$ and $\omega_{F(f)}$ are the $\phi=\pi$ and $\phi=0$ brane-localized kinetic terms for the $U(1)_D$-uncharged electroweak doublet (singlet) fermion field, $T_3^F$ denotes the weak isospin of the electroweak doublet fermion, and $Q^F$ is the fermion's electromagnetic charge. Meanwhile, for the diboson decay channels $Z^{(1)} \rightarrow W^+ W^-$ and $Z^{(1)} \rightarrow Z h$ have widths simply given by
\begin{align}
    \Gamma_{Z^{(1)} \rightarrow Z h} = \frac{1}{(2 s_w^2-1)^2}\Gamma_{Z^{(1)} \rightarrow W^+ W^-} = \frac{(g_L^2 + g_Y^2) R^{-1}}{96 \pi}.
\end{align}
In Figure \ref{fig4}, we display the cross section for Drell-Yan production of a $Z^{(1)}$ which subsequently decays into a dilepton final state, either $e^+ e^-$ or $\mu^+ \mu^-$, for a variety of choices of model parameters, with the 95\% CL exclusion limit from a search of 139 $\textrm{fb}^{-1}$ of data at the $\sqrt{s}=13 \; \textrm{TeV}$ LHC \cite{ATLAS:2019erb} depicted for comparison.\footnote{We should note that if our parameter choices in Figure \ref{fig4} are taken seriously, then the universality of the brane-localized kinetic terms would mean that kinetic mixing would vanish-- this is in precise analogy to the vanishing of kinetic mixing for degenerate-mass portal matter fields in 4 dimensions. However, this universality is useful in exploring the numerics of our model constraints, since the essential features of the total cross sections are not altered if the universality conditions are relaxed.} In Figure \ref{fig5}, we depict the same cross sections, but assuming $\sqrt{s} = 14 \; \textrm{TeV}$ and with the limits expected from a null result with 3 $\textrm{ab}^{-1}$ at the HL-LHC \cite{ATLAS:2018tvr}. We first note that the partial width of the $Z^{(1)}$ into portal matter fields significantly reduces the branching fraction of this particle into dilepton final states, and hence the cross section for the process sought in these searches: For example, if every SM field is part of an $SU(2)_D$ triplet (so $n_Q = n_L = 3$, with $n_Q$ and $n_L$ defined as in Table \ref{table:fermionModelContent}), the relevant cross section for this search is reduced by as much as two orders of magnitude from the prediction when $n_Q=n_L=0$, that is, when there are no portal matter fields in the model. We can also see that our choice of the SM fermion brane-localized kinetic terms has a tremendous effect on the expected Drell-Yan cross section-- this is to be expected, since we have previously noted that the zero-mode fermions will couple to $Z^{(1)}$ with a strength scaled by $\sqrt{2}(\tau_{F0} - \omega_F)/(\pi + \tau_{F0} + \omega_F)$ relative to the 4-dimensional gauge coupling, so we would naively expect the Drell-Yan cross section to scale as the square of this factor. Even without a finely-tuned degeneracy between $\omega_F$ and $\tau_{F0}$, we see that this scaling can alter our predicted cross sections by more than an order of magnitude: If $(\omega_F,\tau_{F0})= (3/2,2)$, the expected cross section is approximately 100 times smaller than the result if $(\omega_F,\tau_{F0})= (1/2,2)$, translating to a lower limit on the compactification scale $R^{-1}$ of $\sim 2.2 \; \textrm{TeV}$ in the former case and $\sim 4 \; \textrm{TeV}$ in the latter, with the effects of the portal matter fields on the $Z^{(1)} \rightarrow ll$ branching fraction ignored.

While the results of Figures \ref{fig4} and \ref{fig5} hardly represent a comprehensive probe of the model parameters, it is useful to note that there certainly exist broad regions of parameter space or which this constraint is very weak indeed. Even without fine tuning of the brane-localized kinetic terms, there are choices of these terms such that the constraint on $R^{-1}$ for these searches is below 1 TeV; these anemic constraints stem in large part because the additional decay channels of $Z^{(1)}$ into portal matter dramatically reduce the branching fraction of $Z^{(1)}$ into dileptons. While the exclusion limits can be significantly improved by null results from the HL-LHC, it is clear in Figure \ref{fig5} that even if these constraints are realized, there still exist significant regions of parameter space in which a null result does not exclude the theory, even for compactification scales under 2.5 TeV.
Because these decay channels (or similar ones) will remain open for \emph{any} massive Kaluza-Klein mode of an SM gauge boson, we can expect similar effects in other searches, such as those for a Kaluza-Klein $W$ boson decaying leptonically or for $Z^{(1)}$ decaying to an SM $Z$ and a Higgs. By introducing brane-localized kinetic terms to the SM gauge fields, which will reduce the mass of the Kaluza-Klein modes relative to the compactification scale $R^{-1}$, it is possible to kinematically disallow some of these decay channels, but only if these terms are large for the SM gauge fields and small for the portal matter fermions. Given the sheer number of portal matter fields introduced in the model (even if only a single generation of quarks is embedded in an $SU(2)_D$ triplet, the model will feature 8 color triplet portal matter fields), in the absence of an underlying mechanism keeping all portal matter brane terms small while allowing large SM gauge boson brane terms it is unlikely that all of the portal matter fermions will have appropriate brane-localized kinetic terms to disallow decay of $Z^{(1)}$ into them. Furthermore, as mentioned in Section \ref{section:gauges}, the mass splitting that leads to this suppression is not unique to a flat extra-dimensional geometry, and may in fact be more severe in a warped context.

\begin{figure}
    \centerline{\includegraphics[width=3.5in]{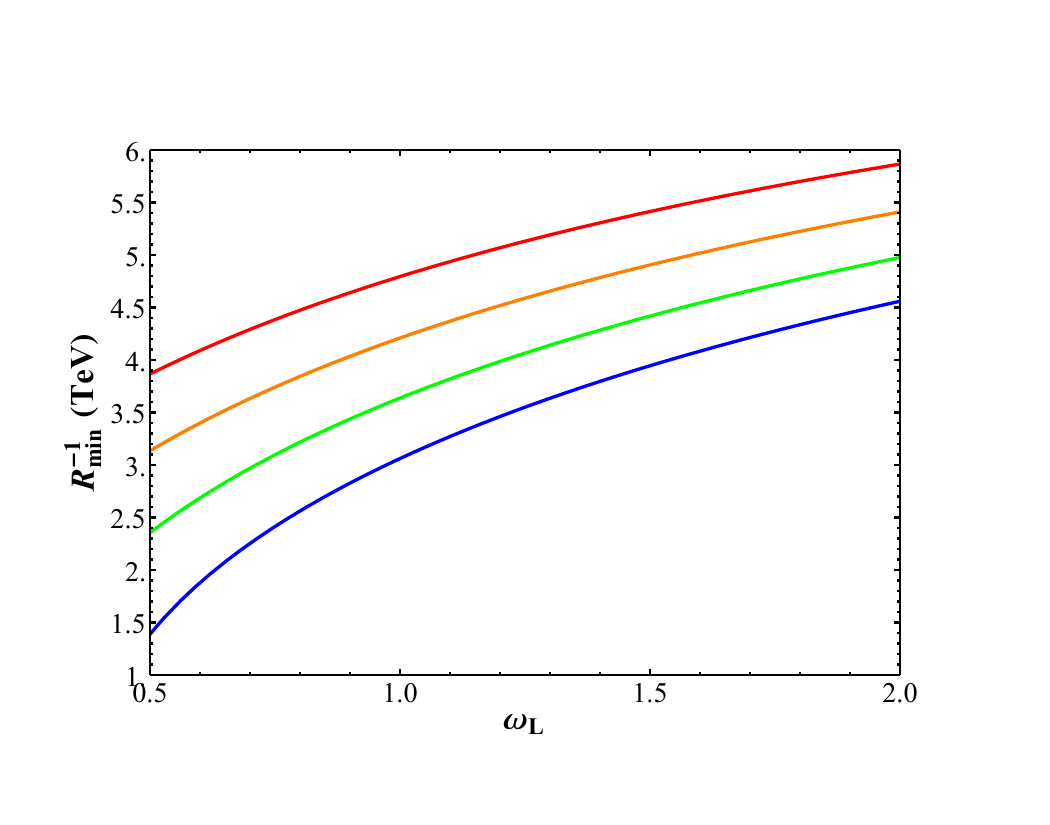}
    \hspace{-0.75cm}
    \includegraphics[width=3.5in]{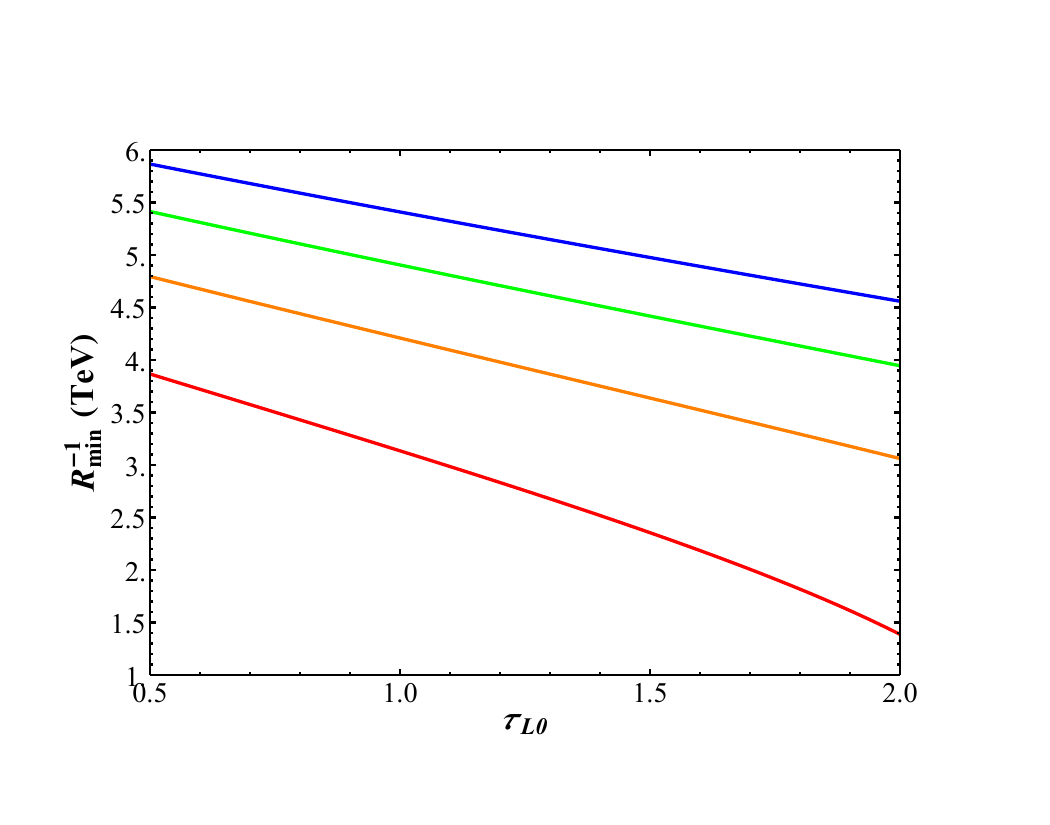}}
    \caption{(Left): Minimum compactification scale $R^{-1}$ in order for the $\rho$ parameter to stay within $2\sigma$ of the global fit value \cite{ParticleDataGroup:2020ssz} given in Eq.(\ref{eq:RhoValue}), plotted as a function of $\omega_L$ for various choices of $\tau_{L0}$ (defined as in Table \ref{table:fermionModelContent}). From the top down: $\tau_{L0} =$ 1/2, 1, 3/2, 2.  (Right): Same as left, but with the constraint plotted as a function of $\tau_{L0}$ for various choices of $\omega_L$. From the top down: $\omega_L =$ 2, 3/2, 1, 1/2.}
    \label{fig6}
\end{figure}

While the constraints on the compactification scale from conventional searches for Kaluza-Klein gauge bosons are severely weakened in our portal matter scenario, the same can not be said for constraints arising from precision low-energy measurements. It therefore behooves us to at least estimate the severity with which these measurements might constrain our model. As our model, with its arbitrarily imposed flavor universality and flat geometry is meant only to be a semi-realistic construction, it is likely not especially useful, and certainly beyond the scope of this paper, to do a global fit to electroweak and Higgs precision measurements in this setup.\footnote{In our simple construction the fact that all bulk fermions share universal vanishing bulk masses means that the usual harsh constraints from flavor-changing four-fermion operators, \eg those considered in \cite{Csaki:2010az}, are severely suppressed, for example.} Instead, we may get a feel for the limits imposed from these constraints by simply considering the expected $\rho$ parameter shift that our model of extra dimensions might generate. To compute this, we can adapt the techniques of \cite{Delgado:2007ne} to integrate out the heavy Kaluza-Klein modes of the theory and express the new physics as effective dimension-6 operators. In the Warsaw basis \cite{Grzadkowski:2010es} of dimension-6 operators, the $\rho$ parameter is then given by \cite{Bagnaschi:2022whn,Kribs:2020jgn}
\begin{align}\label{eq:GeneralRhoFormula}
    \rho - 1 = -\frac{2 s_w c_w}{c_w^2-s_w^2} \frac{v^2 R^2}{4} \bigg( \frac{c_w}{s_w} C_{HD} + \frac{s_w}{c_w} (4 C^{(3)}_{Hl}-2 C_{ll}) + 4 C_{HWB}\bigg),
\end{align}
using the operator definitions of \cite{Bagnaschi:2022whn}. In Appendix \ref{appendix:EDSMEFT}, we find that in our construction the operators in Eq.(\ref{eq:GeneralRhoFormula}) are given by
\begin{align}\label{eq:d6OPerators}
\begin{matrix}
    C_{HD} = - \frac{g_Y^2 \pi^2}{6}, & C^{(3)}_{Hl} = \frac{g_Y^2 \pi^2 (\tau_{L0}-2 \omega_L)}{6(\pi + \tau_{L0} + \omega_L)}, & C_{ll} = -\frac{g_L^2 \pi^2 (\tau_{L0}^2 - \tau_{L0} \omega_L + \omega_L^2)}{12 (\pi + \tau_{L0} + \omega_L)^2}, & C_{HWB} = 0,
\end{matrix}
\end{align}
where $\omega_L$ and $\tau_{L0}$ are the $\phi=0$ and $\phi=\pi$ brane-localized kinetic terms for the SM charged lepton fields, respectively, as in Table \ref{table:fermionModelContent}. Additional contributions to these operators from the massive Kaluza-Klein tower modes of the fermion fields are ignored here, since these will be severely suppressed for light fermions (specifically, they should only emerge at $O(m_{e,\mu}^2 R^2)$ for these measurements). These results give us the expression
\begin{align}
    \rho - 1 = \frac{g_Y^{2}}{(g_L^2-g_Y^{2})}\frac{g_L^2 v^2 R^2 \pi^2}{12(\pi+\tau_{L0}+\omega_L)} \bigg( 2 \pi - 4 \tau_{L0} + 8 \omega_L -\frac{\pi^2-3 \tau_{L0} \omega_L}{\pi+\tau_{L0}+\omega_L}\bigg),
\end{align}
or in terms of the observables $G_F$, $m_Z$, and $\alpha(m_Z^2)$
\begin{align}\label{eq:FinalRhoExpression}
    \rho - 1 = \frac{mZ^2 R^2 \pi^3 \alpha}{3 \sqrt{2} \sqrt{G_F m_Z^2 (G_F m_Z^2 - 2 \sqrt{2} \pi \alpha)}} \bigg( \frac{2 \pi - 4 \tau_{L0}+8 \omega_L}{(\pi+\tau_{L0}+\omega_L)}-\frac{\pi^2-3 \tau_{L0} \omega_L}{(\pi+\tau_{L0}+\omega_L)^2} \bigg).
\end{align}
We can then compare the result in Eq.(\ref{eq:FinalRhoExpression}) to the current global fit constraint \cite{ParticleDataGroup:2020ssz}\footnote{This global fit does \emph{not} account for the recent CDF II result \cite{CDF:2022hxs}, which claims significant tension between a direct measurement of the $W$ boson mass and the prediction of the SM. While it is clear that our toy construction here might explain this discrepancy, the recency of the result and the tension with existing ATLAS \cite{ATLAS:2017rzl} and LHCb \cite{LHCb:2021bjt} measurements of the $W$ boson mass leads us to err on the side of caution and refrain from considering how this measurement might influence our model phenomenology. A detailed exploration of the utility of extra dimensions to explain the $W$ boson mass shift may be of interest, but it is beyond the scope of this paper.}
\begin{align}\label{eq:RhoValue}
    \rho = 1.00038 \pm 0.00020.
\end{align}
In Figure \ref{fig6}, we depict the minimum compactification scale $R^{-1}$ in order to stay within $2 \sigma$ of the fit value given in Eq.(\ref{eq:RhoValue}) as a function of the brane-localized kinetic terms $\tau_{L0}$ and $\omega_L$. We can see that the constraints on $R^{-1}$ stemming from the $\rho$ parameter are usually significantly stronger than the corresponding constraints from Drell-Yan production of $Z^{(1)}$. While $Z^{(1)}$ production yields constraints no stronger than $R^{-1} \gsim 4 \; \textrm{TeV}$, and generally gives results that are much weaker, the $\rho$ parameter offers constraints that can be as severe as $R^{-1} \gsim 6 \; \textrm{TeV}$. This is unsurprising, given the substantial suppression of the $Z^{(1)} \rightarrow ll$ cross section due to the existence of portal matter states. In fact, of the benchmark $\tau_{F0}$ and $\omega_F$ values considered in Figure \ref{fig4}, we find that the current constraint from on $R^{-1}$ from $Z^{(1)}$ production only outperforms the $\rho$ parameter constraint in one region of parameter space: When $\tau_{F0}$ is large and $\omega_F$ is small (among our benchmark points, $(\omega_F, \tau_{F0}) = (1/2,3/2)$ and $(1/2,2)$ are the only points for which the $Z^{(1)}$ constraint is the harsher one. The projected future HL-LHC constraint only might add the point $(\omega_F, \tau_{F0}) = (1,2)$ to this list, depending on the number of generations with portal matter modes).
Of course, we should note that because $\omega_F$ and $\tau_{F0}$ in Figures \ref{fig4} and \ref{fig5} are artificially assumed to take on universal values for all SM fermion fields, while $\omega_L$ and $\tau_{L0}$ only apply to the $SU(2)_L$ doublet leptons, any attempt to naively consider these constraints in combination must be greeted skeptically at best. Given that our constraint from the $\rho$ parameter merely acts as an estimate of the constraints from a global electroweak fit, this skepticism is especially merited.
Ignoring caution and naively requiring our choices of brane-localized kinetic terms to satisfy both the $Z^{(1)}$ search and the $\rho$ parameter constraint, however, we find that in the corners of parameter space where the $\rho$ constraint is exceptionally weak ($\lsim 3 \; \textrm{TeV}$, the $Z^{(1)}$ constraints will invariably be near their harshest, limiting the parameter space to $R \gsim 3 \; \textrm{TeV}$. We can roughly estimate, then, that constraints stemming from ``conventional'' SM-only probes of extra dimensions will generally lead to constraints of $R^{-1} \gsim 3.0-6.0 \; \textrm{TeV}$.

\subsection{Portal Matter and $SU(2)_D$ Bosons: Decay}\label{section:PMDecay}
Having addressed the constraints arising in our construction from familiar probes of extra dimensions, it is now of interest to consider how states associated with portal matter and the $SU(2)_D$ gauge sector will appear experimentally. Before moving into production mechanisms specifically, we will discuss what new states are most likely to be phenomenologically important and discuss the dominant decay channels for them. In addition to the familiar SM particles, our construction has a number of new exotics. At the low GeV scale, the simplified kinetic mixing portal dark matter model we have extended requires a dark photon $A_D$, a dark Higgs $h_D$, and some sort of dark matter particle (about which we have remained vague, since it has limited influence on the phenomenology we wish to explore in this work). In our model, $A_D$ is simply the lightest state of the Kaluza-Klein tower of the gauge boson field $X$ described in Section \ref{section:gauges}-- because this field plays the role of the dark photon, from here on out we will refer to this zero mode with the suggestive name $A_D$. The dark Higgs $h_D$, meanwhile, is simply the real scalar field emerging from the brane-localized scalar $\Phi_D$ in Section \ref{section:gauges}.

The collider signals associated with these fields will be more or less identical to that which has already been discussed for analogous particles in other portal matter papers such as \cite{Rizzo:2018vlb,Rueter:2019wdf}. The specifics of the likely decay channels for the dark photon and the dark Higgs, which we shall find might be copiously produced in collider experiments, are heavily dependent on the relative values of low-energy parameters in the model (specifically the relative masses of the dark photon, the dark Higgs, and the dark matter) which have no significant numerical effect on either the production or decay of the heavier exotic states on which this work focuses. As such, we shall limit our discussion to simply summarizing that depending on the selection of low-energy parameters, the dark photon will generally either decay invisibly (or be long-lived), leaving only missing $E_T$ at the LHC, or it will decay into a pair of light SM fermions, for example $e^+ e^-$. The dark Higgs, meanwhile, will likely decay promptly into a pair of dark photons (one or both of which might be virtual), and from there leave only missing energy or sets of fermion pairs, depending on the decay properties of the  dark photon itself. 

\begin{figure}
    \centerline{\includegraphics[width=3.5in]{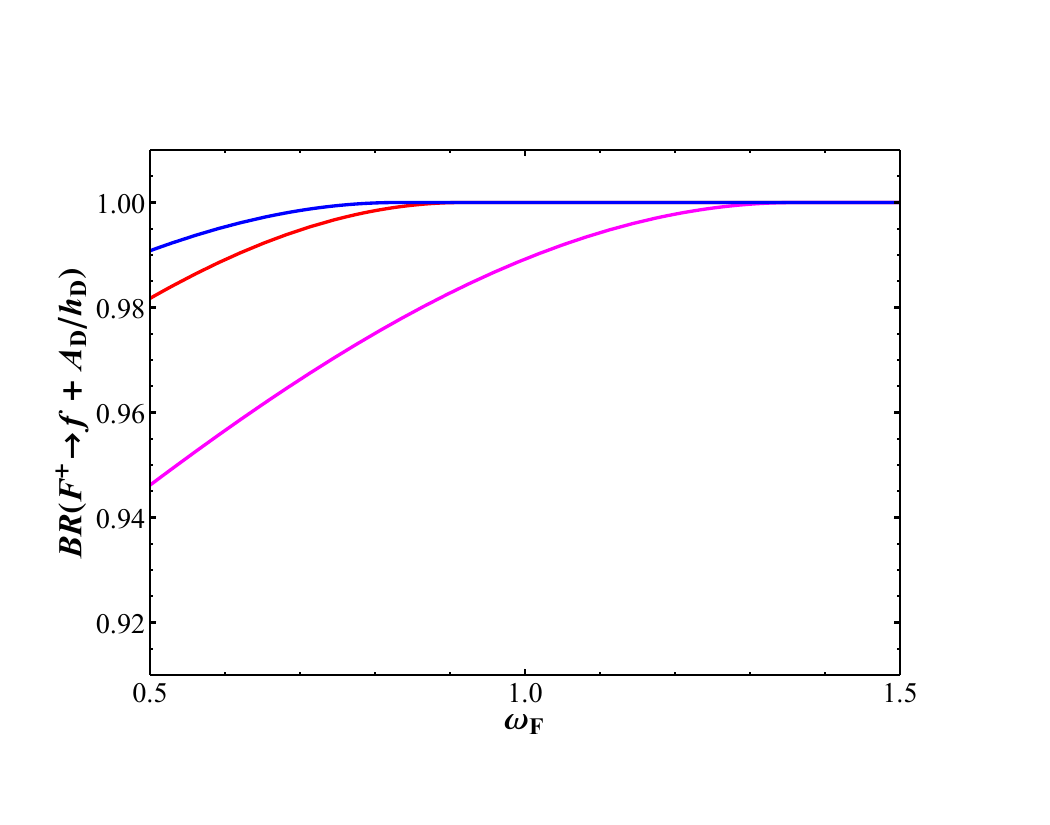}
    \hspace{-0.75cm}
    \includegraphics[width=3.5in]{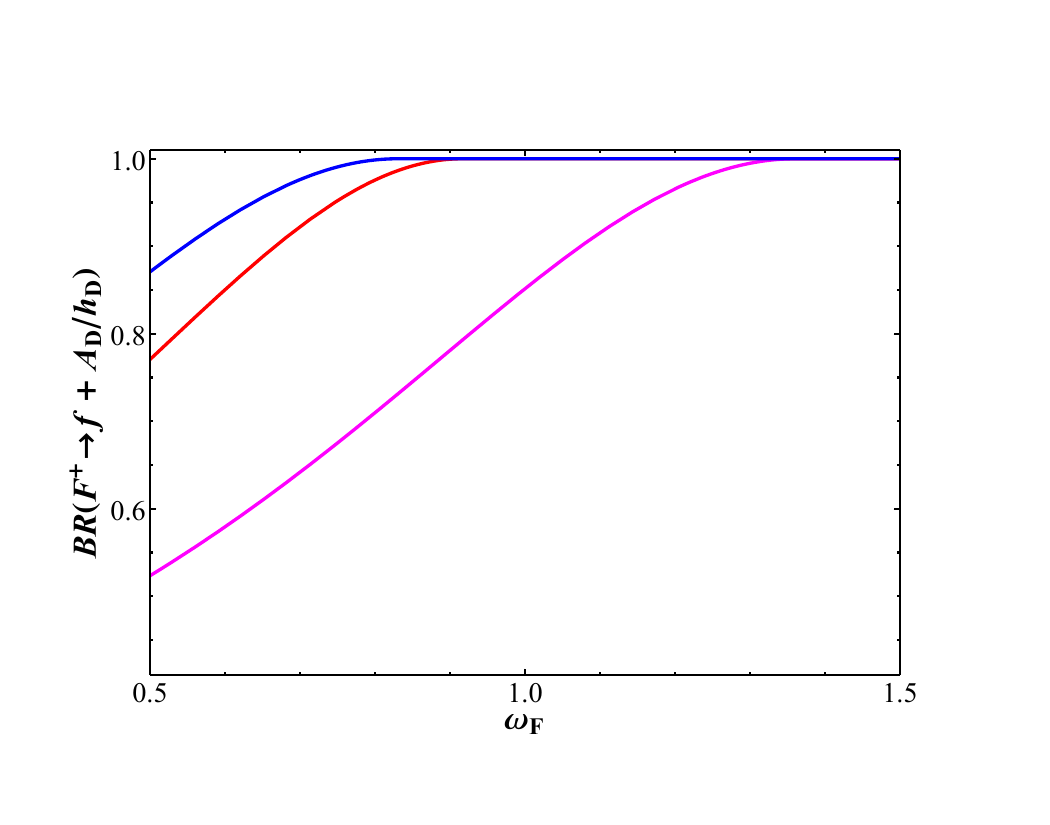}}
    \caption{(Left): The branching faction of a portal matter fermion field $F^+$ with brane-localized kinetic terms $\omega_F$ at $\phi=0$ and $\tau_{F+}$ at $\phi=\pi$ for decay via emission of a dark Higgs or a dark photon with brane-localized kinetic terms $\omega_X$ at $\phi=0$ and $\tau_X$ at $\phi=\pi$, computed from the partial widths given in Eqs.(\ref{eq:PMToDPWidth}) and (\ref{eq:PMToIWidth}) as a function of $\omega_F$ for various choices of other brane-localized kinetic terms. In all cases we assume that $g_D = g_L$, $\tau_X = 1$, and $\tilde{y}^F_+ = 1$, where $\tilde{y}^F_+/\sqrt{2}$ is the dark Higgs Yukawa coupling between $F^+$ and the SM fermion with which it mixes,. Choices of other brane-localized kinetic terms are $(\tau_{F+},\omega_X) = $(1/2,3/2) (Red), (1/2,2) (Magenta), (1,2) (Red) (Right): Same as left, but with $\tilde{y}^F_+ = 1/4$. Note that in both cases, the charts are invariant under the interchange $\tau_{F+} \leftrightarrow \omega_F$.}
    \label{fig7}
\end{figure}

At the TeV scale, we have introduced a much broader range of exotics: There are Kaluza-Klein modes for all SM fields, plus those associated with portal matter and the $SU(2)_D$ gauge bosons. Because of the portal matter and the $I^\pm$ bosons' boundary conditions in the bulk, these fields' lightest Kaluza-Klein modes are overwhelmingly likely to be the least massive TeV-scale exotic particles appearing in the model-- they must have masses less than half the compactification scale $R^{-1}$, and so barring very large brane-localized kinetic terms for the SM fermions, the dark photon, or the SM gauge bosons, $I^\pm$ and portal matter will dominate our low TeV-scale phenomenology. 
In \cite{Rizzo:2018vlb,Rueter:2019wdf} it was found that portal matter fermions, in contrast to vector-like fermions with no additional dark charge, will overwhelmingly decay into SM fermions via the emission of a dark photon $A_D$ or a dark Higgs $h_D$, rather than via the emission of an SM gauge boson. This preference occurs because, while the coupling of a dark photon to a portal matter field and an SM fermion is exceptionally weak (in our setup, $O(v_D R) \lesssim 10^{-3}$, where $v_D$ is the small brane-localized dark Higgs vev and $R$ is the compactification scale), the longitudinal mode of $A_D$ will, because of the dark photon's small mass, enhance this interaction strength by a factor roughly cancelling the suppression of the coupling. We find that in our construction, the partial width of a fermion with $U(1)_D$ charge $+1$ and brane-localized kinetic terms $\omega_F$ and $\tau_{F+}$ and a mass $m_{F+}$ to an SM fermion $f$ plus a dark photon $A_D$ or a dark Higgs $h_D$ will be given by
\begin{align}\label{eq:PMToDPWidth}
    &\Gamma_{F^+ \rightarrow A_D f} = \Gamma_{F^+ \rightarrow h_D F} = \frac{(\tilde{y}^F_+)^2}{32 \pi} x_F R^{-1},\;\;\; x_F = m_{F+} R, \;\;\; \tilde{y}^F_+ \equiv \sqrt{\frac{1+x_F^2 \omega_F^2}{1+x_F^2 \tau_{F+}^2 }} \frac{N^{F_+} y^F_+}{\sqrt{\pi + \tau_{F0} + \omega_F}},\\
    &N^{F_+} \equiv \bigg[ \frac{2 (1 + x_F^2 \tau_{F+}^2)}{\pi (1 + x_F^2 \omega_F^2) (1 + x_n^2 \tau_{F+}^2)+ (\tau_{F+} + \omega_F)(1 + x_F^2 \tau_{F+} \omega_F)} \bigg]^{1/2} , \nonumber
\end{align}
where $\tau_{F0}$ is the $\phi=\pi$ brane-localized kinetic term for the SM fermion $F$, and all small masses (those of the SM fermion, the dark photon, and the dark Higgs) have been allowed to trend to 0. The parameter $\tilde{y}^F_+$ is a more physical normalization of the brane-localized Yukawa coupling $y^F_+$ first described in Eqs.(\ref{eq:fermionAction}) and (\ref{eq:yFDefs})-- physically, $\tilde{y}^F_+/\sqrt{2}$ is the Yukawa coupling between the lightest Kaluza-Klein mode of the portal matter field $F^+$ and its corresponding SM fermion $f$. The expression in Eq.(\ref{eq:PMToDPWidth}) is trivially generalizable to any other portal matter field in the model, simply by substituting analogous parameters. If the lightest mode of the $I^\pm$ field is lighter than a given portal matter field, there is an additional possible decay channel for the portal matter, namely, by emitting an $I^+$ or $I^-$ gauge boson, depending on the $U(1)_D$ charge of the portal matter itself. We find that the width for this process is
\begin{align}\label{eq:PMToIWidth}
    &\Gamma_{F^+ \rightarrow I^+ f} = \frac{g_D^2}{16 \pi} \bigg( \frac{\pi+\tau_X + \omega_X}{\pi + \tau_{F0} + \omega_F} \bigg) \frac{2 \lambda^2 (N^{F_+})^2}{\pi (1+x_I^2 \omega_X^2)+ \omega_X} \frac{x_I^2}{x_F^3} R^{-1}(x_F^2+2 x_I^2),\\
    &\lambda \equiv \bigg((\omega_X-\omega_F) + \frac{x_F}{x_I} \tau_{F+} \frac{(1 + x_I^2 \omega_X^2)(1+ x_F^2 \omega_F^2)}{\pi(-1 + x_F^2 \tau_{F+} \omega_F)} \sin(\pi x_F) \sin(\pi x_I) \bigg), \;\; x_I = m_I R, \nonumber
\end{align}
where $g_D$ is the effective 4-dimensional coupling constant of the dark photon, $m_I$ is the mass of the $I^\pm$ boson, and $\omega_X$ and $\tau_X$ are the brane-localized kinetic terms for the $SU(2)_D$ gauge boson fields, defined in Section \ref{section:gauges}. The parameters $x_F$, $\tau_{F+}$, $\omega_F$, and $N^{F_+}_n$ are defined here just as they were in Eq.(\ref{eq:PMToDPWidth}). The portal matter decay via a dark photon or dark Higgs emission has been well-explored in, \eg, \cite{Rizzo:2018vlb,Kim:2019oyh,Rueter:2019wdf}, however here the presence of another channel, namely decay via $I^\pm$ emission, may complicate our analysis. It therefore behooves us to consider if the branching fraction for portal matter into an $I^\pm$ boson might be significant in any regime of the parameter space we consider. In Figure \ref{fig7}, we depict the branching fraction of a given portal matter field to either a dark photon or a dark Higgs field for different choices of the various fermion and gauge boson brane-localized kinetic terms. Notably, the parameter space of choices of $O(1)$ brane-localized kinetic terms such that the $I^\pm$ boson is lighter than the portal matter fields is somewhat narrow; of the benchmark values $\omega_X =$ 1/2, 1, 3/2, 2 we consider, only $\omega_X =$ 3/2 and 2 have choices of other brane terms with $(\tau_{F+},\omega_F) \geq 1/2$ such that the decay $F^+ \rightarrow I^+ F$ is kinematically accessible. This is unsurprising given the fact that there are two brane terms, $\tau_{F+}$ and $\omega_F$, which both reduce the mass of the portal matter field as they get larger, but only one such term, $\omega_X$, which accomplishes the same for $I^\pm$. Of course, our assumption that the brane-localized kinetic terms for these fields must all be $O(1)$ and random is an arbitrary choice, so it is entirely feasible that some underlying mechanism in the UV theory might favor $\omega_X$ to be significantly larger than $\tau_{F+}$ and $\omega_F$; since the region of parameter space where the portal matter can decay via on-shell $I^\pm$ is narrow, but not absurdly finely tuned, we cannot realistically conjecture about the probability of this arrangement of particle masses emerging.

Secondly, we note that at least in the parameter space we have considered for our broad phenomenological survey, the branching fraction for portal matter decay via an $I^\pm$ boson is usually subdominant to the decay via dark photon/dark Higgs emission, but for particularly small $\tilde{y}^F_+$, it can compete with the more dominant decay channel-- for $\tilde{y}^F_+ = 1/4$, we see in in some regions of parameter space in Figure \ref{fig7} the decay via $I^\pm$ will account for as much as $\sim 50\%$ of portal matter decays. In contrast to the case of dark photon and dark Higgs emission, however, it does not appear terribly feasible to arrange a scenario in which the $I^\pm$ emission decay dominates the portal matter decay width, without either assuming that some of the fermion brane-localized kinetic terms are much smaller than $O(1)$ or that the dark gauge boson brane-localized kinetic terms are much larger-- even if $\tilde{y}^F_+ = 1/4$ and $\tau_{F+},\omega_F=0$, the branching fraciton of $F^+ \rightarrow f + A_D/h_D$ remains at least $10\%$ as long as $\omega_X \lesssim 7$ (assuming, as we have in Figure \ref{fig7}, that $g_D= g_L$). 
Hence, depending on the decay properties of the $I^\pm$ boson, there exist regions of parameter space where the decay channel $F^+ \rightarrow I^+ F$ can be large enough to potentially yield an interesting collider signature from portal matter production (especially if, for example, $I^\pm$ decays entirely visibly and therefore permits reconstruction of its mass), but engineering a scenario in which this portal matter decay channel overwhelmingly dominates over dark Higgs and dark photon emission, and hence the more familiar portal matter collider signatures discussed in \cite{Rizzo:2018vlb,Kim:2019oyh,Rueter:2019wdf} might be rendered inapplicable, likely requires unnatural parameter choices.

Meanwhile, the $I^\pm$ gauge bosons' dominant decay channels closely follow the phenomenology of the dark-charged gauge bosons discussed in \cite{Rueter:2019wdf}, referred to in that work as $W_I$. In particular, if kinematically accessible the overwhelmingly dominant decay channels will be decay into an SM fermion and one of its corresponding portal matter fields. For the decay of $I^+$ into a positively $U(1)_D$-charged portal matter field $F^+$ and the corresponding SM antifermion $\overline{f}$, the partial width is (again ignoring the mass of the SM fermion)
\begin{align}
    \Gamma_{I^+ \rightarrow F^+ \overline{f}} = \frac{\lambda^2 g_D^2}{24 \pi} \bigg( \frac{2 (N^{F_+})^2}{\pi (1 + x_I^2 \omega_X^2) + \omega_X}\bigg) \bigg( \frac{\pi + \tau_X + \omega_X}{\pi + \tau_{F0} + \omega_F}\bigg) \bigg( 1 + \frac{x_F^2}{2 x_I^2}\bigg)x_I R^{-1},
\end{align}
where all symbols are defined as in Eqs.(\ref{eq:PMToDPWidth}) and (\ref{eq:PMToIWidth}), and the generalization to a decay to the antiparticle of portal matter with negative $U(1)_D$ charge and an SM fermion is trivial, as is the generalization to decays of $I^-$ field. We note that an equivalent  In the regime in which this class of decay channel is kinematically accessible for \emph{any} portal matter field, the branching fraction of $I^\pm$ to portal matter-SM fermion pair states will be close to 100\%-- this is because the other possible tree-level decay channels are either suppressed by tiny portal matter-SM fermion mixing (in the case of $I^+ \rightarrow \overline{f} f$) or suppressed by three-body phase space (in the case of $I^+ \rightarrow A_D \overline{f} f$ from an intermediate virtual heavier Kaluza-Klein mode of a portal matter fermion field). If there are no portal matter fields light enough for this decay channel to be kinematically accessible, then much like the $W_I$ field in \cite{Rueter:2019wdf} the dominant decay channel will be a three-body decay, $I^+ \rightarrow A_D \overline{f} f$, via the exchange of a virtual portal matter field. Care must be taken to properly compute this decay width, because the virtual portal matter field exchange will involve an exchange over the field's \emph{entire} Kaluza-Klein tower, and so during the computation we must invoke the infinite sums given in Eq.(\ref{eq:FermionSums}). We arrive at the partial width for a given SM fermion species $f$ given by (ignoring the mass of the SM fermions and the dark photon or dark Higgs)
\begin{align}
    &\Gamma_{I^+ \rightarrow \overline{f} f A_D}= \Gamma_{I^+ \rightarrow \overline{f} f h_D} = \frac{x_I R^{-1}}{192 \pi^3} \bigg( \frac{\pi+\tau_X+\omega_X}{\pi(1+x_I^2 \omega_X^2)+\omega_X} \bigg) \int_0^{x_I^2} d x_{23} \int_0^{x_I^2-x_{23}} d x_{12} \mathcal{F}_{I^+ \rightarrow \overline{f}f A_D},\\
    &\mathcal{F}_{I^+ \rightarrow \overline{f}f A_D} \equiv \bigg(1 -\frac{x_{23}+x_{12}}{x_I^2}+ \frac{x_{12} x_{23}}{2 x_I^4} \bigg) \bigg\{ \frac{1}{(\pi+\tau_{f0} + \omega_f)^2} \bigg(\frac{y^f_-}{x_{23}-x_I^2} \mathcal{G}_{f-}(x_{23}) -\frac{y^f_+}{x_{12}-x_I^2} \mathcal{G}_{f+}(x_{12})\bigg)^2 + f \rightarrow F \bigg\}, \nonumber\\
    &\mathcal{G}_{f,F\pm}(x_{ij}) \equiv \frac{x_I (\omega_X-\omega_{f,F}) + (1 + x_I^2 \omega_X^2) \sin(\pi x_I) (- \cos(\pi \sqrt{x_{ij}}) + \sqrt{x_{ij}} \omega_{f,F} \sin(\pi \sqrt{x_{ij}}))}{(-1 + \sqrt{x_{ij}} \tau_{f\pm,F\pm} \omega_{f,F})\cos(\pi \sqrt{x_{ij}})- \sqrt{x_{ij}}(\tau_{f\pm,F\pm} + \omega_{f,F}) \sin(\pi \sqrt{x_{ij}})} \nonumber,
\end{align}
where $y^{f(F)}_\pm$, $\tau_{f(F)\pm}$ and $\omega_{f(F)}$ denote the brane-localized dark Higgs Yukawa couplings, the $\phi=\pi$, and $\phi=0$ brane-localized kinetic terms respectively for the $SU(2)_L$ singlet (doublet) portal matter fields corresponding to the SM fermion $f$ (note that for the sake of brevity when writing this expression we have not used the normalized Yukawa couplings $\tilde{y}^{f,F}_\pm$ used in Eq.(\ref{eq:PMToDPWidth}) here), while $\tau_X$ and $\omega_X$ denote the $\phi=\pi$ and $\phi=0$ brane-localized kinetic terms for the $SU(2)_D$ gauge fields. The integration variable $x_{12} (x_{23})$ corresponds to the invariant mass of the combined four-momentum for the final-state SM fermion (antifermion) with the dark photon or dark Higgs.

With our knowledge of the dominant decay channels of both the lightest $I^\pm$ and portal matter modes, we can now comment holistically about the possible collider signals of both particles. For a given portal matter particle $F^\pm$, the signature present over all of the model parameter space will feature decay into a corresponding SM fermion $f$ via the emission of a dark photon or a dark Higgs. As discussed in detail in \eg, \cite{Rizzo:2018vlb,Kim:2019oyh,Rueter:2019wdf}), this decay features a single a high-$p_T$ SM fermion (or, for color-triplet portal matter, a jet) of the SM species that the portal matter mixes with, plus either missing energy or a highly boosted lepton-jet, depending on the preferred decay channel of the dark photon and dark Higgs. In the event that the $I^\pm$ boson is less massive than the portal matter field in question, it is feasible that an $O(1)$ fraction of portal matter produced will decay via emission of the heavy $I^\pm$. The $I^\pm$ field will then decay (either via intermediate emission of a real portal matter field, if one exists that is lighter than the $I^\pm$, or by virtual portal matter exchange if not) into a dark photon or dark Higgs and a particle-antiparticle pair of SM fermions of some species $q$, which may or may not be the same species as $f$-- the decay chain will therefore take the form
\begin{align}
    F^\pm \rightarrow I^\pm + f \rightarrow A_D/h_D + f + q\overline{q}.
\end{align}
This signature may be quite atypical and distinctive: For color-triplet $f$ and $q$ we might anticipate a single portal matter field to produce a dark photon/dark Higgs (and hence either missing energy or a highly-collimated pair of light SM fermions) plus multiple jets. Given that scenarios in which an $O(1)$ fraction of portal matter decays might occur through this channel, a detailed study of such a decay channel and its signal may be merited.

On its own, the signal of a promptly produced $I^\pm$ boson will closely match the behavior of the $W_I$ boson described in \cite{Rueter:2019wdf}, producing either an on-shell portal matter field (which in turn would decay into a dark Higgs or dark photon plus an SM fermion) or proceeding in a 3-body decay to produce a dark Higgs or dark photon plus a particle-antiparticle pair of the same species of SM fermions.

\subsection{Portal Matter and $SU(2)_D$ Bosons: Production}\label{section:PMProduction}

Having addressed how the most phenomenologically relevant TeV-scale exotics, namely portal matter and $I^\pm$ bosons, decay in our model, we can move on to a survey of the various collider production mechanisms for these fields. 

\begin{figure}
    \centering
    \includegraphics[width=3.5in]{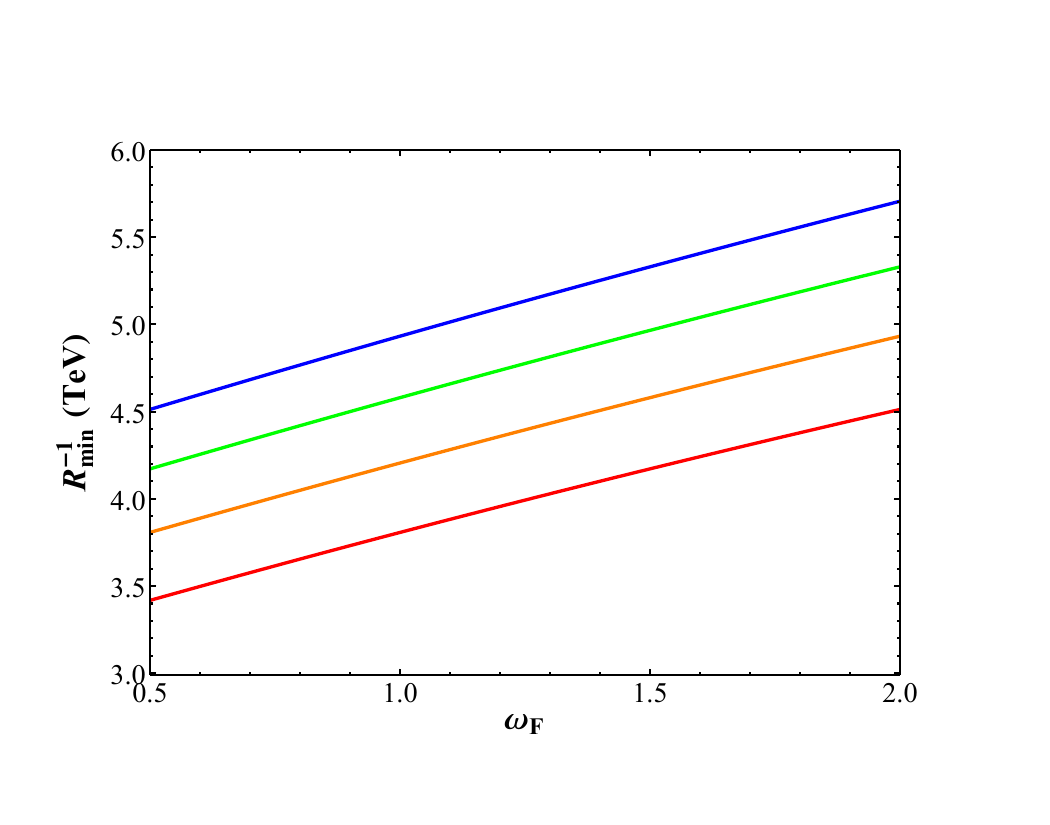}
    \caption{The minimum compactification scale $R^{-1}$ assuming that a portal matter field is constrained to have mass greater than 1.3 TeV, plotted as a function of the portal matter's $\phi=0$ brane-localized kinetic term $\omega_F$, for various choices of the $\phi=\pi$ brane-localized kinetic term $\tau_{F \pm}$. From the top down: $\tau_{F \pm} = $2, 3/2, 1, 1/2. Note that the mass spectrum of the portal matter modes is invariant under the exchange $\omega_F \leftrightarrow \tau_{F \pm}$.}
    \label{fig8}
\end{figure}

Perhaps the most obvious channel, portal matter pair production from either QCD (in the case of color-triplet portal matter) or electroweak interactions (for leptonic portal matter), has been well-explored in the perfectly analogous scenarios of \cite{Rizzo:2018vlb,Kim:2019oyh,Rueter:2019wdf,Guedes:2021oqx,OsmanAcar:2021plv}, and summarized well in \cite{Rizzo:2022qan}, so we see no need to perform the analysis again here-- we shall simply quote some results and comment on their relevance in our context. In both the leptonic and quark-like portal matter cases, the pair production cross sections scale quite strongly with the portal matter mass \cite{Rizzo:2018vlb,Guedes:2021oqx}, so for pair production searches we are especially concerned with the very lightest portal matter states that are present, even in the event of a large ensemble of portal matter states in the model, as long as there are $O(1)$ differences in brane-localized kinetic terms (and hence masses $O(1)$ proportional differences in masses) between portal matter states. In the event that the dark Higgs and dark photon decay invisibly (or are long-lived enough to decay outside of the detector), direct contact can be made with squark and slepton searches for jets or leptons and missing energy to find these lightest portal matter particles. For color-triplet portal matter, these results generally constrain the portal matter mass $m_{Q\pm} \gsim 1.3-1.5 \; \textrm{TeV}$ \cite{Rizzo:2018vlb,Kim:2019oyh}, while for the case of leptonic portal matter the searches of \cite{Guedes:2021oqx,OsmanAcar:2021plv} yield constraints on leptonic portal mass of $m^L_\pm \gsim 0.9-1.1 \; \textrm{TeV}$, provided that the portal matter mixes with first- or second- generation charged leptons. Critically, because portal matter fields will be much lighter than SM Kaluza-Klein tower modes, even these general results can offer constraints on our model parameter space comparable to or even exceeding those we have found from more conventional extra dimensions probes in Section \ref{section:SMExtraDimensions}. To get a (very rough) sense of this, in Figure \ref{fig8} we depict the constraint on $R^{-1}$ assuming that a portal matter field with $\phi=0$ and $\phi=\pi$ brane-localized kinetic terms of $\omega_F$ and $\tau_{F \pm}$, respectively, must be constrained to a mass of $\geq 1.3 \; \textrm{TeV}$.

The constraints on $R^{-1}$ arising from Figure \ref{fig8} are clearly numerically comparable to those emerging from the constraint on the $\rho$ parameter, and depending on relative choices of the brane-localized kinetic terms for the color-triplet portal matter and the $SU(2)_L$ doublet SM fermions that contribute to the tree-level $\rho$ correction, in many regions of parameter space these constraints can be stronger.\footnote{Of course, a true global fit to various precision measurements would be sensitive to a wider variety of SM fermion brane-localized kinetic terms than simply those of the isodoublet leptons. However since we note that a wide variety of a priori uncorrelated brane-localized kinetic terms would enter this analysis, of which only the $\phi=0$ brane-localized kinetic terms are shared by any portal matter fields, there are presumably analogous regions of parameter space in a more robust analysis for which the portal matter searches will be the stronger constraint.} Furthermore, we note that the constraints depicted here would come from the \emph{lightest} color-triplet portal matter fields in the model, and hence can imply global constraints on brane-localized kinetic terms for all quark-like portal matter fields, since larger brane terms will invariably decrease the mass of a given Kaluza-Klein mode. Of course, our model construction involves a far broader range of possible production mechanisms for the states associated with the portal matter sector than simple quark and lepton pair production. We shall now further explore some of these.

\begin{figure}
    \centerline{\includegraphics[width=3.5in]{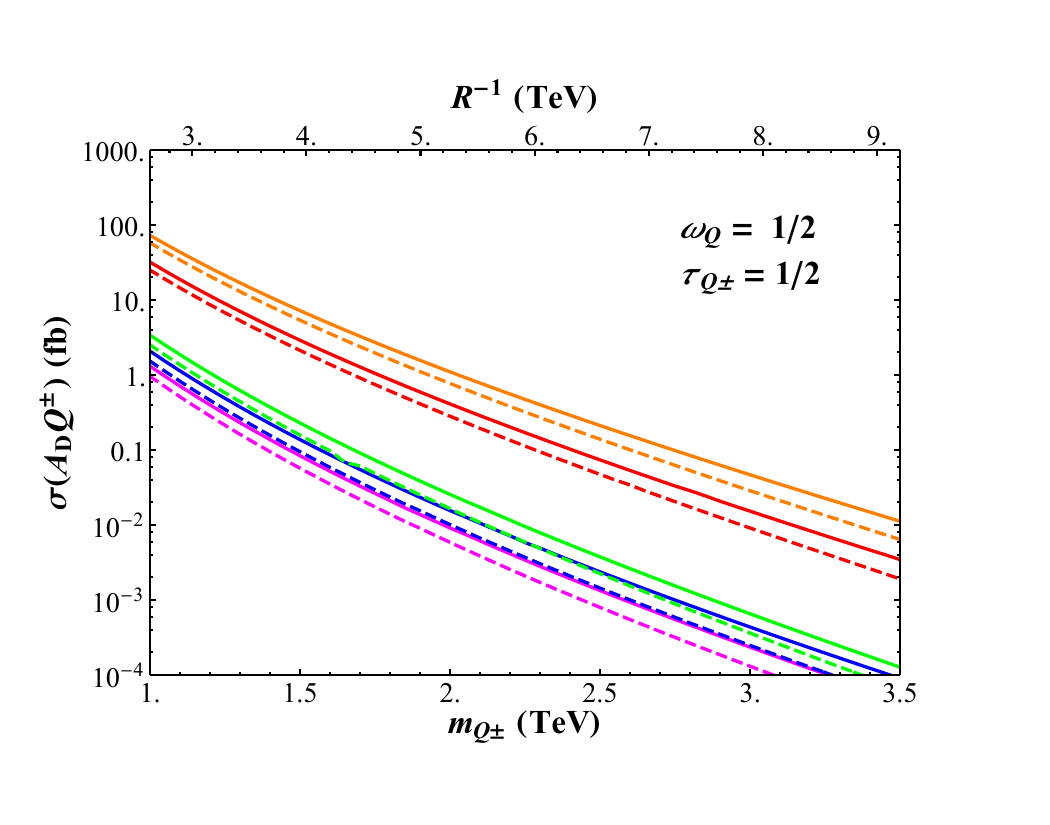}
    \hspace{-0.75cm}
    \includegraphics[width=3.5in]{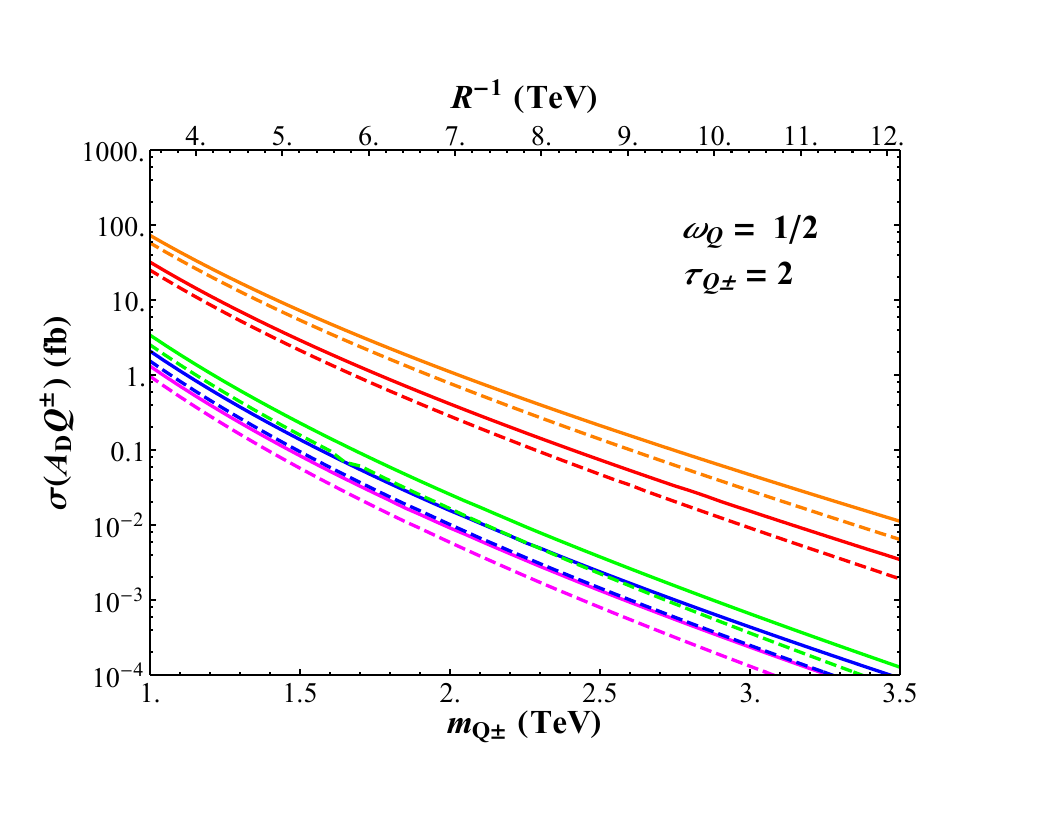}}
    \vspace*{-1.5cm}
    \centering
    \includegraphics[width=3.5in]{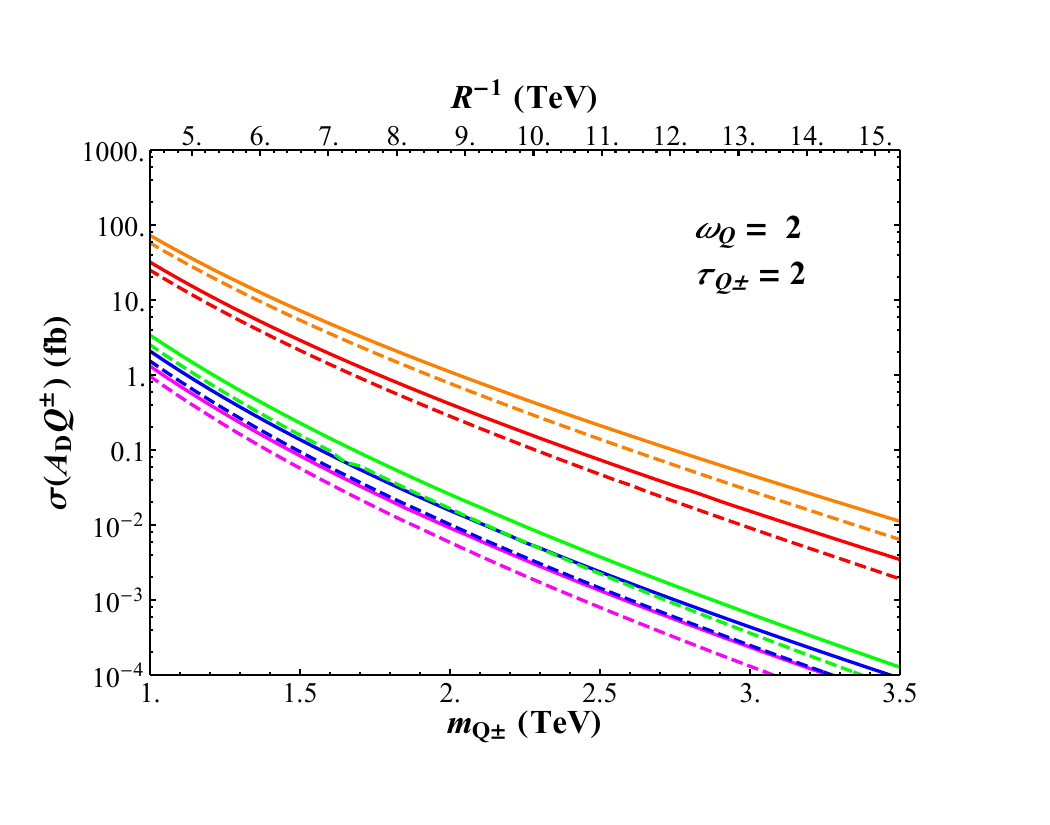}
    \caption{The total scattering cross section for the process $qg \rightarrow Q^\pm A_D/h_D$ at the LHC at $\sqrt{s} = 14 \; \textrm{TeV}$ (solid) and $\sqrt{s} = 13 \; \textrm{TeV}$ (dashed), assuming $\tilde{y}^Q_\pm = 1$, where $\tilde{y}^Q_\pm$ is defined as $\tilde{y}^F_\pm$ is in Eq.(\ref{eq:PMToDPWidth}). Cross sections are given from the top down assuming the portal matter field is coupled to the $u$ (orange), $d$ (red) $s$ (green), $c$ (blue), $b$ (magenta). Brane-localized kinetic terms for the portal matter field are listed on the chart, and the compactification scale $R^{-1}$ is depicted on the upper x-axis of the chart-- note that our subscript $Q$ implies we are considering a weak isospin doublet portal matter field (see Table \ref{table:fermionModelContent}), but these results are agnostic to whether the quark in question is an isospin doublet or singlet.}
    \label{fig9}
\end{figure}

In contrast to portal matter pair production, the remaining collider production mechanisms we consider for portal matter states and $I^\pm$ bosons will involve $SU(2)_D$ gauge interactions, and will therefore be much more model-dependent. Many of the production mechanisms we consider will be familiar to readers of \cite{Rueter:2019wdf} for that model's portal matter and its $W_I$ boson. However, the multiplicity of portal matter fields in our construction, combined with the fact that many process amplitudes are significantly modified by the effects of embedding the theory in an extra dimension, make it worthwhile to re-evaluate these processes within our new model. We begin with a look at production of a single portal matter field and a dark photon or a dark Higgs. In the limit the dark photon/dark Higgs mass can be ignored, we find a leading order cross section of
\begin{align}
    &\sigma (q g \rightarrow Q^\pm A_D) = \frac{(\tilde{y}^Q_\pm)^2 \alpha_s (s-m_{Q\pm}^2)}{192 s^2} \mathcal{F}_{qg \rightarrow Q^\pm A_D},\\
    &\mathcal{F}_{qg \rightarrow Q^\pm A_D} \equiv\int d \cos\theta \bigg\{ -2 m_{Q\pm}^2 \bigg( \frac{1}{s} + \frac{1}{u'} \bigg) \bigg( 1 + \frac{m_{Q\pm}^2}{u'} \bigg) - \frac{t^2}{s u'} \bigg\}, \nonumber\\
    &t = (m_{Q\pm}^2-s)\sin^2 \bigg( \frac{\theta}{2} \bigg), \;\;\; u' = -m_{Q\pm}^2 + (m_{Q\pm}^2-s)\cos^2 \bigg( \frac{\theta}{2} \bigg), \nonumber
\end{align}
where $m_{Q\pm}$ is the mass of the $Q^\pm$ portal matter field, and $\tilde{y}^Q_\pm$ is defined as $\tilde{y}^F_+$ in Eq.(\ref{eq:PMToDPWidth}). Notably, this amplitude depends explicitly \emph{only} on the coupling constant $\tilde{y}^Q_+$ and the portal matter mass-- all explicit dependence on the brane-localized kinetic terms in the model are simply absorbed into the definition of $\tilde{y}^Q_\pm$. The only change in the physics that occurs when the brane-localized kinetic terms of the theory are adjusted for a fixed portal matter mass and $\tilde{y}^Q_\pm$ is the relationship between the portal matter mass and the compactification scale of the extra dimension. In Figure \ref{fig9}, we depict the total cross section for the single production of a given portal matter mode that mixes with various SM fermions at the $\sqrt{s} = 13 \; \textrm{TeV}$ and $\sqrt{s} = 14 \; \textrm{TeV}$. Notably, for $O(1)$ selections of the portal matter's brane-localized kinetic terms, large compactification scales can still produce production cross sections for this process as high as the multi-fb level. In fact, referencing the results of \cite{Rizzo:2018vlb,Rueter:2019wdf}, we see that for reasonable $O(1)$ values of $\tilde{y}^F_\pm$ the production cross section for single portal matter fields is actually comparable to that of the QCD production of portal matter pairs, albeit heavily dependent on which SM field the portal matter mixes with. It is possible, then, that comparable limits on the portal matter parameter space might be placed using using monojet searches at the LHC, assuming dark photons and dark Higgses will decay invisibly or outside of the detector. Furthermore, as in the case of the portal matter pair production, the low mass of the portal matter fields relative to the compactification scale in turn allows for relatively mild limits on the portal matter field masses to translate to much more robust limits on $R^{-1}$, albeit the precise relationship between these two mass scales is heavily dependent on the brane-localized kinetic terms.

Much like \cite{Rueter:2019wdf}, there are further signals in the model which stem from the production of the new gauge boson $I^\pm$. The simplest of these is merely the analogue of the $g q \rightarrow Q^\pm A_D/h_D$ process discussed above, with an $I^\pm$ boson produced instead of a dark photon or dark Higgs. The production cross section for these events is given by
\begin{align}\label{eq:gqToQI}
    &\sigma(qg \rightarrow Q^\pm I^\mp) = \frac{g_D^2 \alpha_s \beta_{QI} (s-m_{Q\pm}^2-m_I^2)}{96 s^2} (C_{QI}^\pm)^2 \mathcal{F}_{qg \rightarrow Q^\pm I^\mp},\\
    &\mathcal{F}_{qg \rightarrow Q^\pm I^\mp} \equiv \bigg( 2 + \frac{m_{Q\pm}^2}{m_I^2} \bigg) \bigg\{ 2 \Delta m^2 \bigg( \frac{1}{s}+\frac{1}{u'}\bigg)\bigg( 1+\frac{m_{Q\pm}^2}{u'}\bigg)-\frac{2 m_I^2 \Delta m^2}{s u'} - \frac{(m_I^2-t)^2}{s u'} + \frac{4 m_I^2}{(m_{Q\pm}^2 + 2 m_I^2)}\bigg\}, \nonumber\\
    &\Delta m^2 \equiv m_{Q\pm}^2-m_I^2, \;\;\; \beta_{QI} \equiv \sqrt{1 - \frac{4 m_{Q\pm}^2 m_I^2}{(s-m_{Q\pm}^2-m_I^2)^2}},\nonumber\\
    &u' \equiv -m_{Q\pm}^2-\frac{(s-m_{Q\pm}^2-m_I^2)}{2}(1 + \beta_{QI} \cos\theta), \;\;\; t \equiv -\frac{(s-m_{Q\pm}^2-m_I^2)}{2}(1 - \beta_{QI} \cos\theta), \nonumber
\end{align}
where $m_I$ is the mass of the lightest $I^\pm$ boson. The dimensionless constant $C^\pm_{QI}$ is simply the coupling constant of the $I^\mp$ boson to a vertex featuring $q$ and $Q^\pm$ in units of the 4-dimensional dark photon coupling $g_D$. It depends on the bulk profiles of the $I^\mp$ boson and the $Q^\pm$ portal matter fermion, and is given by
\%
\begin{align}\label{eq:CQIDef}
    &C^\pm_{QI} = \frac{\sqrt{2}N^{Q_\pm}}{\sqrt{\pi (1 + x_I^2 \omega_X^2) + \omega_X}} \sqrt{\frac{\pi + \tau_X + \omega_X}{\pi+ \tau_{Q0} + \omega_Q}} \frac{m_I^2}{\Delta m^2} \mathcal{C}^\pm_{QI},\\
    &\mathcal{C}^\pm_{QI} \equiv \bigg\{ (\omega_X - \omega_Q) - \frac{m_{Q\pm} \tau_{Q\pm}}{m_I} \bigg( \frac{(1 + m_I^2 R^2 \omega_X^2)(1 + m_{Q\pm}^2 R^2 \omega_Q^2)}{(1 - m_{Q\pm}^2 R^2 \tau_{Q\pm} \omega_Q)} \bigg) \sin (\pi m_{Q \pm} R) \sin (\pi m_I R) \bigg\} \nonumber,
\end{align}
where $\omega_X$ and $\tau_X$ denote the $\phi=0$ and $\phi=\pi$ and brane-localized kinetic terms for the $SU(2)_D$ gauge fields, and $\omega_Q$ denotes the $\phi=0$ brane-localized kinetic term for the fermion $Q$, and $\tau_{Q \pm(0)}$ denotes the brane-localized kinetic term for the portal matter (SM) fermion $Q$. $N^{Q_\pm}$ is defined in exact analogy to $N^{F_+}$ in Eq.(\ref{eq:PMToDPWidth}). In spite of its complicated form, the constant $C^\pm_{QI}$ in Eq.(\ref{eq:CQIDef}) plays only a limited role in altering the value of the cross section in Eq.(\ref{eq:gqToQI}), because its numerical magnitude is invariably within $\sim 25\%$ of unity for $O(1)$ selections of brane-localized kinetic terms.

\begin{figure}
    \centerline{\includegraphics[width=3.5in]{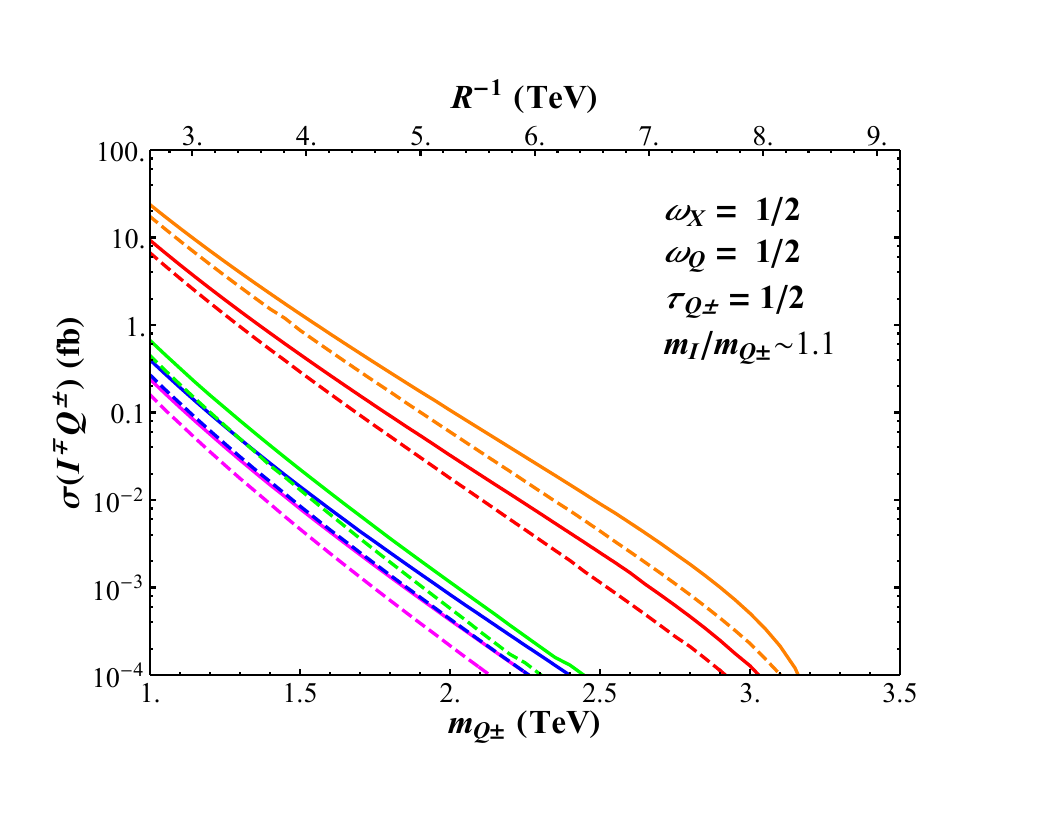}
    \hspace{-0.75cm}
    \includegraphics[width=3.5in]{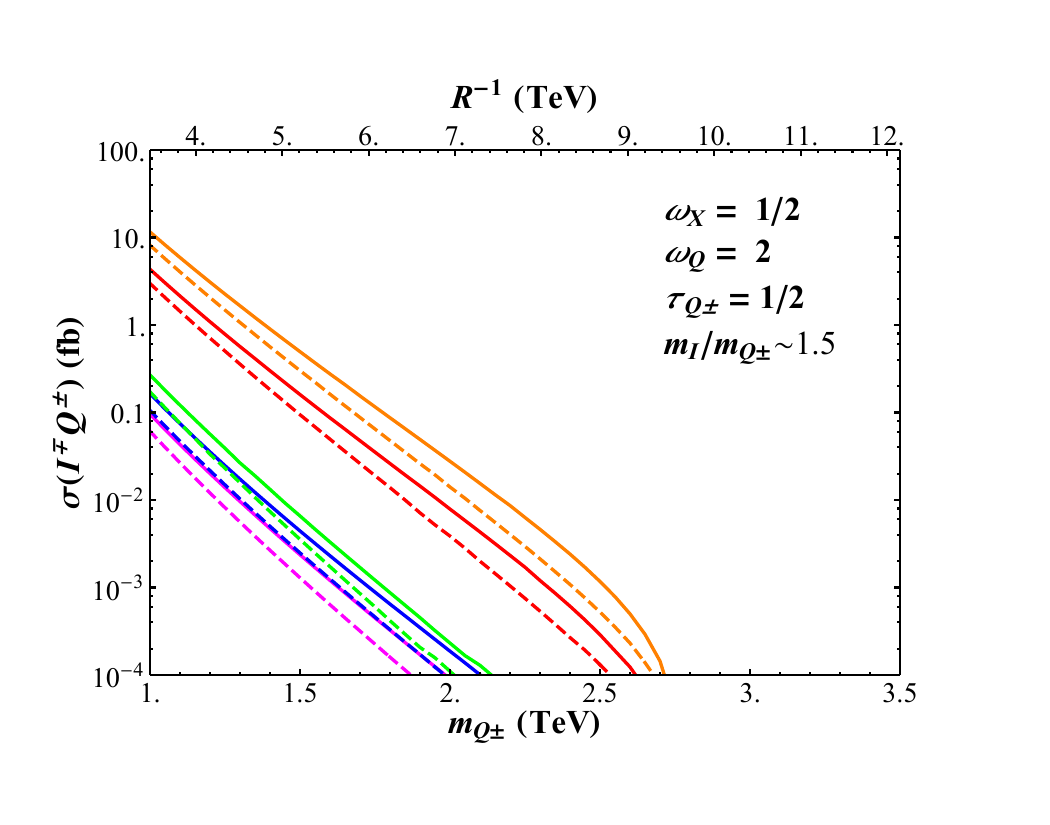}}
    \vspace*{-1cm}
    \centerline{\includegraphics[width=3.5in]{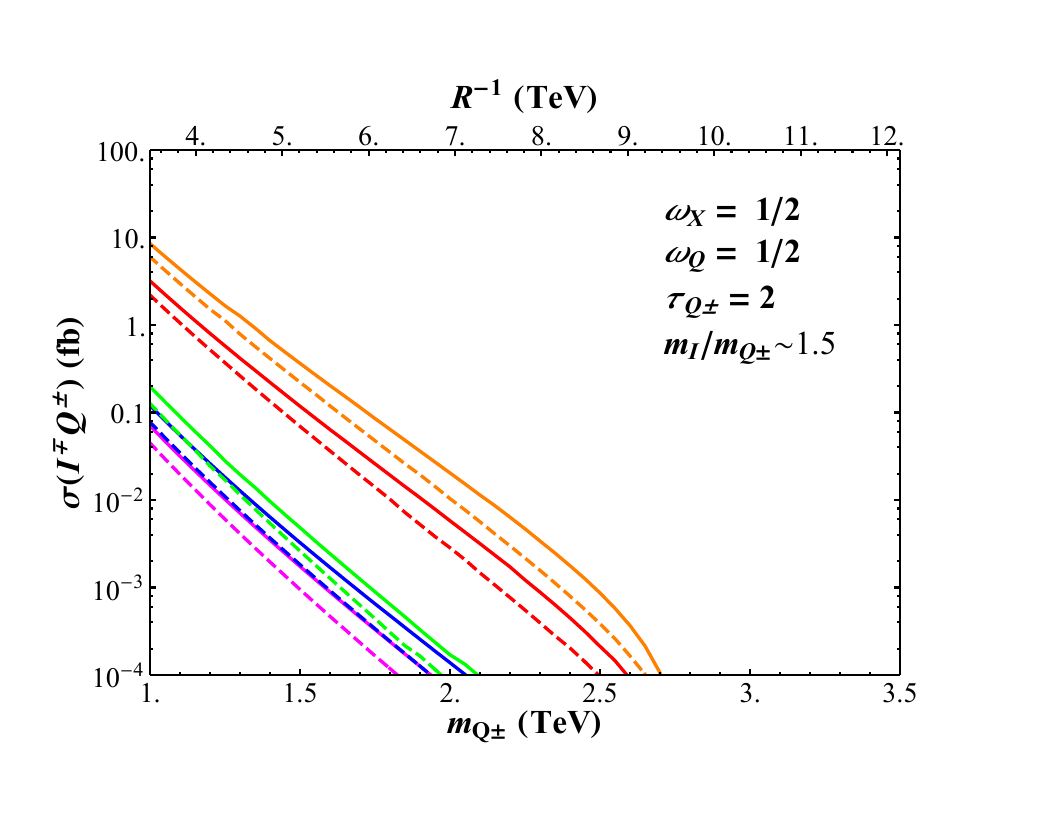}
    \hspace{-0.75cm}
    \includegraphics[width=3.5in]{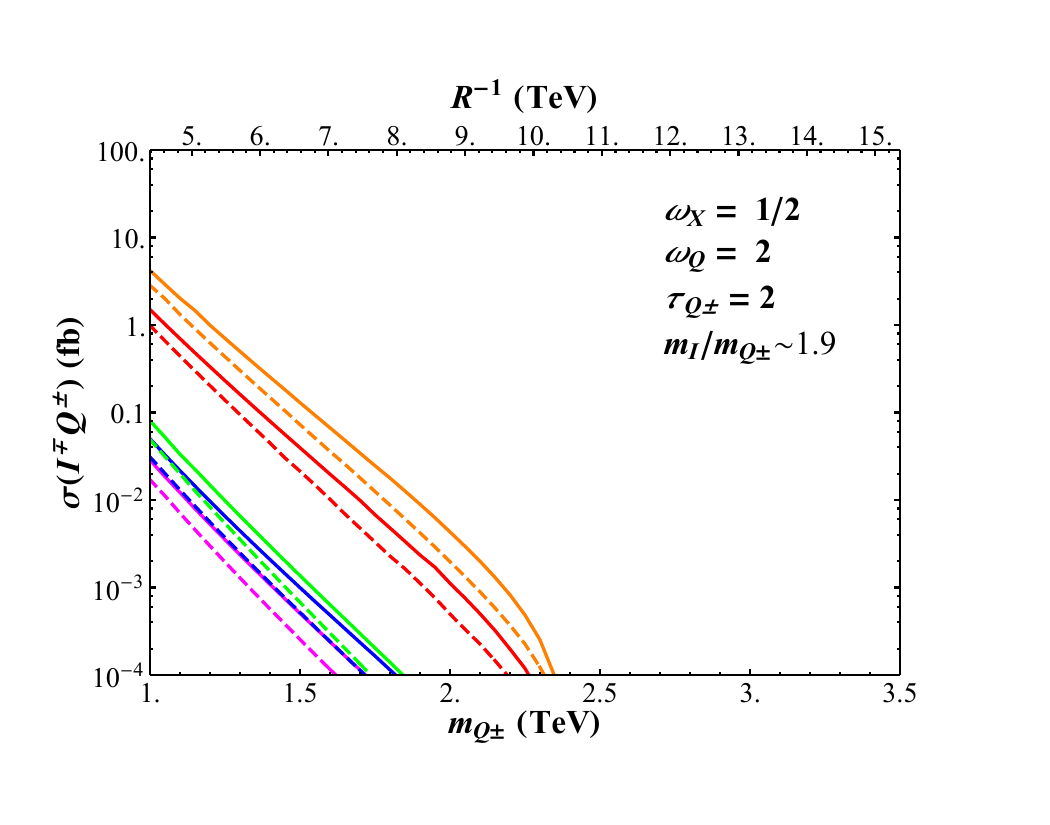}}
    \vspace*{-0.75cm}
    \caption{The total scattering cross section for the process $qg \rightarrow Q^\pm I^\mp$ at the LHC at $\sqrt{s} = 14 \; \textrm{TeV}$ (solid) and $\sqrt{s} = 13 \; \textrm{TeV}$ (dashed), assuming $g_D = g_L$ and $\omega_X = 1/2$. Cross sections are given from the top down assuming the portal matter field is coupled to the $u$ (orange), $d$ (red) $s$ (green), $c$ (blue), $b$ (magenta). Brane-localized kinetic terms for the portal matter and $I$ fields are listed on the chart, and the compactification scale $R^{-1}$ is depicted on the upper x-axis of the chart. The brane-localized kinetic terms $\tau_X$ and $\tau_{Q0}$, as appearing in Eq.(\ref{eq:gqToQI}), only multiplicatively rescale the cross sections by a factor $(\pi+\tau_X+\omega_X)/(\pi+\tau_{Q0}+\omega_Q) \sim 1$ (assuming $O(1)$ brane terms), so they are assumed to be 0 here.}
    \label{fig10}
\end{figure}

\begin{figure}
    \centerline{\includegraphics[width=3.5in]{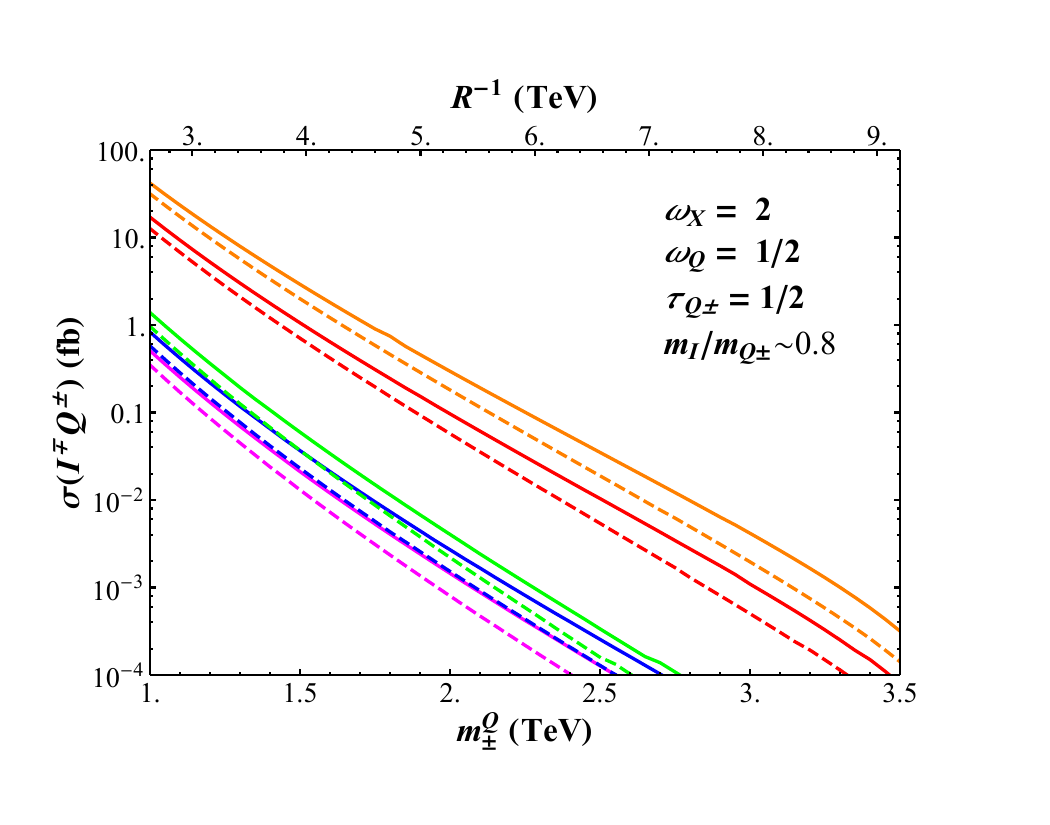}
    \hspace{-0.75cm}
    \includegraphics[width=3.5in]{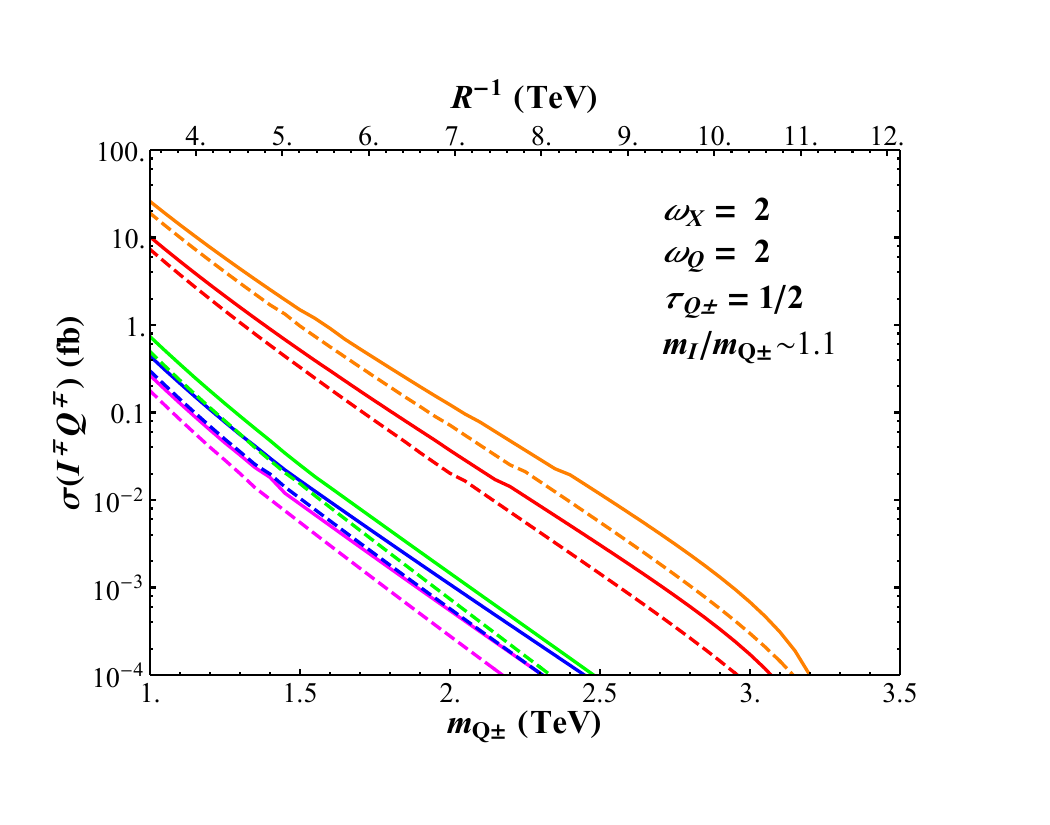}}
    \vspace*{-1cm}
    \centerline{\includegraphics[width=3.5in]{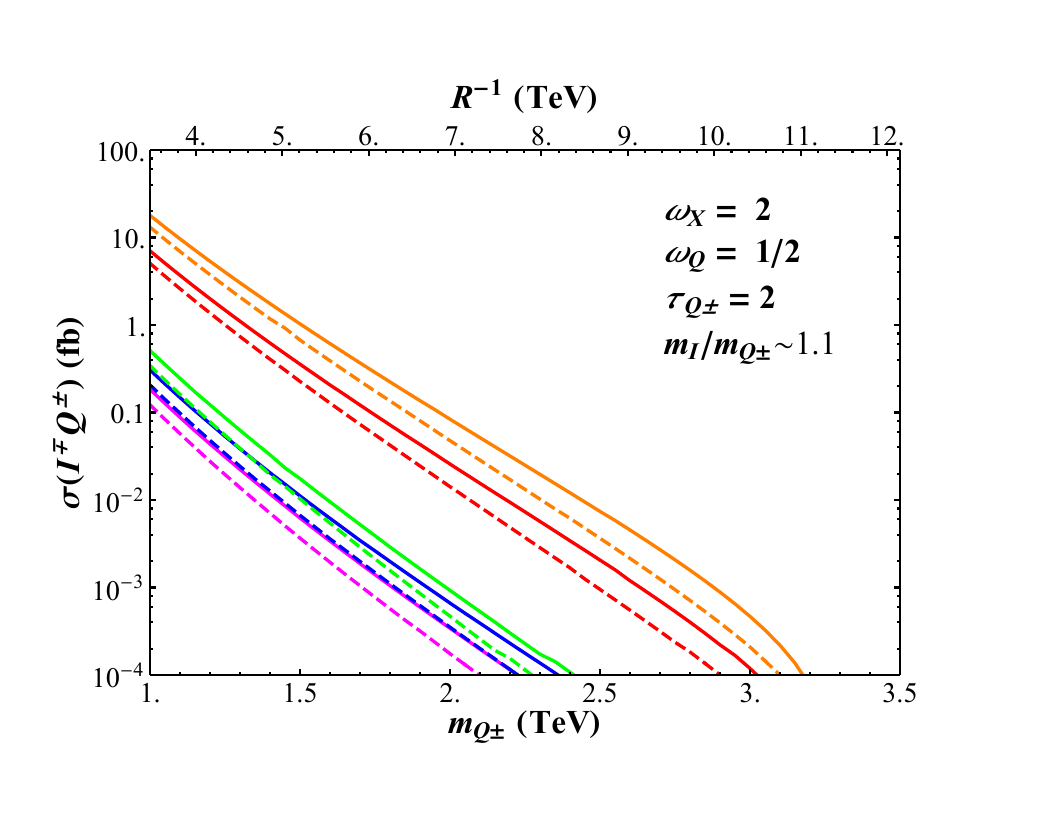}
    \hspace{-0.75cm}
    \includegraphics[width=3.5in]{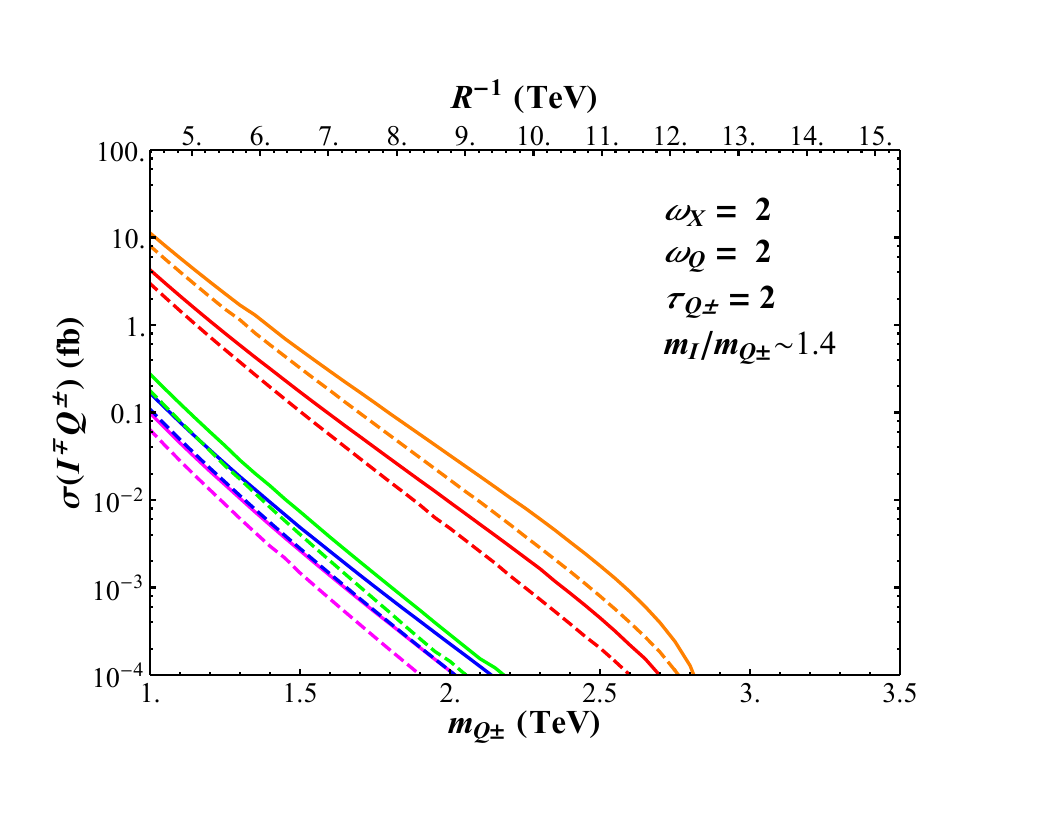}}
    \vspace*{-0.75cm}
    \caption{As in Figure \ref{fig10}, but for different brane-localized kinetic term choices. The total scattering cross section for the process $qg \rightarrow Q^\pm I^\mp$ at the LHC at $\sqrt{s} = 14 \; \textrm{TeV}$ (solid) and $\sqrt{s} = 13 \; \textrm{TeV}$ (dashed), assuming $g_D = g_L$ and $\omega_X = 2$. Cross sections are given from the top down assuming the portal matter field is coupled to the $u$ (orange), $d$ (red) $s$ (green), $c$ (blue), $b$ (magenta). Brane-localized kinetic terms for the portal matter and $I$ fields are listed on the chart, and the compactification scale $R^{-1}$ is depicted on the upper x-axis of the chart.}
    \label{fig11}
\end{figure}

We depict the total cross section for the process $qg \rightarrow Q^\pm I^\mp$ in Figures \ref{fig10} and \ref{fig11} at the LHC with $\sqrt{s} = 13 \; \textrm{TeV}$ and $\sqrt{s} = 13 \; \textrm{TeV}$. As we can see in these Figures, the cross section for this process depends modestly on selections of brane-localized kinetic terms-- varying by as much as an order of magnitude for different choices in parameter space. It should be noted, however, that this variation is almost entirely due to the variation in the relationship between the $I$ boson and the portal matter masses-- as we discussed previously the effective coupling constant between the $I$ bosons, portal matter, and an SM fermion depends only quite weakly on the choice of brane-localized kinetic terms.

If there is a kinematically accessible two-body decay for the $I^\pm$ gauge boson into a portal matter-SM fermon pair, then the experimental signature for this process can closely resemble that of portal matter pair production\footnote{If not, then the dominant decay channel for the $I$ bosons will be a three-body decay via an off-shell portal matter field, leading to the same possible final states as decay through on-shell portal matter.}-- with the exception that the two portal matter fields produced may be associated with different SM fields, and if $I^\pm$ decays hadronically there will be an extra jet associated with the $I^\pm$ decay.\footnote{As noted in \cite{Rueter:2019wdf}, this might easily be mistaken for QCD ISR.} For example, if the $I^\pm$ boson decays into leptonic portal matter and the dark photons/dark Higgses decay invisibly or outside of the detector, the final state might consist of an electron-positron or muon-antimuon pair, a jet (which depending on the identity of the promptly produced portal matter field, might be $b-$tagged or $t-$tagged), and missing energy. In the event that the dark photons and/or dark Higgses emitted by the decaying portal matter and $I^\pm$ decay visibly, it may be possible to reconstruct the $I^\pm$ mass peak, which as noted in \cite{Rueter:2019wdf} can drastically reduce backgrounds. A detailed study of the experimental constraints on this channel is far beyond the scope of this paper, but may be of interest, especially given the fact that both this work and \cite{Rueter:2019wdf} predict production of this type when simple extensions to the minimal model of portal matter presented in \cite{Rizzo:2018vlb} are constructed.

\begin{figure}
    \centerline{\includegraphics[width=3.5in]{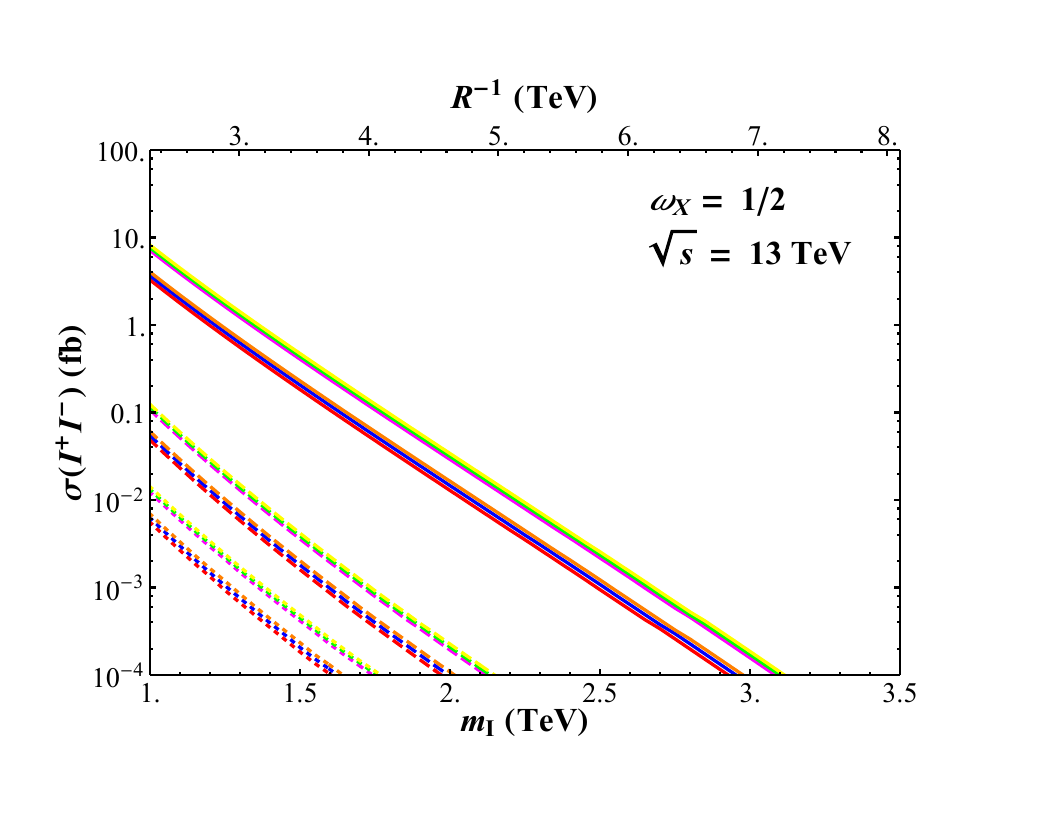}
    \hspace{-0.75cm}
    \includegraphics[width=3.5in]{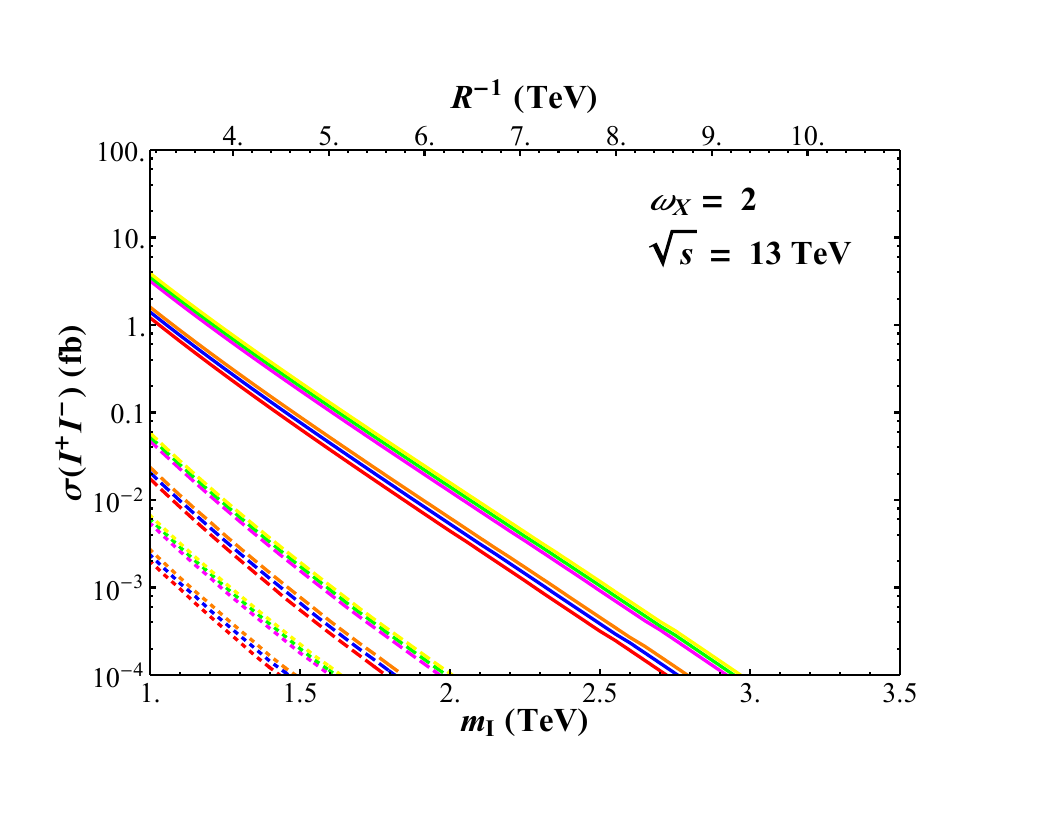}}
    \vspace*{-1cm}
    \centerline{\includegraphics[width=3.5in]{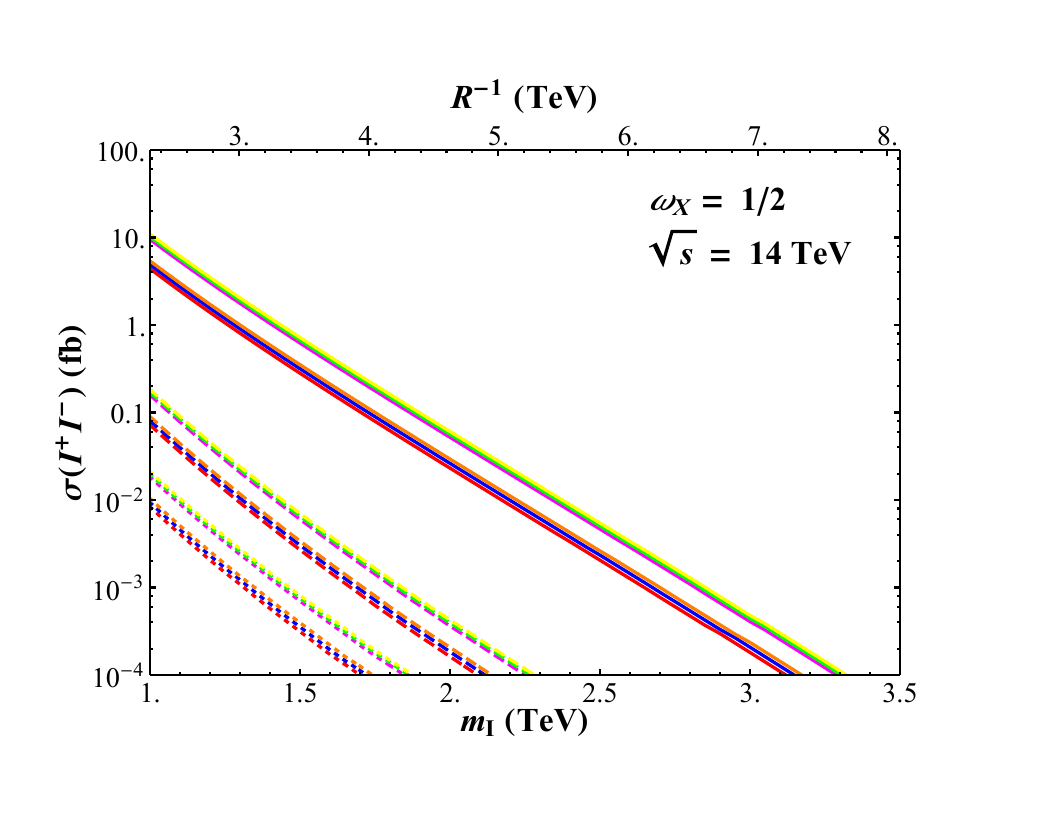}
    \hspace{-0.75cm}
    \includegraphics[width=3.5in]{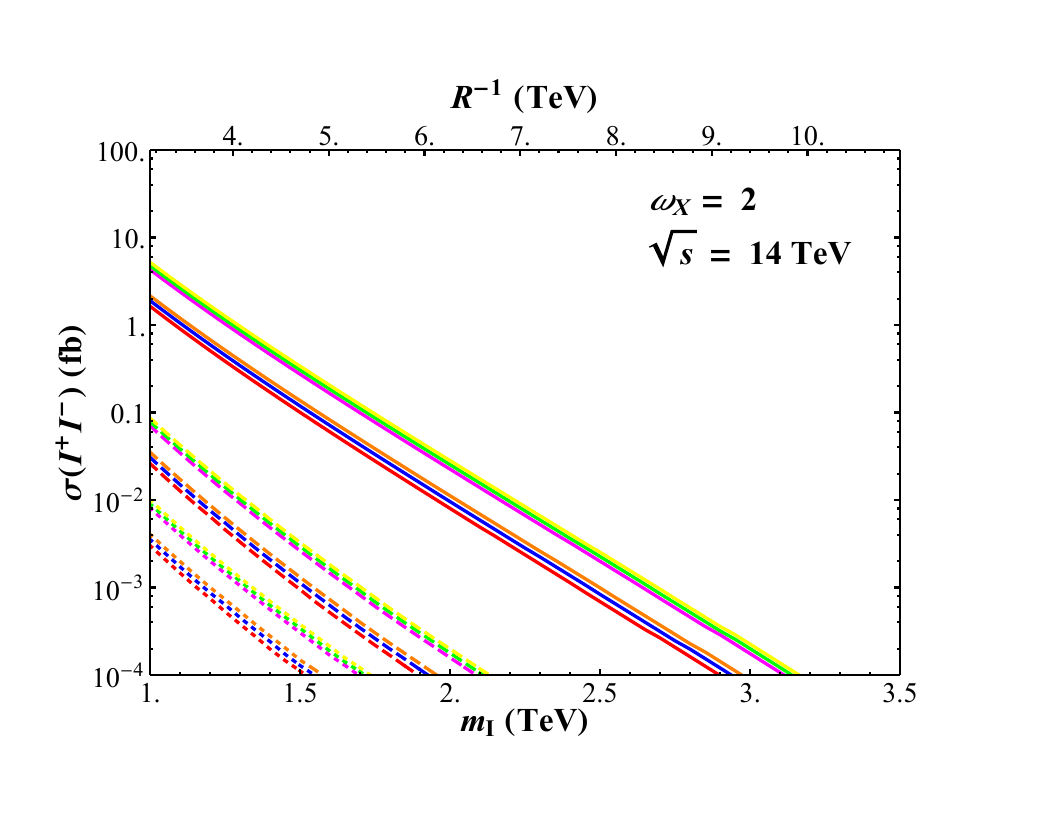}}
    \vspace*{-0.75cm}
    \caption{The production cross section for the process $q \overline{q} \rightarrow I^+ I^-$ at the LHC at $\sqrt{s}=13 \; \textrm{TeV}$ (Top) and $\sqrt{s}=14 \; \textrm{TeV}$ (Bottom). Only one generation of quarks at a time is assumed to be in portal matter multiplets, either the first (Solid), the second (Dashed), or the third (Dotted). The fermion brane-localized kinetic terms are assumed to be degenerate between all species, so that in reference to Table \ref{table:fermionModelContent}, we have assumed $\tau_{Q0,+,-}=\tau_{u0,+,-}=\tau_{d0,+,-}$ and $\omega_Q=\omega_u=\omega_d$. Different choices of brane-localized kinetic terms are depicted, with $(\omega_Q, \tau_{Q+},\tau_{Q_-}) =$ (1/2,1/2,1/2) (Red), (2,1/2,1/2) (Magenta), (1/2,1/2,2) (Blue), (2,1/2,2) (Green), (1/2,2,2) (Orange), (2,2,2) (Yellow), while the choice of $\omega_X$ is depicted on the chart. We assume $\tau_{Q0}=\tau_X=0$, since these parameters simply rescale the amplitude by an $\sim 1$ factor, and $g_D = g_L$.}
    \label{fig12}
\end{figure}

In addition to being produced with a single portal matter field, there are two processes by which $I^\pm$ bosons might be produced without a prompt portal matter field, either as an $I^+ I^-$ pair or in association with a single dark photon or dark Higgs. We shall consider pair production first. In contrast to the analogous process in \cite{Rueter:2019wdf} (pair production of $W_I$ bosons, in their notation), $I^+ I^-$ pair production receives contributions from $t$-channel exchanges of the entire tower of Kaluza-Klein portal matter states; as a result, the expression for the cross section becomes quite complicated. Another interesting distinction between the $I^+ I^-$ production process in our model and that of the analogous $W_I$ boson pairs in \cite{Rueter:2019wdf} stems from how the cross section maintains unitarity: In the case of the model in \cite{Rueter:2019wdf}, the amplitude of $W_I$ pair production from the $s$-channel exchange of a heavy $U(1)_D$-neutral gauge boson, $Z_I$, interfered destructively with the amplitude contribution from $t$-channel exchange of portal matter fermions. In our construction, however, the amplitude contribution from the $s$-channel exchange of the only suitable gauge bosons in the model, $A_D$ and its associated Kaluza-Klein tower, are negligibly small due to the heavy suppression of $A_D$'s coupling to SM fermion pairs (taking place at leading order only through kinetic mixing). Instead, unitarity is maintained by destructive interference between the different portal matter fields which mix with a given SM fermion: Those with a $U(1)_D$ charge of $+1$ and those with a $U(1)_D$ charge of $-1$ interfere and prevent the cross section from growoing uncontrollably. We arrive at the expression
\begin{align}
    &\sigma(q \overline{q} \rightarrow I^+ I^-) = \frac{g_D^4 \sqrt{1-\frac{4 m_I^2}{s}}}{128 \pi s} \int d \cos\theta \, \mathcal{F}_{q \overline{q} \rightarrow I^+ I^-},\\
    &\mathcal{F}_{q \overline{q} \rightarrow I^+ I^-} \equiv \bigg(\frac{\pi + \tau_X + \omega_X}{\pi + \tau_{Q0} + \omega_Q}\bigg)^2 \bigg\{ (A_Q^2 + 2 B_Q^2 + 2 C_Q^2) \bigg(\frac{t u}{m_I^4} - 1 \bigg) + \frac{4 s}{m_I^2}(A_Q^2-B_Q^2+C_Q^2) \bigg\} + Q \rightarrow q, \nonumber\\
    &A_{Q,q} \equiv t \mathcal{H}_{Q,q-}(\sqrt{t R^2}) - u \mathcal{H}_{Q,q+}(\sqrt{u R^2}),\;\;\; B_{Q,q} \equiv m_I^2 \big( \mathcal{H}_{Q,q-}(\sqrt{t R^2}) - \mathcal{H}_{Q,q+}(\sqrt{u R^2}) \big), \nonumber\\
    &C_{Q,q} \equiv m_I^2 \big( \mathcal{H}_{Q,q-}(\sqrt{t R^2}) + \mathcal{H}_{Q,q+}(\sqrt{u R^2}) \big), \nonumber
\end{align}
for the $I^+ I^-$ pair production cross section, where the subscript $Q(q)$ refers to brane-localized kinetic terms associated with the weak isospin doublet (singlet) SM and portal matter fields, $s$, $t$, and $u$ are the Mandelstam variables, $\theta$ is the center-of-mass scattering angle, and the function $\mathcal{H}_{Q,q \pm}$ is given by
\begin{align}
    &\mathcal{H}_{Q,q \pm} (\sqrt{t R^2} ) \equiv \bigg( \frac{1}{t - m_I^2} + \frac{2(-2 m_I^2 + t)(-\omega_X + \omega_{Q,q})}{(t - m_I^2)^2 (\pi(1 + m_I^2 R^2 \omega_X^2) + \omega_X)} + \frac{2 m_I^2}{(t - m_I^2)^2} \tilde{\mathcal{H}}_{Q,q \pm} (\sqrt{t R^2}) \bigg) \nonumber\\
    &\tilde{\mathcal{H}}_{Q,q \pm} (\sqrt{t R^2}) \equiv \frac{\mathcal{A} + \mathcal{B}\, \alpha_{\sqrt{t R^2}}(\tau_{Q,q\pm},\pi) + \mathcal{C} \, \beta_{\sqrt{t R^2}}(\omega_{Q,q},\pi)}{t R^2 (\pi(1 + x_I^2 \omega_X^2) + \omega_X) \big[ \sqrt{t R^2} \tau_{Q,q \pm} \alpha_{\sqrt{t R^2}}(\omega_{Q,q},\pi) -\beta_{\sqrt{t R^2}}(\omega_{Q,q},\pi)\big]},\\
    &\mathcal{A} \equiv 2 x_I t R^2 (1 + x_I^2 \omega_X^2)(\omega_{Q,q} - \omega_X ) \sin (\pi x_I), \;\;\; \mathcal{B} \equiv x_I^2 (\omega_{Q,q} - \omega_X )^2, \;\;\; \mathcal{C} \equiv t R^2 \tau_{Q,q \pm} (1 + x_I^2 \omega_X^2) \nonumber,
\end{align}
where the functions $\alpha$ and $\beta$ are defined in Eq.(\ref{eq:FGdef}), in Section \ref{section:fermions}, and $x_I \equiv m_I R$. The total production cross section for this process at the LHC at $\sqrt{s} = 13 \; \textrm{TeV}$ and $\sqrt{s} = 14 \; \textrm{TeV}$ is depicted for various choices of brane-localized kinetic terms and different generations of quarks embedded in portal matter multiplets in Figure \ref{fig12}. In contrast to the $W_I^\dagger W_I$ production process in described in \cite{Rueter:2019wdf}, we find that the cross sections for pair production of this form are comparable to the results from production of a single $I^\pm$ associated with a portal matter field (likely because the absence of an $s$-channel amplitude in our setup signficantly alters the cross section computation), but lack a region of parameter space in which there might be a resonant enhancement of the production rate. Instead, we see that selecting differing brane-localized kinetic terms can have at most an $O(1)$ effect on the predicted cross section. It is also interesting to note that for larger $\omega_X$, the cross section for this process is reduced, but the compactification scale $R^{-1}$ associated with a given value of $m_I$ increases. Hence, at $\sqrt{s} = 13 \; \textrm{TeV}$, if we assume that the entire first generation of quarks takes on universal brane-localized kinetic terms of $(\omega_Q,\tau_{Q+},\tau_{Q-})=(2,1/2,2)$, a constraint on this channel of $\leq 1 \; \textrm{fb}$ would translate to a bound $m_I \gsim 1.3 \; \textrm{TeV}$ if $\omega_X = 1/2$, but only  $m_I \gsim 1.2 \; \textrm{TeV}$ if $\omega_X = 2$. However, these bounds translate to a constraint of $R^{-1} \gsim 3.1 \textrm{TeV}$ for $\omega_X=1/2$ and $R^{-1} \gsim 3.8 \textrm{TeV}$ for $\omega_X=2$, making the weaker constraint on $m_I$ the stronger constraint on the compactification scale. Assuming the decay of $I^\pm$ into on-shell portal matter is open, the signal for $I^+ I^-$ pair production can closely match the signals for single $I^+$ production in association with a portal matter field or portal matter pair production, albeit with additional jets and/or high-$p_T$ leptons depending on whether the $I^\pm$ bosons decay hadronically or leptonically. As before, if the dark photons or dark Higgses decay visibly, it may be possible to reconstruct the $I^\pm$ mass peaks and therefore drastically reduce backgrounds for these searches.

The final process which we consider is another diboson signal, in this case the production of a single $I^\pm$ boson and a dark photon or dark Higgs. As in the case for $I^+ I^-$ pair production, we see here that the contribution to the amplitude from $t$-channel exchanges of portal matter must include exchanges of all Kaluza-Klein modes of the portal matter fields. Unlike the pair production cross section, however, here there does exist a significant $s$-channel amplitude that contributes to the cross section-- specifically, an $s$-channel exchange of an $I^\pm$ boson. For the total cross section, we arrive at the expression
\begin{align}\label{eq:qqToIA}
    &\sigma(q \overline{q} \rightarrow I^\pm A_D) = \sigma(q \overline{q} \rightarrow I^+ h_D) = \frac{g_D^2 (s - m_I^2)}{128 \pi s^2} \frac{(\pi + \tau_X + \omega_X)}{(\pi(1 + x_I^2 \omega_X^2) + \omega_X)} \int d\cos\theta \, \mathcal{F}_{q \overline{q} \rightarrow I^\pm A_D},\\
    &\mathcal{F}_{q \overline{q} \rightarrow I^\pm A_D} \equiv \frac{(2 m_I^2 s + t u)}{(\pi + \tau_{Q0} + \omega_Q)^2} \bigg[ \frac{y^Q_+}{(t - m_I^2)} \mathcal{J}_{Q_+} (\sqrt{t R^2}) - \frac{y^Q_-}{(u - m_I^2)} \mathcal{J}_{Q_-} (\sqrt{R^2 u}) \bigg] + Q \rightarrow q \nonumber,\\
    &\mathcal{J}_{Q_\pm} (\sqrt{t R^2}) \equiv \frac{x_I (\omega_X - \omega_Q) - (1 + x_I^2 \omega_X^2) \sin(x_I \pi) \beta_{\sqrt{t R^2}}(\omega_Q,\pi)}{\sqrt{t R^2} \tau_{Q\pm} \alpha_{\sqrt{t R^2}}(\omega_Q, \pi)-\beta_{\sqrt{t R^2}}(\omega_Q, \pi)} \nonumber
\end{align}
where $s$, $t$, and $u$ are still the Mandelstam variables, and $\theta$ is the center-of-mass scattering angle. The subscripts and superscripts $Q(q)$ again denote brane-localized kinetic terms and Yukawa couplings associated with the weak isospin doublet (singlet) fermion fields-- in the notation of Table \ref{table:fermionModelContent}, the sub- or superscript $q$ will denote either $d$ or $u$, depending on whether the initial-state quarks of the process are up-like or down-like. It is critical to note that the brane-localized Yukawa couplings in Eq.(\ref{eq:qqToIA}) are the \emph{unnormalized} Yukawa couplings first defined in Eq.(\ref{eq:yFDefs}), and not the parameters $\tilde{y}^{Q,q}_\pm$, which first appear in Eq.(\ref{eq:PMToDPWidth}) and correspond to the Yukawa couplings of a dark Higgs with a single portal matter field and its corresponding SM field. We have used the unnormalized $y^{Q,q}_\pm$ in Eq.(\ref{eq:qqToIA}) solely for the sake of brevity, since using the normalized quantities would needlessly complicate our expression, but in our numerical computations of the cross sections, we shall use the normalized quantities $\tilde{y}^{Q,q}_\pm$ defined as in Eq.(\ref{eq:PMToDPWidth}) for the sake of consistency with our earlier results.

\begin{figure}
    \centerline{\includegraphics[width=3.5in]{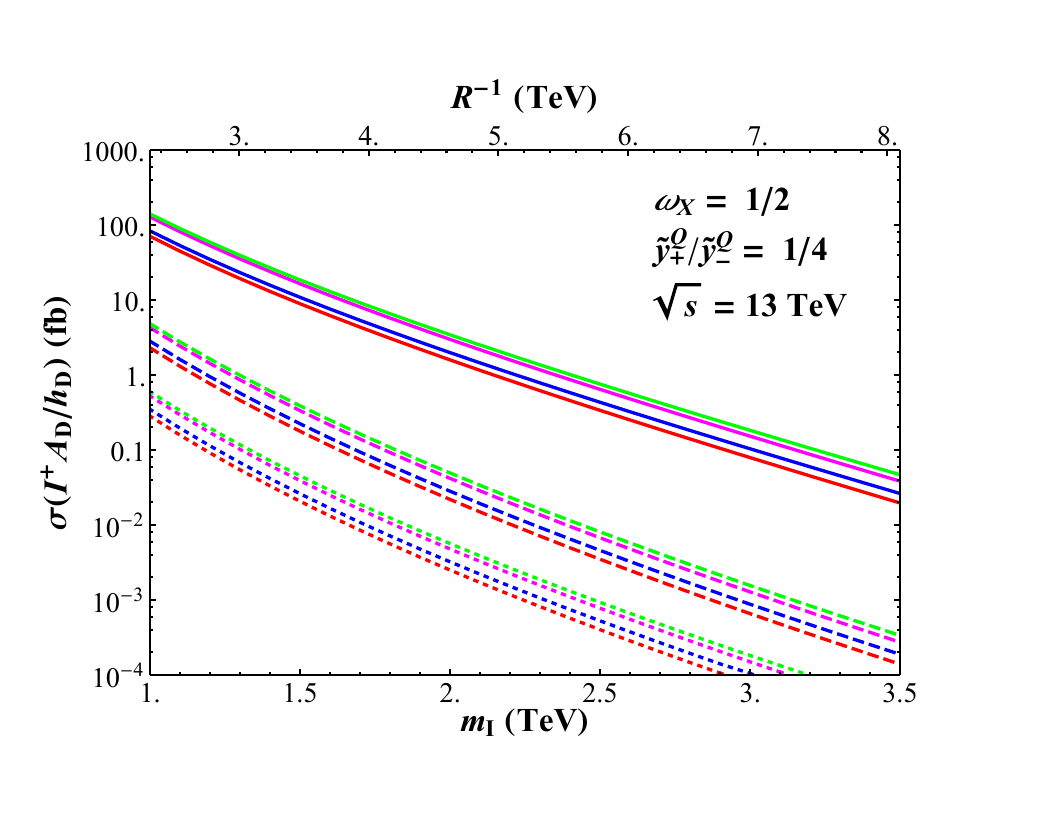}
    \hspace{-0.75cm}
    \includegraphics[width=3.5in]{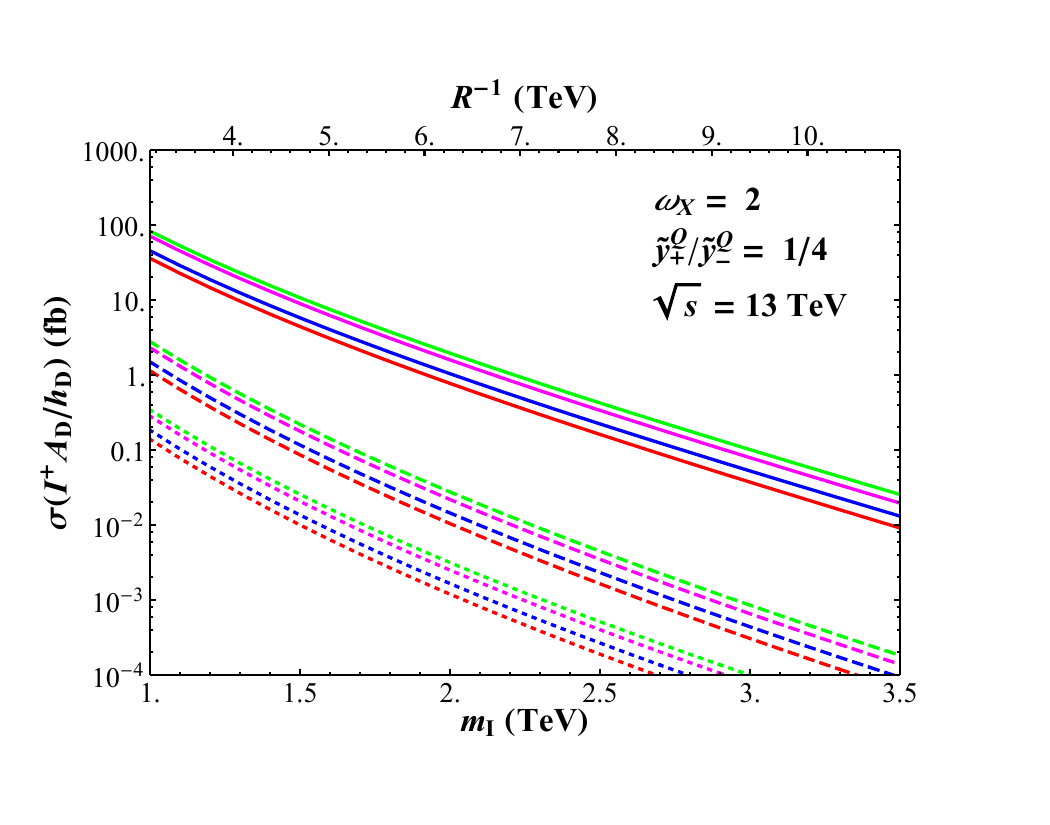}}
    \vspace*{-1cm}
    \centerline{\includegraphics[width=3.5in]{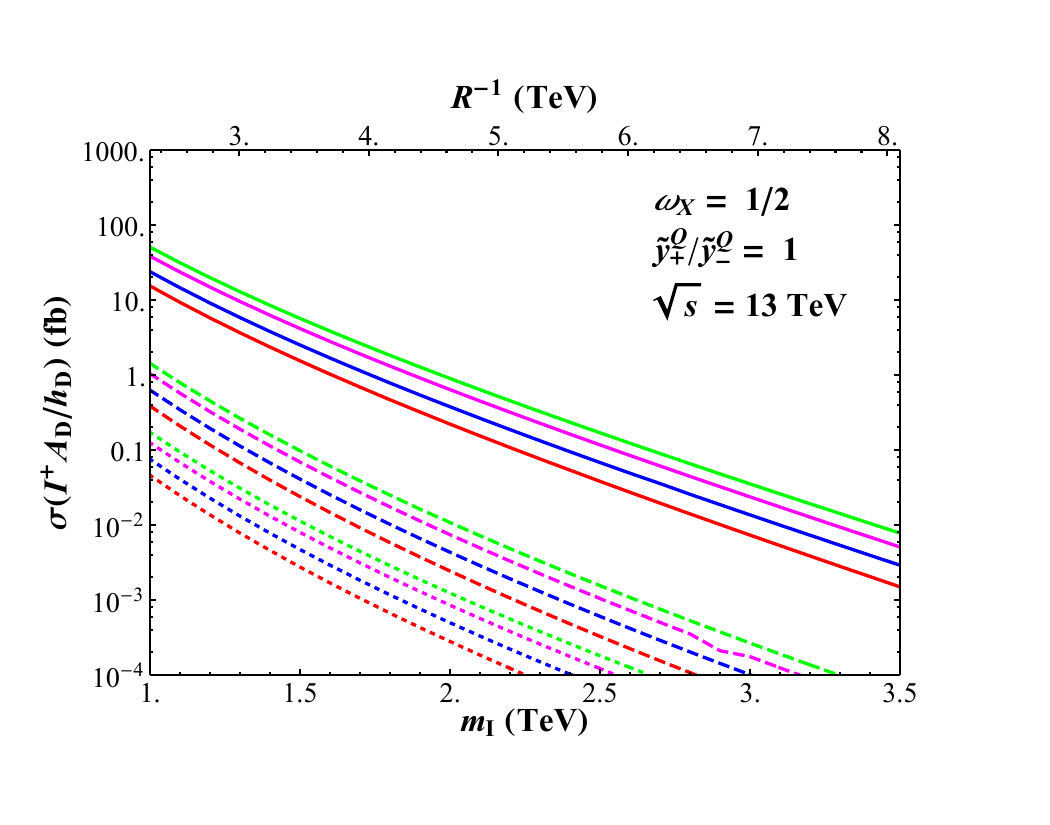}
    \hspace{-0.75cm}
    \includegraphics[width=3.5in]{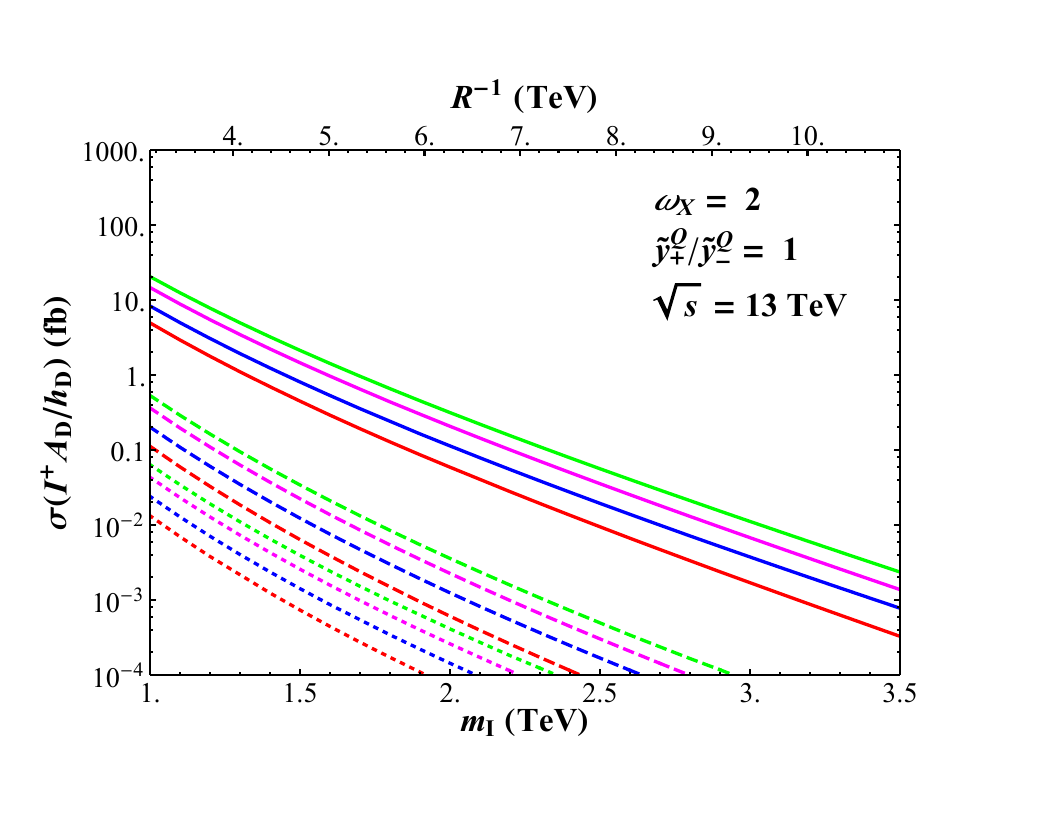}}
    \vspace*{-1cm}
    \centerline{\includegraphics[width=3.5in]{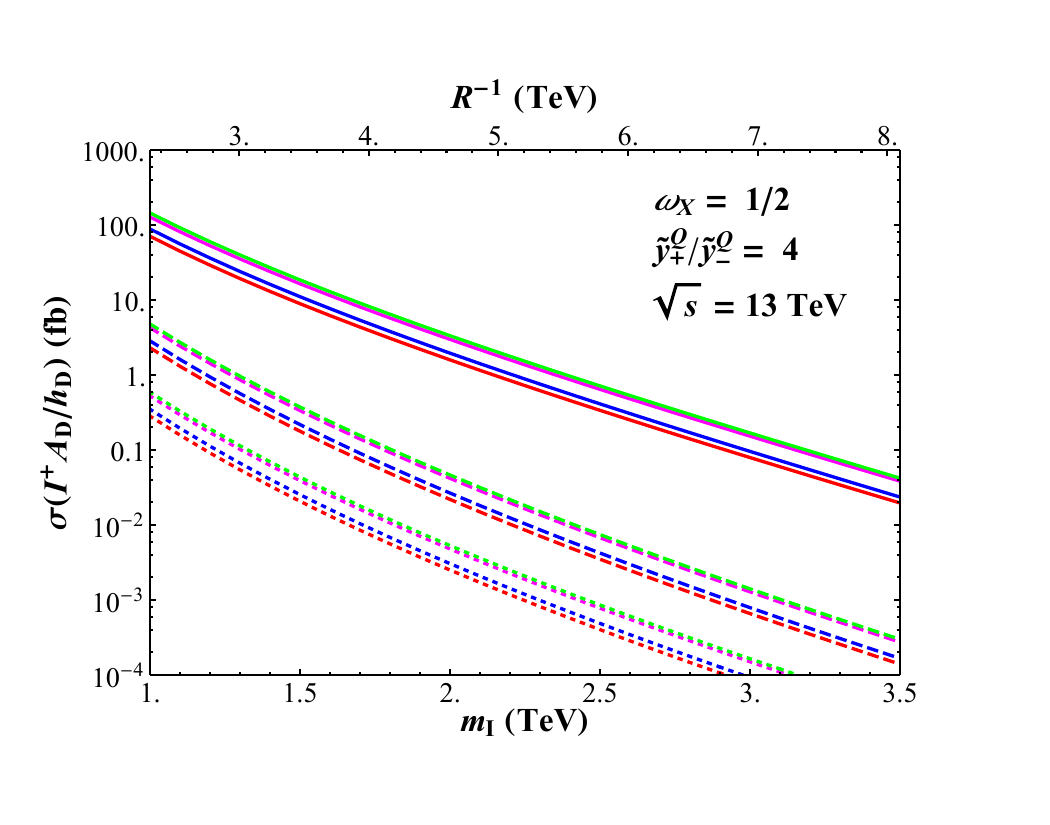}
    \hspace{-0.75cm}
    \includegraphics[width=3.5in]{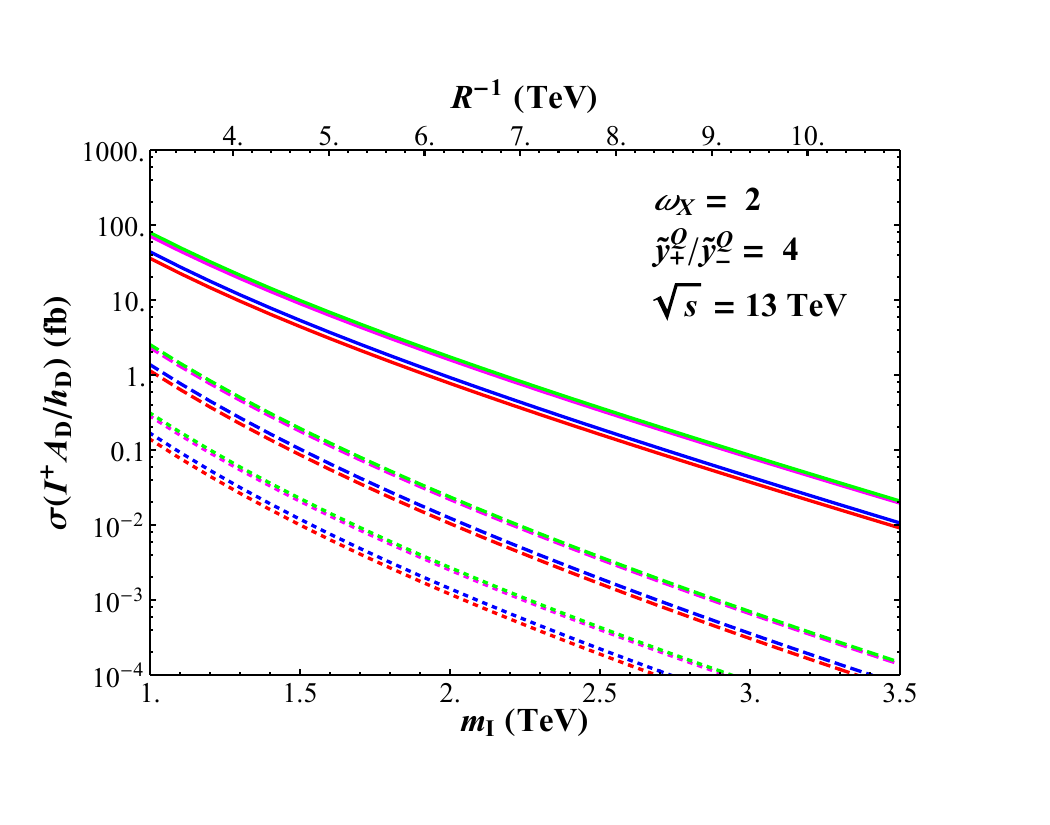}}
    \vspace*{-0.75cm}
    \caption{The total scattering cross section for the process $q\overline{q} \rightarrow I^+ A_D/h_D$ at the LHC at $\sqrt{s} = 13 \; \textrm{TeV}$, assuming $g_D = g_L$, $\tau_{Q,u,d0}=\tau_X=0$, degeneracy between weak isosinglet and isodoublet brane-localized kinetic terms (as in Figure \ref{fig12}) and Yukawa couplings, and with the quantity $\sqrt{(\tilde{y}^{Q,u,d}_+)^2 + (\tilde{y}^{Q,u,d}_-)^2}$ normalized to unity. Portal matter brane-localized kinetic terms are assumed to be $(\omega_Q, \tau_{Q+}, \tau_{Q-}) =$ (1/2,1/2,1/2) (Red), (2,1/2,1/2) (Magenta), (1/2,1/2,2) (Blue), (2,1/2,2) (Green). Only one generation of quarks at a time is assumed to be embedded in portal matter multiplets, either the first generation (Solid), the second generation (Dashed), or the third generation (Dotted).}
    \label{fig13}
\end{figure}

\begin{figure}
    \centerline{\includegraphics[width=3.5in]{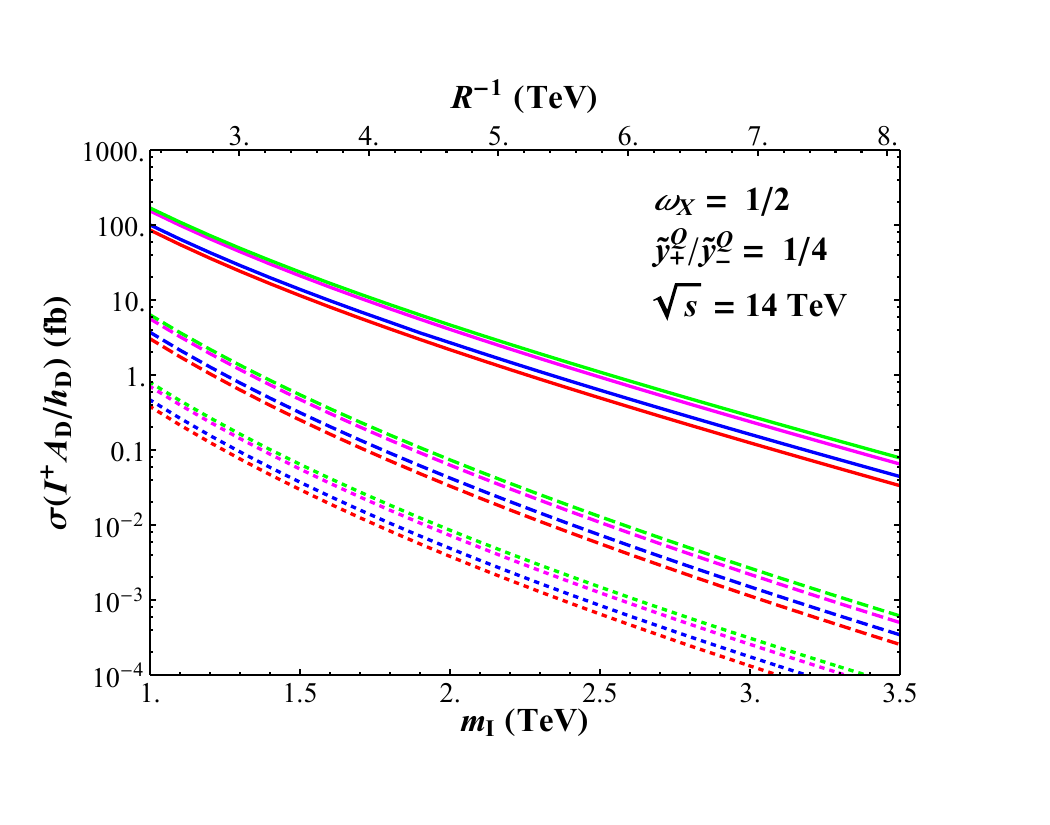}
    \hspace{-0.75cm}
    \includegraphics[width=3.5in]{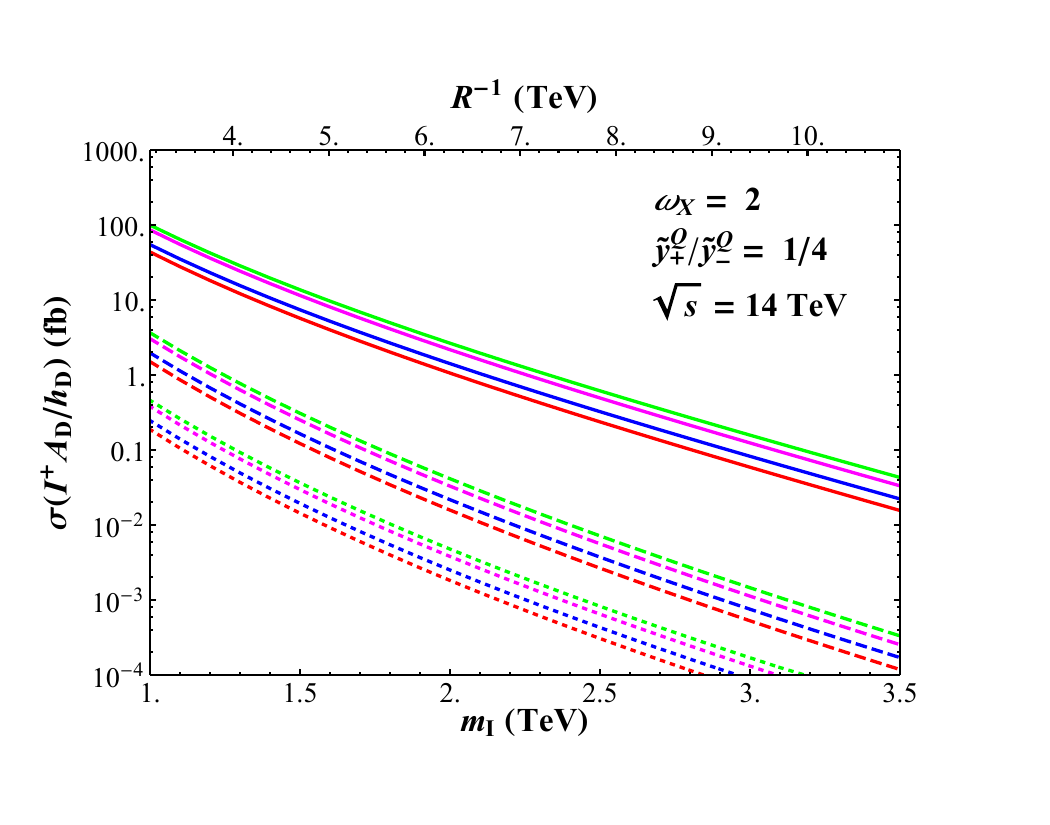}}
    \vspace*{-1cm}
    \centerline{\includegraphics[width=3.5in]{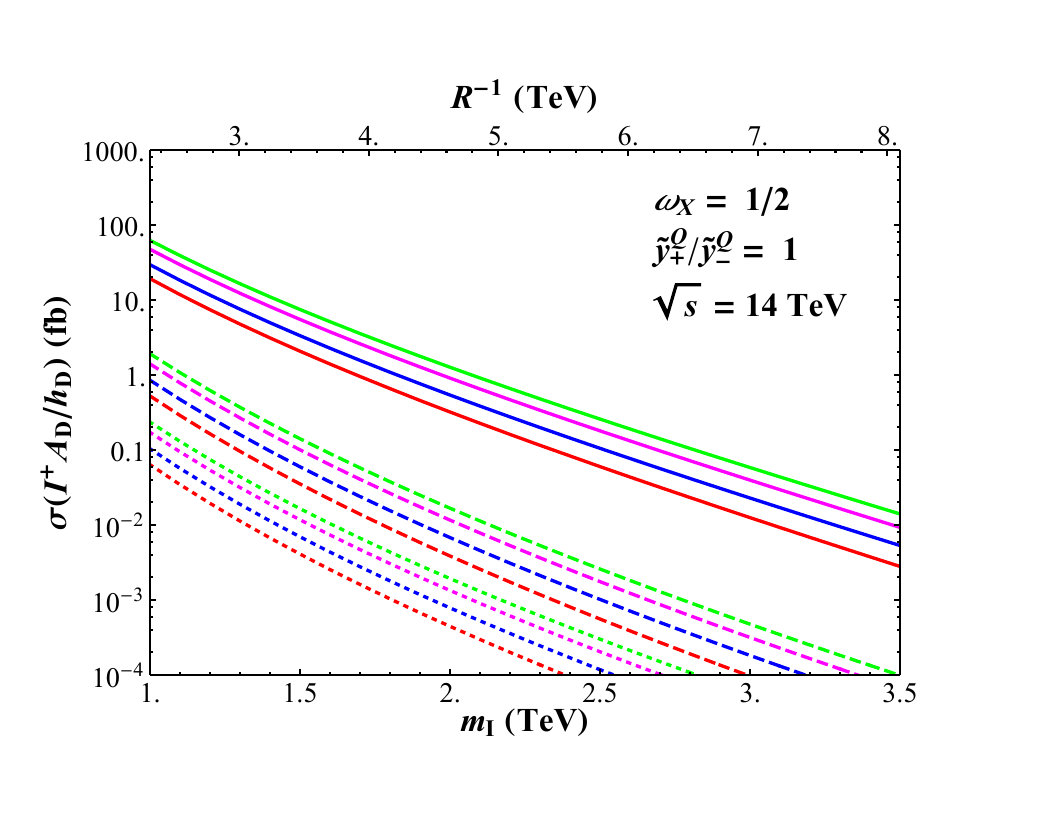}
    \hspace{-0.75cm}
    \includegraphics[width=3.5in]{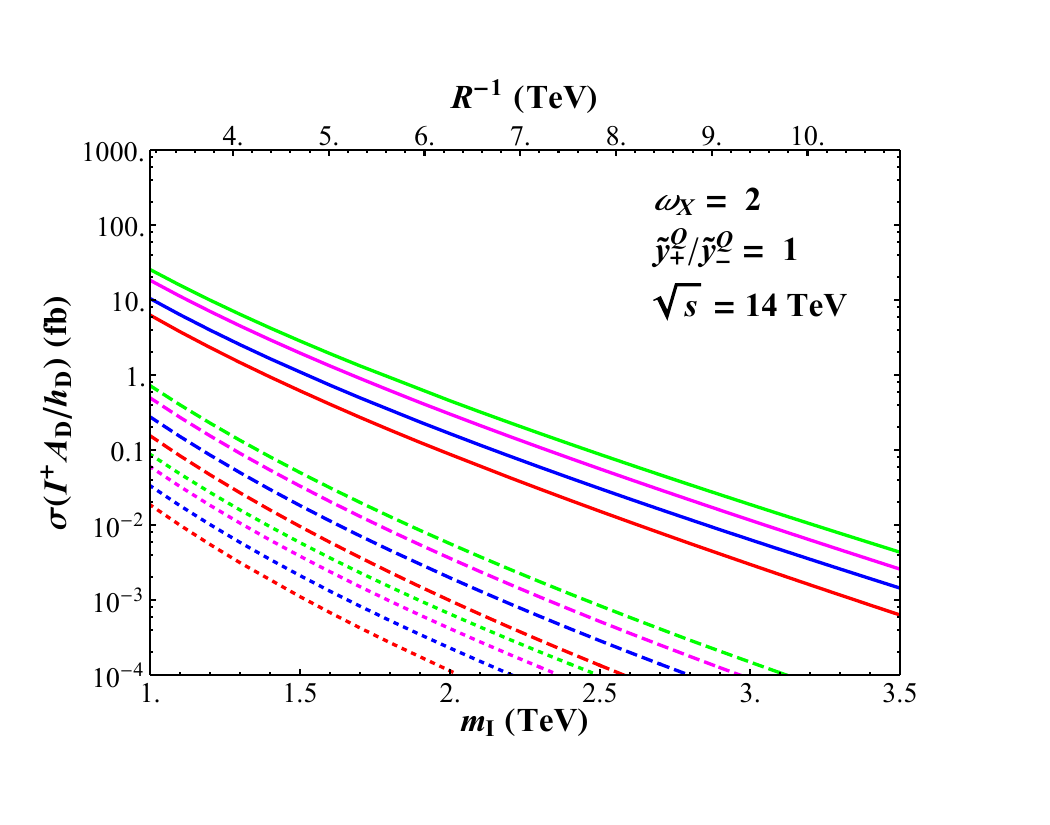}}
    \vspace*{-1cm}
    \centerline{\includegraphics[width=3.5in]{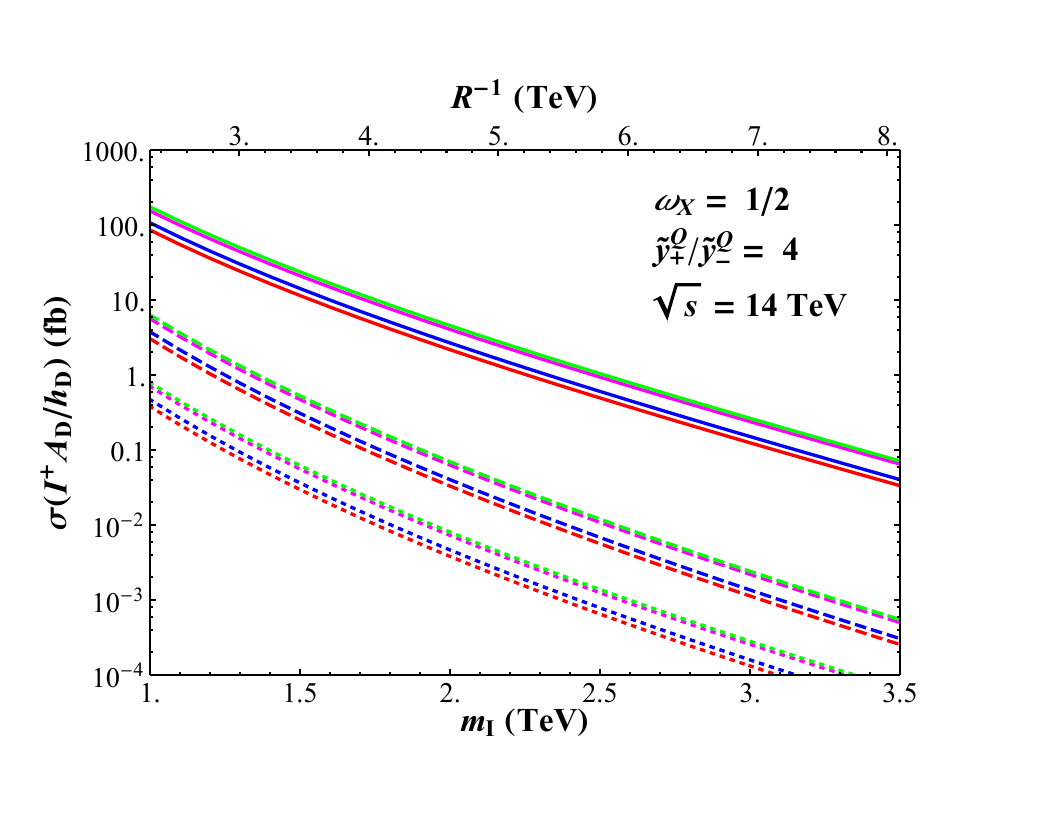}
    \hspace{-0.75cm}
    \includegraphics[width=3.5in]{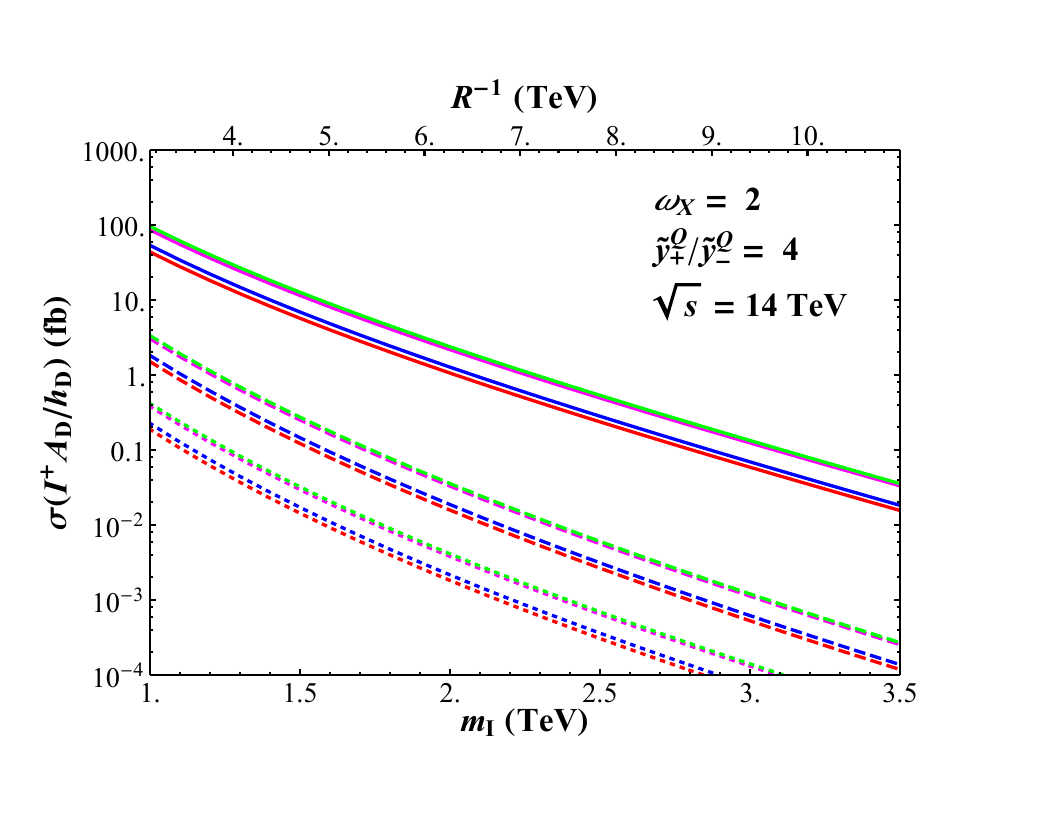}}
    \vspace*{-0.75cm}
    \caption{Same as Figure \ref{fig13}, but with the center-of-mass energy assumed to be $\sqrt{s} = 14 \; \textrm{TeV}$.}
    \label{fig14}
\end{figure}

In Figures \ref{fig13} and \ref{fig14}, we depict the LHC cross section for the $I^+ A_D$ production process for a variety of selections of brane-localized kinetic terms at $\sqrt{s} = 13 \; \textrm{TeV}$ and $\sqrt{s} = 14 \; \textrm{TeV}$. The signal associated with this process will again depend heavily on the decay properties of the $I^\pm$ boson and the dark photon/dark Higgs, but if we assume that the dark photon and dark Higgs decay invisibly, then the expected dominant decay channels of the $I^+$ suggest that constraints on this channel might arise from a monojet or mono-$Z'$ search, depending on whether the $I^+$ decays into an SM quark or leptonic final state. Interestingly, we find that this channel can represent the largest production cross section featuring $I^\pm$ gauge bosons of those we have considered so far, outpacing both the production associated with a single portal matter field and the $I^+ I^-$ pair production-- we can see in Figure \ref{fig14} that for some regions of parameter space when the first generation is embedded in a portal matter multiplet, we might anticipate thousands of signal events at the HL-LHC for compactification scales as high as $R^{-1} \approx 8 \; \textrm{TeV}$.
It is also interesting to note that the cross section is somewhat dependent on the relative values of the brane-localized Yukawa couplings $\tilde{y}^Q_+$ and $\tilde{y}^Q_-$ (and their equivalents for weak isospin singlet fields, $\tilde{y}^q_+$ and $\tilde{y}^q_-$). In particular, we see that the cross section is generally diminished by an $O(1)$ factor when we assume $\tilde{y}^Q_+ = \tilde{y}^Q_-$, indicating that the differently $U(1)_D$-charged portal matter states are interfering destructively with one another. In general, however, we see that the brane-localized kinetic terms and Yukawa couplings have a fairly muted effect on the total cross section; we do not see any set of parameter choices affording more than an order several enhancement/suppression of the predicted cross section for a given value of $m_I$. The primary effect of the brane-localized kinetic terms is altering the mass spectrum of the $I$ bosons with respect to the compactification scale, which can have a profound effect on whether a constraint from a search in this channel might compete with or dominate over more conventional extra dimensions searches discussed in Section \ref{section:SMExtraDimensions}.

\section{Discussion and Conclusions}\label{section:Conclusion}

In this paper, we have aimed to emphasize the utility of generating phenomenologically feasible portal matter for kinetic mixing/vector portal dark matter models using extra dimensions. First, we noted that if a generic dark gauge group $\mathcal{G}_D$ is broken by orbifold boundary conditions, then we would expect that a bulk fermion embedded in a representation of $\mathcal{G}_D$ will be split into Kaluza-Klein towers featuring a chiral zero-mode and those with only heavy vector-like states, and that these two classes of towers will differ in their quantum numbers under $\mathcal{G}_D$. We proceeded to discuss the techniques one might employ to construct a model with the vector-like modes acting as portal matter to facilitate kinetic mixing between the SM hypercharge and a dark $U(1)_D \subset \mathcal{G}_D$ in generic terms, before specializing to a specific minimal model.

In our minimal model, SM fermions are embedded in multiplets of a dark gauge group $SU(2)_D$ in a single flat extra dimension. Orbifold boundary conditions then break $SU(2)_D$ down to the dark photon's gauge group $U(1)_D$, while simultaneously projecting out light chiral states for portal matter fermions. We have demonstrated that this construction will lead to finite and calculable kinetic mixing in the low-energy limit of the theory, and further posited the conditions for such finite and calculable kinetic mixing in \emph{any} 5-dimensional theory. We then proceeded to consider the phenomenological signatures and constraints of our minimal construction. Notably, we found that the portal matter and dark sector gauge bosons in our model will have dramatically lighter lowest-lying massive Kaluza-Klein modes than those of the SM fields. In our flat setup, we can intuitively see this mass splitting as a consequence of these fields having different boundary conditions at the two branes in the model: In the absence of brane-localized kinetic terms, bulk fields that satisfy Dirichlet or von Neumann boundary conditions at both branes have as their lowest-lying Kaluza-Klein modes fields with sinusoidal bulk wave functions with a wavelength equal to the size of the extra dimension, leading to a mass equal to the compactification scale $R^{-1}$, while those that satisfy Dirichlet boundary conditions on one brane and von Neumann boundary conditions on the other will have a lightest Kaluza-Klein mode with a bulk wavefunction such that the extra dimension spans only half of its wavelength, and hence a mass of $R^{-1}/2$. As a result of this mass splitting, we find that searches for new TeV-scale particles associated with the portal matter sector will offer comparable or even more stringent constraints than the searches for SM Kaluza-Klein modes or low-energy precision experiments which normally characterize probes of extra dimensions. We therefore find that in our model, it is entirely feasible that the \emph{first} experimental evidence which might emerge of an extra dimension would be the portal matter sector. Furthermore, we have argued that a similar, even exaggerated mass splitting between the portal matter sector and the SM Kaluza-Klein modes will remain in a much broader class of 5-dimensional constructions within this paradigm, including models with a warped extra dimension, so the harshest constraints on compactification scales emerging from portal matter rather than SM searches is a fairly robust prediction of setups with Kaluza-Klein portal matter: It is purely a consequence of geometry.\footnote{A reader may notice that in the construction of \cite{Wojcik:2020wgm}, it was also found that the portal matter sector particles would be significantly lighter than the other extended gauge symmetry incorporated with it, an $SU(3)$ flavor symmetry. In this case, however, such a splitting was simply a consequence of the large mass and mixing hierarchies present in the quark flavor sector forcing the characteristic scales of different heavy particles associated with the flavor sector to large values-- and the assumption that the portal matter fields would acquire masses at the TeV scale. In our extra-dimensional construction here, the mass splitting we observe between the portal matter sector and the SM Kaluza-Klein towers is a much more generic prediction.}

There are a variety of directions in which this work might be continued. Perhaps the simplest extension of our construction here would be to recreate the model in a warped context. As mentioned in Section \ref{section:modelSetup}, a warped realization of this paradigm would resolve both the problem of stabilizing the bulk at the TeV scale and the low scale at which the flat-space model requires UV completion, at the expense of considerably more complicated bulk profiles for all involved fields. Fully realizing this construction in a warped bulk would present other challenges as well. First, because our orbifold boundary conditions are exhausted in breaking $SU(2)_D$ and generating chiral fermions, we would be unable to relax significant constraints from electroweak precision parameters by proposing a bulk custodial symmetry as in \cite{Casagrande:2010si}, unless we were to assume that the custodial symmetry were broken entirely by transplanckian vev's localized on the UV brane. Furthermore, in contrast to our flat space construction, the difference in the natural mass scales at the two branes in a warped setup would require us to localize both the SM Higgs and the dark Higgs on the same brane, suggesting a moderate $O(10^{-(3-4)})$ tuning of the mixing parameter between the two fields to satisfy phenomenological constraints. The warped scenario would also introduce qualitatively similar but still somewhat intriguing phenomenology compared to our flat setup. As noted in Section \ref{section:recipe}, in the likely event that $SU(2)_D$ is broken on the TeV-brane in such a construction, the mass splitting between the lightest portal matter and dark gauge boson Kaluza Klein modes and those of SM fields is substantially more pronounced than in the flat case. In the absence of brane-localized kinetic terms, the $I^\pm$ bosons in our model would have a mass only $\sim 1/10$ that of the lightest SM Kaluza-Klein gauge bosons. In principle, such a mass splitting could allow the portal matter sector to be probed even for exceptionally large compactification scales: If, for example, we assume that the Kaluza-Klein gluons must have a mass of $\gsim 21 \; \textrm{TeV}$, as would be expected in RS models without some sort of flavor protection and bulk masses motivated by the SM fermion mass hierarchy \cite{Csaki:2008zd}, the fields associated with the portal matter sector might still provide new resonances at the scale of only $O(1-2 \; \textrm{TeV})$. A more careful exploration of how to realize our construction in an RS context may be merited, to see if these optimistic speculations are borne out.

Outside of repeating this work in a warped extra dimension, the other most obvious extension to the model presented here would be to expand the dark gauge group $SU(2)_D$, which could allow it to encompass other possible BSM gauge symmetries, such as flavor. As noted in Section \ref{section:recipe}, enlarging this gauge group would present novel model-building difficulties which are conveniently sidestepped in the minimal construction, in particular how to reduce the rank of the dark gauge group. Although we conjecture briefly on solutions to these issues, addressing these difficulties explicitly in a model which incorporates a larger dark gauge group, in particular one which includes another well-motivated BSM gauge symmetry, would represent an interesting continuation of our work here.

Finally, apart from extending the model-building work presented here, it would also be useful to consider the phenomenology of the heavier Kaluza-Klein modes, such as the massive modes of the $U(1)_D$-neutral fermion towers, which we have glossed over here. While their immediate phenomenological consequences will be more muted than those of particles associated with the portal sector, it would be of interest to consider in detail the production modes, branching fractions, and expected mass spectra of these fields, since their appearance could be enormously valuable in experimentally differentiating between a 4-dimensional portal matter construction and various realizations of Kaluza-Klein portal matter.

As it stands, the use of Kaluza-Klein modes to generate kinetic mixing is a natural but up until now seldom explored avenue in which phenomenologically feasible portal matter might be generated. In addition, such constructions demonstrate novel and distinct phenomenology from both conventional theories with large extra dimensions and minimal constructions of portal matter, meriting further exploration into the capabilities, limits, and distinctive signatures of this class of models.

\section*{Acknowledgements}
The author would like to thank Lisa Everett and Thomas Rizzo for discussions related to this work.
This work was supported by the U.S. Department of Energy under the contract number DE-SC0017647.

\appendix
\section{Fermion Kaluza-Klein Towers}\label{appendix:FermionTowers}
Here we present the complete set of fermion Kaluza-Klein towers using the notation discussed in Section \ref{section:fermions}.
\begin{align}\label{eq:Q0Tower}
    &\mathbf{F^0:} \nonumber\\
    &\begin{matrix}
        \mathbf{C}^{F}_n (\phi) = \begin{pmatrix} 1\\ \frac{-\mu^F_+}{1 + x_n^2 \tau_{F+} \tau_{F0}}\\ \frac{-\mu^F_-}{1 + x^2_n \tau_{F-} \tau_{F0}}\end{pmatrix} N^{F_0}_n \beta_{x_n} (\omega_F, \phi), & \mathbf{S}^{F}_n (\phi) = \begin{pmatrix} 1\\ \frac{-\mu^F_+}{1 + x_n^2 \tau_{F+} \tau_{F0}}\\ \frac{-\mu^F_-}{1 + x^2_n \tau_{F-} \tau_{F0}}\end{pmatrix} N^{F_0}_n \alpha_{x_n} (\omega_F, \phi),\\
        \mathbf{C}^{f}_n (\phi) = \begin{pmatrix} 1\\ 0\\ 0\end{pmatrix} \frac{\mu_h N^{F_0}_n \beta_{x_n} (\tau_{f0},\pi-\phi)}{\alpha_{x_n} (\omega_f,\pi) + x_n \tau_{f0} \beta_{x_n} (\omega_f,\pi)}, & \mathbf{S}^{f}_n (\phi) = \begin{pmatrix} 1\\ 0\\ 0\end{pmatrix} \frac{\mu_h N^{F_0}_n \alpha_{x_n} (\tau_{f0},\pi-\phi)}{\alpha_{x_n} (\omega_f,\pi) + x_n \tau_{f0} \beta_{x_n}(\omega_f,\pi)},
    \end{matrix}\\
    &N^{F_0}_n =  \bigg[ \frac{2 (1 + x_n^2 \tau_{F0}^2)}{(\tau_{F0} + \omega_F)(1 + x_n^2 \tau_{F0} \omega_F)+ \pi (1 + x_n^2 \tau_{F0}^2)(1 + x_n^2 \omega_F^2)} \bigg]^{1/2} \nonumber,\\
    &x_n \; \textrm{such that} \;\; \alpha_{x_n} (\omega_F, \pi) = - x_n \tau_{F0} \beta_{x_n} (\omega_F,\pi). \nonumber
\end{align}
%
\begin{align}\label{eq:Q+TowerAppendix}
    &\mathbf{F^+:} \nonumber\\
    &\begin{matrix}
        \mathbf{C}^F_n = \begin{pmatrix}\frac{\mu^F_+}{1 + x_n^2 \tau_{F0} \tau_{F+}}\\ 1\\ 0 \end{pmatrix} N^{F_+}_n \beta_{x_n} (\omega_F, \phi), & \mathbf{S}^F_n (\phi) = \begin{pmatrix} \frac{\mu^F_+}{1 + x_n^2 \tau_{F0} \tau_{F+}}\\ 1\\ 0 \end{pmatrix} N^{F_+}_n \alpha_{x_n} (\omega_F, \phi),\\
        \mathbf{C}^f_n (\phi) = -\begin{pmatrix} 0\\ 1\\ 0\end{pmatrix} \frac{\mu_h N^{F_+}_n \alpha_{x_n} (\tau_{f+}, \pi-\phi)}{\beta_{x_n} (\omega_f,\pi)-xn \tau_{f+} \alpha_{x_n} (\omega_f,\pi)}, & \mathbf{S}^f_n (\phi) = \begin{pmatrix} 0\\ 1\\ 0\end{pmatrix} \frac{\mu_h N^{F_+}_n \beta_{x_n} (\tau_{f+}, \pi-\phi)}{\beta_{x_n}(\omega_f,\pi) - x_n \tau_{f+} \alpha_{x_n} (\omega_f, \pi)}
    \end{matrix},\\
    &N^{F_+}_n = \bigg[ \frac{2 (1 + x_n^2 \tau_{F+}^2)}{\pi (1 + x_n^2 \omega_F^2) (1 + x_n^2 \tau_{F+}^2)+ (\tau_{F+} + \omega_F)(1 + x_n^2 \tau_{F+} \omega_F)} \bigg]^{1/2},  \nonumber\\
    &x_n \; \textrm{such that} \;\; \beta_{x_n} (\omega_F,\pi) = x_n \tau_{F+} \alpha_{x_n} (\omega_F,\pi), \nonumber
\end{align}
%
\begin{align}\label{eq:Q-Tower}
    &\mathbf{F^-:} \nonumber\\
    &\begin{matrix}
        \mathbf{C}^F_n = \begin{pmatrix}\frac{\mu^F_-}{1 + x_n^2 \tau_{F0} \tau_{F-}}\\ 0\\ 1 \end{pmatrix} N^{F_-}_n \beta_{x_n} (\omega_F, \phi), & \mathbf{S}^F_n (\phi) = \begin{pmatrix} \frac{\mu^F_-}{1 + x_n^2 \tau_{F0} \tau_{F-}}\\ 0\\ 1 \end{pmatrix} N^{F_-}_n \alpha_{x_n} (\omega_F, \phi),\\
        \mathbf{C}^f_n (\phi) = -\begin{pmatrix} 0\\ 0\\ 1\end{pmatrix} \frac{\mu_h N^{F_-}_n \alpha_{x_n} (\tau_{f-}, \pi-\phi)}{\beta_{x_n} (\omega_f,\pi)-xn \tau_{f-} \alpha_{x_n} (\omega_f,\pi)}, & \mathbf{S}^f_n (\phi) = \begin{pmatrix} 0\\ 0\\ 1\end{pmatrix} \frac{\mu_h N^{F_-}_n \beta_{x_n} (\tau_{f-}, \pi-\phi)}{\beta_{x_n}(\omega_f,\pi) - x_n \tau_{f-} \alpha_{x_n} (\omega_f, \pi)}
    \end{matrix},\\
    &N^{F_-}_n = \bigg[ \frac{2 (1 + x_n^2 \tau_{F-}^2)}{\pi (1 + x_n^2 \omega_F^2) (1 + x_n^2 \tau_{F-}^2)+ (\tau_{F-} + \omega_F)(1 + x_n^2 \tau_{F-} \omega_F)} \bigg]^{1/2},  \nonumber\\
    &x_n \; \textrm{such that} \;\; \beta_{x_n} (\omega_F,\pi) = x_n \tau_{F-} \alpha_{x_n} (\omega_F,\pi), \nonumber
\end{align}
%
\begin{align}\label{eq:q0Tower}
    &\mathbf{f}^0: \nonumber\\
    &\begin{matrix}
        \mathbf{C}^{F}_n (\phi) = - \begin{pmatrix} 1\\ 0\\ 0 \end{pmatrix} \frac{\mu_h N^{f_0}_n \beta_{x_n} (\tau_{F0}, \pi-\phi)}{\alpha_{x_n}(\omega_F,\pi) + x_n \tau_{F0} \beta_{x_n} (\omega_F,\pi)}, &\mathbf{S}^{F}_n (\phi) = \begin{pmatrix} 1\\ 0\\ 0 \end{pmatrix} \frac{\mu_h N^{f_0}_n \alpha_{x_n} (\tau_{F0}, \pi-\phi)}{\alpha_{x_n}(\omega_F,\pi) + x_n \tau_{F0} \beta_{x_n} (\omega_F,\pi)}\\
        \mathbf{C}^{f}_n (\phi) = -\begin{pmatrix} 1\\ \frac{\mu^f_+}{1+x_n^2 \tau_{f+} \tau_{f0}}\\ \frac{\mu^f_-}{1+x_n^2 \tau_{f-} \tau_{f0}}\end{pmatrix} N^{f_0}_n \beta_{x_n} (\omega_f, \phi), & \mathbf{S}^{f}_n (\phi) = \begin{pmatrix} 1\\ \frac{\mu^f_+}{1+x_n^2 \tau_{f+} \tau_{f0}}\\ \frac{\mu^f_-}{1+x_n^2 \tau_{f-} \tau_{f0}}\end{pmatrix} N^{f_0}_n \alpha_{x_n} (\omega_f, \phi),
    \end{matrix}\\
    &N^{f_0}_n = \bigg[ \frac{2 (1 + x_n^2 \tau_{f0}^2)}{\pi (1 + x_n^2 \tau_{f0}^2)(1 + x_n^2 \omega_f^2) + (\tau_{f0} + \omega_f)(1 + x_n^2 \tau_{f0} \omega_f)}\bigg]^{1/2} \nonumber\\
    &x_n \; \textrm{such that} \; \; \alpha_{x_n} (\omega_f, \pi) = - x_n \tau_{f0} \beta_{x_n} (\omega_f, \pi). \nonumber
\end{align}
%
\begin{align}\label{eq:q+Tower}
    &\mathbf{f}^+: \nonumber\\
    &\begin{matrix}
        \mathbf{C}^F_n (\phi) = \begin{pmatrix} 0\\ 1\\ 0\end{pmatrix} \frac{\mu_h N^{f_+}_n \alpha_{x_n} (\tau_{F+}, \pi-\phi)}{\beta_{x_n} (\omega_F,\pi)-x_n \tau_{F+} \alpha_{x_n} (\omega_F, \pi)}, & \mathbf{S}^F_n (\phi) = \begin{pmatrix} 0\\ 1\\ 0\end{pmatrix} \frac{\mu_h N^{f_+}_n \beta_{x_n} (\tau_{F+}, \pi-\phi)}{\beta_{x_n} (\omega_F,\pi)-x_n \tau_{F+} \alpha_{x_n} (\omega_F, \pi)}\\
        \mathbf{C}^f_n (\phi) = \begin{pmatrix} \frac{\mu^f_+}{1 + x_n^2 \tau_{f+} \tau_{f0}}\\ -1\\ 0\end{pmatrix} N^{f_+}_n \beta_{x_n} (\omega_f, \phi), & \mathbf{S}^f_n (\phi) = \begin{pmatrix} \frac{-\mu^f_+}{1 + x_n^2 \tau_{f0} \tau_{f+}}\\ 1\\ 0\end{pmatrix} N^{f_+}_n \alpha_{x_n} (\omega_f, \phi)
    \end{matrix}\\
    &N^{f_+}_n = \bigg[ \frac{2 (1 + x_n^2 \tau_{f+}^2)}{\pi (1 + x_n^2 \omega_f^2) (1 + x_n^2 \tau_{f+}^2)+ (\tau_{f+} + \omega_f)(1 + x_n^2 \tau_{f+} \omega_f)} \bigg]^{1/2},  \nonumber\\
    &x_n \; \textrm{such that} \;\; \beta_{x_n} (\omega_f,\pi) = x_n \tau_{f+} \alpha_{x_n} (\omega_f,\pi), \nonumber
\end{align}
%
\begin{align}\label{eq:q-Tower}
    &\mathbf{f}^-: \nonumber\\
    &\begin{matrix}
        \mathbf{C}^F_n (\phi) = \begin{pmatrix} 0\\ 0\\ 1\end{pmatrix} \frac{\mu_h N^{f_-}_n \alpha_{x_n} (\tau_{F-}, \pi-\phi)}{\beta_{x_n} (\omega_F,\pi)-x_n \tau_{F-} \alpha_{x_n} (\omega_F, \pi)}, & \mathbf{S}^F_n (\phi) = \begin{pmatrix} 0\\ 0\\ 1\end{pmatrix} \frac{\mu_h N^{f_-}_n \beta_{x_n} (\tau_{F-}, \pi-\phi)}{\beta_{x_n} (\omega_F,\pi)-x_n \tau_{F-} \alpha_{x_n} (\omega_F, \pi)}\\
        \mathbf{C}^f_n (\phi) = \begin{pmatrix} \frac{\mu^f_-}{1 + x_n^2 \tau_{f-} \tau_{f0}}\\ 0\\ -1\end{pmatrix} N^{f_-}_n \beta_{x_n} (\omega_f, \phi), & \mathbf{S}^f_n (\phi) = \begin{pmatrix} \frac{-\mu^f_-}{1 + x_n^2 \tau_{f0} \tau_{f-}}\\ 0\\ 1\end{pmatrix} N^{f_-}_n \alpha_{x_n} (\omega_f, \phi)
    \end{matrix}\\
    &N^{f_-}_n = \bigg[ \frac{2 (1 + x_n^2 \tau_{f-}^2)}{\pi (1 + x_n^2 \omega_f^2) (1 + x_n^2 \tau_{f-}^2)+ (\tau_{f-} + \omega_f)(1 + x_n^2 \tau_{f-} \omega_f)} \bigg]^{1/2},  \nonumber\\
    &x_n \; \textrm{such that} \;\; \beta_{x_n} (\omega_f,\pi) = x_n \tau_{f-} \alpha_{x_n} (\omega_f,\pi), \nonumber
\end{align}

\section{Computing the SMEFT Operators from the Extra Dimension}\label{appendix:EDSMEFT}

This Appendix provides more detail on how we determined the value of the dimension-6 operators in Eq.(\ref{eq:d6OPerators}) arising from the extra-dimensional physics in our model that affect the electroweak $\rho$ parameter.

Referencing our work in Section \ref{section:gauges} to compute the SM gauge boson bulk profiles, as well as using the expression of Eq.(\ref{eq:qZeroModeTower}) for the fermion zero mode profile, it is straightforward to compute the effective action for the SM gauge bosons and their interaction terms with the $SU(2)_L$ doublet light leptons by integrating over the extra dimension. Ignoring the effects of heavy fermion Kaluza-Klein modes (since for the light SM leptons we consider here, these will only enter at $O(m_{e,\mu}^2 R^2)$, which will be subdominant to the $O(m_{W,Z}^2 R^2)$ effects of the gauge boson Kaluza-Klein modes), we arrive at an effective four-dimensional action,
\begin{align}\label{eq:EDEWAction}
    S &= -\frac{1}{4}Z_{\mu \nu}^2 - \frac{1}{2} W^+_{\mu \nu} W^-_{\mu \nu} - \frac{1}{4} A_{\mu \nu}^2 + \frac{g_L^2 v^2}{4} \bigg( 1- \frac{g_L^2 v^2 R^2 \pi^2}{12} \bigg) W^+_\mu W^-_\mu +\frac{(g_L^2 + g_Y^2)v^2}{8} \bigg( 1 - \frac{(g_L^2 + g_Y^2) v^2 R^2 \pi^2}{12}\bigg) Z_\mu^2 \nonumber\\
    &+(g_L^2+g_Y^2)^{\frac{1}{2}}(T_3^F-s_w^2 Q^F) J^{Z_L}_\mu Z_\mu \bigg( 1 + \frac{(g_L^2+g_Y^2)v^2 R^2 \pi^2 (\tau_{L0}-2 \omega_L)}{24 (\pi + \tau_{L0} + \omega_L)} \bigg)\\
    &+g_L (J^{W^+}_\mu W^+ + h.c.) \bigg( 1 + \frac{g_L^2 v^2 R^2 \pi^2 (\tau_{L0}-2 \omega_L)}{24 (\pi + \tau_{L0} + \omega_L)}\bigg) + C_{ll} \big((\overline{\mu}_L \gamma_\rho e_L) (\overline{\nu}_\mu \gamma_\rho \nu_e) + h.c.\big),\nonumber
\end{align}
where $v$ is the SM Higgs vev, $J^{Z_L}$ and $J^{W^+}$ are the usual currents for left-handed fermion couplings to the SM $Z$ boson and the $W^+$ boson, respectively, and $g_{L(Y)}$ denotes the $SU(2)_L (U(1)_Y)$ 4-dimensional coupling defined in terms of the dimensionful five-dimensional couplings as $g_{L,Y} = g_{L5,Y5}/\sqrt{\pi R}$, and $\omega_L$ and $\tau_{L0}$ are the $\phi=0$ and $\phi=\pi$ brane-localized kinetic terms for the SM charged lepton fields, respectively. Most four-fermion operators have been omitted in Eq.(\ref{eq:EDEWAction}), since they aren't relevant to our analysis, however we will have to compute coefficient of the 4-fermion operator contribution to the Fermi constant, $C_{ll}$. To evaluate this coefficient, we simply need to compute the Fermi constant in the 5-dimensional model, which experiences tree-level contributions from the exchange of the entire $W$ boson Kaluza-Klein tower. Using Eq.(\ref{eq:ZWBosonSum}) to compute the contribution of the entire tower of Kaluza-Klein modes in closed form, we arrive at
\begin{align}\label{eq:EDGF}
    G_F = \frac{1}{\sqrt{2} v^2} \bigg( 1 + \frac{g_L^2 v^2 R^2 \pi^2 (\pi^2 + 3 \pi \tau_{L0} + 3 \tau_{L0}^2)}{12 (\pi + \tau_{L0} + \omega_L)^2}\bigg).
\end{align}
Subtracting the contribution of the SM $W$ boson, we then obtain that the operator coefficient
\begin{align}
    C_{ll} = -\frac{g_L^2 R^2 \pi^2 (\tau_{L0}^2 - \tau_{L0} \omega_L + \omega_L^2)}{12 (\pi + \tau_{L0} + \omega_L)^2}
\end{align}
will reproduce the Fermi constant expression in Eq.(\ref{eq:EDGF}). We can now make direct contact with Warsaw basis operators by following the procedure of \cite{Delgado:2007ne}. Specifically, by redefining the SM Higgs vev as
\begin{align}
    v \rightarrow v \bigg( 1 + \frac{g_L^2 v^2 R^2 \pi^2}{24}\bigg),
\end{align}
the action in Eq.(\ref{eq:EDEWAction}) becomes (up to negligible $O(v^4 R^4)$ corrections)
\begin{align}\label{eq:EDEWAction2}
    S &= -\frac{1}{4}Z_{\mu \nu}^2 - \frac{1}{2} W^+_{\mu \nu} W^-_{\mu \nu} - \frac{1}{4} A_{\mu \nu}^2 + \frac{g_L^2 v^2}{4} W^+_\mu W^-_\mu +\frac{(g_L^2 + g_Y^2)v^2}{8} \bigg( 1 - \frac{g_Y^2 v^2 R^2 \pi^2}{12}\bigg) Z_\mu^2 \nonumber\\
    &+(g_L^2+g_Y^2)^{\frac{1}{2}}(T_3^F-s_w^2 Q^F) J^{Z_L}_\mu Z_\mu \bigg( 1 + \frac{(g_L^2+g_Y^2)v^2 R^2 \pi^2 (\tau_{L0}-2 \omega_L)}{24 (\pi + \tau_{L0} + \omega_L)} \bigg)\\
    &+g_L (J^{W^+}_\mu W^+ + h.c.) \bigg( 1 + \frac{g_L^2 v^2 R^2 \pi^2 (\tau_{L0}-2 \omega_L)}{24 (\pi + \tau_{L0} + \omega_L)}\bigg) +  C_{ll} \big((\overline{\mu}_L \gamma_\rho e_L) (\overline{\nu}_\mu \gamma_\rho \nu_e) + h.c.\big).\nonumber
\end{align}
As was noted in \cite{Delgado:2007ne}, this action is recreated in the Warsaw basis with 
\begin{align}
\begin{matrix}
    C_{HD} = - \frac{g_Y^2 \pi^2}{6}, & C^{(3)}_{Hl} = \frac{g_Y^2 \pi^2 (\tau_{L0}-2 \omega_L)}{6(\pi + \tau_{L0} + \omega_L)}, & C_{ll} = -\frac{g_L^2 \pi^2 (\tau_{L0}^2 - \tau_{L0} \omega_L + \omega_L^2)}{12 (\pi + \tau_{L0} + \omega_L)^2}, & C_{HWB} = 0,
\end{matrix}
\end{align}
where we extract $C_{HD}$ from the $Z$ boson mass shift, $C^{(3)}_{Hl}$ from the $W$ boson coupling shift, and $C_{ll}$ is read directly from our action. The operator $C_{HWB}$ in SMEFT will simply yield kinetic mixing between the photon and the $Z$ boson, and so is absent here. This is, of course, the same result we quote in Eq.(\ref{eq:d6OPerators})


\setlength{\bibsep}{3pt}
\bibliographystyle{JHEP}
\bibliography{main}

\providecommand{\href}[2]{#2}\begingroup\raggedright\begin{thebibliography}{10}

\bibitem{Arcadi:2017kky}
G.~Arcadi, M.~Dutra, P.~Ghosh, M.~Lindner, Y.~Mambrini, M.~Pierre, S.~Profumo,
  and F.~S. Queiroz, {\it {The waning of the WIMP? A review of models,
  searches, and constraints}},  {\em Eur. Phys. J. C} {\bf 78} (2018), no.~3
  203, [\href{http://arxiv.org/abs/1703.07364}{{\tt arXiv:1703.07364}}].

\bibitem{Kawasaki:2013ae}
M.~Kawasaki and K.~Nakayama, {\it {Axions: Theory and Cosmological Role}},
  {\em Ann. Rev. Nucl. Part. Sci.} {\bf 63} (2013) 69--95,
  [\href{http://arxiv.org/abs/1301.1123}{{\tt arXiv:1301.1123}}].

\bibitem{Graham:2015ouw}
P.~W. Graham, I.~G. Irastorza, S.~K. Lamoreaux, A.~Lindner, and K.~A. van
  Bibber, {\it {Experimental Searches for the Axion and Axion-Like Particles}},
   {\em Ann. Rev. Nucl. Part. Sci.} {\bf 65} (2015) 485--514,
  [\href{http://arxiv.org/abs/1602.00039}{{\tt arXiv:1602.00039}}].

\bibitem{Battaglieri:2017aum}
M.~Battaglieri et~al., {\it {US Cosmic Visions: New Ideas in Dark Matter 2017:
  Community Report}},  in {\em {U.S. Cosmic Visions: New Ideas in Dark
  Matter}}, 7, 2017.
\newblock \href{http://arxiv.org/abs/1707.04591}{{\tt arXiv:1707.04591}}.

\bibitem{Alexander:2016aln}
J.~Alexander et~al., {\it {Dark Sectors 2016 Workshop: Community Report}},  8,
  2016.
\newblock \href{http://arxiv.org/abs/1608.08632}{{\tt arXiv:1608.08632}}.

\bibitem{Holdom:1985ag}
B.~Holdom, {\it {Two U(1)'s and Epsilon Charge Shifts}},  {\em Phys. Lett. B}
  {\bf 166} (1986) 196--198.

\bibitem{Holdom:1986eq}
B.~Holdom, {\it {Searching for $\epsilon$ Charges and a New U(1)}},  {\em Phys.
  Lett. B} {\bf 178} (1986) 65--70.

\bibitem{Pospelov:2007mp}
M.~Pospelov, A.~Ritz, and M.~B. Voloshin, {\it {Secluded WIMP Dark Matter}},
  {\em Phys. Lett. B} {\bf 662} (2008) 53--61,
  [\href{http://arxiv.org/abs/0711.4866}{{\tt arXiv:0711.4866}}].

\bibitem{Izaguirre:2015yja}
E.~Izaguirre, G.~Krnjaic, P.~Schuster, and N.~Toro, {\it {Analyzing the
  Discovery Potential for Light Dark Matter}},  {\em Phys. Rev. Lett.} {\bf
  115} (2015), no.~25 251301, [\href{http://arxiv.org/abs/1505.00011}{{\tt
  arXiv:1505.00011}}].

\bibitem{Essig:2013lka}
R.~Essig et~al., {\it {Working Group Report: New Light Weakly Coupled
  Particles}},  in {\em {Community Summer Study 2013}: {Snowmass on the
  Mississippi}}, 10, 2013.
\newblock \href{http://arxiv.org/abs/1311.0029}{{\tt arXiv:1311.0029}}.

\bibitem{Curtin:2014cca}
D.~Curtin, R.~Essig, S.~Gori, and J.~Shelton, {\it {Illuminating Dark Photons
  with High-Energy Colliders}},  {\em JHEP} {\bf 02} (2015) 157,
  [\href{http://arxiv.org/abs/1412.0018}{{\tt arXiv:1412.0018}}].

\bibitem{Planck:2018vyg}
{\bf Planck} Collaboration, N.~Aghanim et~al., {\it {Planck 2018 results. VI.
  Cosmological parameters}},  {\em Astron. Astrophys.} {\bf 641} (2020) A6,
  [\href{http://arxiv.org/abs/1807.06209}{{\tt arXiv:1807.06209}}]. [Erratum:
  Astron.Astrophys. 652, C4 (2021)].

\bibitem{PandaX-II:2020oim}
{\bf PandaX-II} Collaboration, Q.~Wang et~al., {\it {Results of dark matter
  search using the full PandaX-II exposure}},  {\em Chin. Phys. C} {\bf 44}
  (2020), no.~12 125001, [\href{http://arxiv.org/abs/2007.15469}{{\tt
  arXiv:2007.15469}}].

\bibitem{XENON:2018voc}
{\bf XENON} Collaboration, E.~Aprile et~al., {\it {Dark Matter Search Results
  from a One Ton-Year Exposure of XENON1T}},  {\em Phys. Rev. Lett.} {\bf 121}
  (2018), no.~11 111302, [\href{http://arxiv.org/abs/1805.12562}{{\tt
  arXiv:1805.12562}}].

\bibitem{LUX:2016ggv}
{\bf LUX} Collaboration, D.~S. Akerib et~al., {\it {Results from a search for
  dark matter in the complete LUX exposure}},  {\em Phys. Rev. Lett.} {\bf 118}
  (2017), no.~2 021303, [\href{http://arxiv.org/abs/1608.07648}{{\tt
  arXiv:1608.07648}}].

\bibitem{SuperCDMS:2017mbc}
{\bf SuperCDMS} Collaboration, R.~Agnese et~al., {\it {Results from the Super
  Cryogenic Dark Matter Search Experiment at Soudan}},  {\em Phys. Rev. Lett.}
  {\bf 120} (2018), no.~6 061802, [\href{http://arxiv.org/abs/1708.08869}{{\tt
  arXiv:1708.08869}}].

\bibitem{DarkSide:2018kuk}
{\bf DarkSide} Collaboration, P.~Agnes et~al., {\it {DarkSide-50 532-day Dark
  Matter Search with Low-Radioactivity Argon}},  {\em Phys. Rev. D} {\bf 98}
  (2018), no.~10 102006, [\href{http://arxiv.org/abs/1802.07198}{{\tt
  arXiv:1802.07198}}].

\bibitem{Gherghetta:2019coi}
T.~Gherghetta, J.~Kersten, K.~Olive, and M.~Pospelov, {\it {Evaluating the
  price of tiny kinetic mixing}},  {\em Phys. Rev. D} {\bf 100} (2019), no.~9
  095001, [\href{http://arxiv.org/abs/1909.00696}{{\tt arXiv:1909.00696}}].

\bibitem{Rizzo:2018vlb}
T.~G. Rizzo, {\it {Kinetic Mixing and Portal Matter Phenomenology}},  {\em
  Phys. Rev. D} {\bf 99} (2019), no.~11 115024,
  [\href{http://arxiv.org/abs/1810.07531}{{\tt arXiv:1810.07531}}].

\bibitem{Belotsky:2004ga}
K.~Belotsky, D.~Fargion, M.~Y. Khlopov, R.~V. Konoplich, M.~G. Ryskin, and
  K.~I. Shibaev, {\it {May heavy hadrons of the 4th generation be hidden in our
  universe while close to detection?}},
  \href{http://arxiv.org/abs/hep-ph/0411271}{{\tt hep-ph/0411271}}.

\bibitem{Khlopov:2006uv}
M.~Y. Khlopov and C.~A. Stephan, {\it {Composite dark matter with invisible
  light from almost-commutative geometry}},
  \href{http://arxiv.org/abs/astro-ph/0603187}{{\tt astro-ph/0603187}}.

\bibitem{Kim:2019oyh}
J.~H. Kim, S.~D. Lane, H.-S. Lee, I.~M. Lewis, and M.~Sullivan, {\it {Searching
  for Dark Photons with Maverick Top Partners}},  {\em Phys. Rev. D} {\bf 101}
  (2020), no.~3 035041, [\href{http://arxiv.org/abs/1904.05893}{{\tt
  arXiv:1904.05893}}].

\bibitem{Rueter:2019wdf}
T.~D. Rueter and T.~G. Rizzo, {\it {Towards A UV-Model of Kinetic Mixing and
  Portal Matter}},  {\em Phys. Rev. D} {\bf 101} (2020), no.~1 015014,
  [\href{http://arxiv.org/abs/1909.09160}{{\tt arXiv:1909.09160}}].

\bibitem{Wojcik:2020wgm}
G.~N. Wojcik and T.~G. Rizzo, {\it {SU(4) flavorful portal matter}},  {\em
  Phys. Rev. D} {\bf 105} (2022), no.~1 015032,
  [\href{http://arxiv.org/abs/2012.05406}{{\tt arXiv:2012.05406}}].

\bibitem{Dienes:1996zr}
K.~R. Dienes, C.~F. Kolda, and J.~March-Russell, {\it {Kinetic mixing and the
  supersymmetric gauge hierarchy}},  {\em Nucl. Phys. B} {\bf 492} (1997)
  104--118, [\href{http://arxiv.org/abs/hep-ph/9610479}{{\tt hep-ph/9610479}}].

\bibitem{Hebecker:2003jt}
A.~Hebecker and M.~Ratz, {\it {Group theoretical aspects of orbifold and
  conifold GUTs}},  {\em Nucl. Phys. B} {\bf 670} (2003) 3--26,
  [\href{http://arxiv.org/abs/hep-ph/0306049}{{\tt hep-ph/0306049}}].

\bibitem{Antoniadis:1990ew}
I.~Antoniadis, {\it {A Possible new dimension at a few TeV}},  {\em Phys. Lett.
  B} {\bf 246} (1990) 377--384.

\bibitem{Csaki:2004ay}
C.~Csaki, {\it {TASI lectures on extra dimensions and branes}},  in {\em
  {Theoretical Advanced Study Institute in Elementary Particle Physics (TASI
  2002): Particle Physics and Cosmology: The Quest for Physics Beyond the
  Standard Model(s)}}, pp.~605--698, 4, 2004.
\newblock \href{http://arxiv.org/abs/hep-ph/0404096}{{\tt hep-ph/0404096}}.

\bibitem{Csaki:2005vy}
C.~Csaki, J.~Hubisz, and P.~Meade, {\it {TASI lectures on electroweak symmetry
  breaking from extra dimensions}},  in {\em {Theoretical Advanced Study
  Institute in Elementary Particle Physics}: {Physics in D $\geqq$ 4}},
  pp.~703--776, 10, 2005.
\newblock \href{http://arxiv.org/abs/hep-ph/0510275}{{\tt hep-ph/0510275}}.

\bibitem{Kawamura:2013rj}
Y.~Kawamura and T.~Miura, {\it {Classification of Standard Model Particles in
  $E_6$ Orbifold Grand Unified Theories}},  {\em Int. J. Mod. Phys. A} {\bf 28}
  (2013) 1350055, [\href{http://arxiv.org/abs/1301.7469}{{\tt
  arXiv:1301.7469}}].

\bibitem{Agashe:2003zs}
K.~Agashe, A.~Delgado, M.~J. May, and R.~Sundrum, {\it {RS1, custodial isospin
  and precision tests}},  {\em JHEP} {\bf 08} (2003) 050,
  [\href{http://arxiv.org/abs/hep-ph/0308036}{{\tt hep-ph/0308036}}].

\bibitem{Kawamura:2000ev}
Y.~Kawamura, {\it {Triplet doublet splitting, proton stability and extra
  dimension}},  {\em Prog. Theor. Phys.} {\bf 105} (2001) 999--1006,
  [\href{http://arxiv.org/abs/hep-ph/0012125}{{\tt hep-ph/0012125}}].

\bibitem{Georgi:1974sy}
H.~Georgi and S.~L. Glashow, {\it {Unity of All Elementary Particle Forces}},
  {\em Phys. Rev. Lett.} {\bf 32} (1974) 438--441.

\bibitem{Pati:1974yy}
J.~C. Pati and A.~Salam, {\it {Lepton Number as the Fourth Color}},  {\em Phys.
  Rev. D} {\bf 10} (1974) 275--289. [Erratum: Phys.Rev.D 11, 703--703 (1975)].

\bibitem{Randall:1999ee}
L.~Randall and R.~Sundrum, {\it {A Large mass hierarchy from a small extra
  dimension}},  {\em Phys. Rev. Lett.} {\bf 83} (1999) 3370--3373,
  [\href{http://arxiv.org/abs/hep-ph/9905221}{{\tt hep-ph/9905221}}].

\bibitem{Goldberger:1999uk}
W.~D. Goldberger and M.~B. Wise, {\it {Modulus stabilization with bulk
  fields}},  {\em Phys. Rev. Lett.} {\bf 83} (1999) 4922--4925,
  [\href{http://arxiv.org/abs/hep-ph/9907447}{{\tt hep-ph/9907447}}].

\bibitem{Casagrande:2008hr}
S.~Casagrande, F.~Goertz, U.~Haisch, M.~Neubert, and T.~Pfoh, {\it {Flavor
  Physics in the Randall-Sundrum Model: I. Theoretical Setup and Electroweak
  Precision Tests}},  {\em JHEP} {\bf 10} (2008) 094,
  [\href{http://arxiv.org/abs/0807.4937}{{\tt arXiv:0807.4937}}].

\bibitem{Georgi:2000ks}
H.~Georgi, A.~K. Grant, and G.~Hailu, {\it {Brane couplings from bulk loops}},
  {\em Phys. Lett. B} {\bf 506} (2001) 207--214,
  [\href{http://arxiv.org/abs/hep-ph/0012379}{{\tt hep-ph/0012379}}].

\bibitem{Dvali:2000hr}
G.~R. Dvali, G.~Gabadadze, and M.~Porrati, {\it {4-D gravity on a brane in 5-D
  Minkowski space}},  {\em Phys. Lett. B} {\bf 485} (2000) 208--214,
  [\href{http://arxiv.org/abs/hep-th/0005016}{{\tt hep-th/0005016}}].

\bibitem{delAguila:2003kd}
F.~del Aguila, M.~Perez-Victoria, and J.~Santiago, {\it {Some consequences of
  brane kinetic terms for bulk fermions}},  in {\em {38th Rencontres de Moriond
  on Electroweak Interactions and Unified Theories}}, 5, 2003.
\newblock \href{http://arxiv.org/abs/hep-ph/0305119}{{\tt hep-ph/0305119}}.

\bibitem{Muck:2001yv}
A.~Muck, A.~Pilaftsis, and R.~Ruckl, {\it {Minimal higher dimensional
  extensions of the standard model and electroweak observables}},  {\em Phys.
  Rev. D} {\bf 65} (2002) 085037,
  [\href{http://arxiv.org/abs/hep-ph/0110391}{{\tt hep-ph/0110391}}].

\bibitem{Flacke:2008ne}
T.~Flacke, A.~Menon, and D.~J. Phalen, {\it {Non-minimal universal extra
  dimensions}},  {\em Phys. Rev. D} {\bf 79} (2009) 056009,
  [\href{http://arxiv.org/abs/0811.1598}{{\tt arXiv:0811.1598}}].

\bibitem{Gorbunov:2021ccu}
D.~Gorbunov, I.~Krasnov, and S.~Suvorov, {\it {Constraints on light scalars
  from PS191 results}},  {\em Phys. Lett. B} {\bf 820} (2021) 136524,
  [\href{http://arxiv.org/abs/2105.11102}{{\tt arXiv:2105.11102}}].

\bibitem{BNL-E949:2009dza}
{\bf BNL-E949} Collaboration, A.~V. Artamonov et~al., {\it {Study of the decay
  $K^+\to\pi^+\nu \bar\nu$ in the momentum region $140 < P_\pi < 199$ MeV/c}},
  {\em Phys. Rev. D} {\bf 79} (2009) 092004,
  [\href{http://arxiv.org/abs/0903.0030}{{\tt arXiv:0903.0030}}].

\bibitem{Randall:2001gb}
L.~Randall and M.~D. Schwartz, {\it {Quantum field theory and unification in
  AdS5}},  {\em JHEP} {\bf 11} (2001) 003,
  [\href{http://arxiv.org/abs/hep-th/0108114}{{\tt hep-th/0108114}}].

\bibitem{Rizzo:2020ybl}
T.~G. Rizzo and G.~N. Wojcik, {\it {Kinetic mixing, dark photons and extra
  dimensions. Part III. Brane localized dark matter}},  {\em JHEP} {\bf 03}
  (2021) 173, [\href{http://arxiv.org/abs/2006.06858}{{\tt arXiv:2006.06858}}].

\bibitem{Ahmed:2019zxm}
A.~Ahmed, A.~Carmona, J.~Castellano~Ruiz, Y.~Chung, and M.~Neubert, {\it
  {Dynamical origin of fermion bulk masses in a warped extra dimension}},  {\em
  JHEP} {\bf 08} (2019) 045, [\href{http://arxiv.org/abs/1905.09833}{{\tt
  arXiv:1905.09833}}].

\bibitem{Casagrande:2010si}
S.~Casagrande, F.~Goertz, U.~Haisch, M.~Neubert, and T.~Pfoh, {\it {The
  Custodial Randall-Sundrum Model: From Precision Tests to Higgs Physics}},
  {\em JHEP} {\bf 09} (2010) 014, [\href{http://arxiv.org/abs/1005.4315}{{\tt
  arXiv:1005.4315}}].

\bibitem{GrootNibbelink:2001bx}
S.~Groot~Nibbelink, {\it {Dimensional regularization of a compact dimension}},
  {\em Nucl. Phys. B} {\bf 619} (2001) 373--384,
  [\href{http://arxiv.org/abs/hep-th/0108185}{{\tt hep-th/0108185}}].

\bibitem{Appelquist:2000nn}
T.~Appelquist, H.-C. Cheng, and B.~A. Dobrescu, {\it {Bounds on universal extra
  dimensions}},  {\em Phys. Rev. D} {\bf 64} (2001) 035002,
  [\href{http://arxiv.org/abs/hep-ph/0012100}{{\tt hep-ph/0012100}}].

\bibitem{Datta:2010us}
A.~Datta, K.~Kong, and K.~T. Matchev, {\it {Minimal Universal Extra Dimensions
  in CalcHEP/CompHEP}},  {\em New J. Phys.} {\bf 12} (2010) 075017,
  [\href{http://arxiv.org/abs/1002.4624}{{\tt arXiv:1002.4624}}].

\bibitem{ATLAS:2022ooq}
{\bf ATLAS} Collaboration, {\it {Measurements of Higgs boson production by
  gluon$-$gluon fusion and vector-boson fusion using $H\rightarrow W W^*
  \rightarrow e\nu \mu\nu$ decays in $pp$ collisions at $\sqrt{s}=13$ TeV with
  the ATLAS detector}},  \href{http://arxiv.org/abs/2207.00338}{{\tt
  arXiv:2207.00338}}.

\bibitem{CMS:2020djy}
{\bf CMS} Collaboration, A.~M. Sirunyan et~al., {\it {Measurement of the top
  quark Yukawa coupling from $\mathrm{t\bar{t}}$ kinematic distributions in the
  dilepton final state in proton-proton collisions at $\sqrt{s}=$ 13 TeV}},
  {\em Phys. Rev. D} {\bf 102} (2020), no.~9 092013,
  [\href{http://arxiv.org/abs/2009.07123}{{\tt arXiv:2009.07123}}].

\bibitem{Azatov:2010pf}
A.~Azatov, M.~Toharia, and L.~Zhu, {\it {Higgs Production from Gluon Fusion in
  Warped Extra Dimensions}},  {\em Phys. Rev. D} {\bf 82} (2010) 056004,
  [\href{http://arxiv.org/abs/1006.5939}{{\tt arXiv:1006.5939}}].

\bibitem{Csaki:2010az}
C.~Csaki, J.~Heinonen, J.~Hubisz, S.~C. Park, and J.~Shu, {\it {5D UED: Flat
  and Flavorless}},  {\em JHEP} {\bf 01} (2011) 089,
  [\href{http://arxiv.org/abs/1007.0025}{{\tt arXiv:1007.0025}}].

\bibitem{Cheung:2001mq}
K.-m. Cheung and G.~L. Landsberg, {\it {Kaluza-Klein States of the Standard
  Model Gauge Bosons: Constraints from High Energy Experiments}},  {\em Phys.
  Rev. D} {\bf 65} (2002) 076003,
  [\href{http://arxiv.org/abs/hep-ph/0110346}{{\tt hep-ph/0110346}}].

\bibitem{ATLAS:2019erb}
{\bf ATLAS} Collaboration, G.~Aad et~al., {\it {Search for high-mass dilepton
  resonances using 139 fb$^{-1}$ of $pp$ collision data collected at
  $\sqrt{s}=$13 TeV with the ATLAS detector}},  {\em Phys. Lett. B} {\bf 796}
  (2019) 68--87, [\href{http://arxiv.org/abs/1903.06248}{{\tt
  arXiv:1903.06248}}].

\bibitem{ATLAS:2018tvr}
{\bf ATLAS} Collaboration, {\it {Prospects for searches for heavy $Z^\prime$
  and $W^\prime$ bosons in fermionic final states with the ATLAS experiment at
  the HL-LHC}}, .

\bibitem{Delgado:2007ne}
A.~Delgado and A.~Falkowski, {\it {Electroweak observables in a general 5D
  background}},  {\em JHEP} {\bf 05} (2007) 097,
  [\href{http://arxiv.org/abs/hep-ph/0702234}{{\tt hep-ph/0702234}}].

\bibitem{Grzadkowski:2010es}
B.~Grzadkowski, M.~Iskrzynski, M.~Misiak, and J.~Rosiek, {\it {Dimension-Six
  Terms in the Standard Model Lagrangian}},  {\em JHEP} {\bf 10} (2010) 085,
  [\href{http://arxiv.org/abs/1008.4884}{{\tt arXiv:1008.4884}}].

\bibitem{Bagnaschi:2022whn}
E.~Bagnaschi, J.~Ellis, M.~Madigan, K.~Mimasu, V.~Sanz, and T.~You, {\it {SMEFT
  analysis of m$_{W}$}},  {\em JHEP} {\bf 08} (2022) 308,
  [\href{http://arxiv.org/abs/2204.05260}{{\tt arXiv:2204.05260}}].

\bibitem{Kribs:2020jgn}
G.~D. Kribs, X.~Lu, A.~Martin, and T.~Tong, {\it {Custodial symmetry violation
  in the SMEFT}},  {\em Phys. Rev. D} {\bf 104} (2021), no.~5 056006,
  [\href{http://arxiv.org/abs/2009.10725}{{\tt arXiv:2009.10725}}].

\bibitem{ParticleDataGroup:2020ssz}
{\bf Particle Data Group} Collaboration, P.~A. Zyla et~al., {\it {Review of
  Particle Physics}},  {\em PTEP} {\bf 2020} (2020), no.~8 083C01.

\bibitem{CDF:2022hxs}
{\bf CDF} Collaboration, T.~Aaltonen et~al., {\it {High-precision measurement
  of the W boson mass with the CDF II detector}},  {\em Science} {\bf 376}
  (2022), no.~6589 170--176.

\bibitem{ATLAS:2017rzl}
{\bf ATLAS} Collaboration, M.~Aaboud et~al., {\it {Measurement of the $W$-boson
  mass in pp collisions at $\sqrt{s}=7$ TeV with the ATLAS detector}},  {\em
  Eur. Phys. J. C} {\bf 78} (2018), no.~2 110,
  [\href{http://arxiv.org/abs/1701.07240}{{\tt arXiv:1701.07240}}]. [Erratum:
  Eur.Phys.J.C 78, 898 (2018)].

\bibitem{LHCb:2021bjt}
{\bf LHCb} Collaboration, R.~Aaij et~al., {\it {Measurement of the W boson
  mass}},  {\em JHEP} {\bf 01} (2022) 036,
  [\href{http://arxiv.org/abs/2109.01113}{{\tt arXiv:2109.01113}}].

\bibitem{Guedes:2021oqx}
G.~Guedes and J.~Santiago, {\it {New leptons with exotic decays: collider
  limits and dark matter complementarity}},  {\em JHEP} {\bf 01} (2022) 111,
  [\href{http://arxiv.org/abs/2107.03429}{{\tt arXiv:2107.03429}}].

\bibitem{OsmanAcar:2021plv}
A.~Osman~Acar, O.~E. Delialioglu, and S.~Sultansoy, {\it {A search for the
  first generation charged vector-like leptons at future colliders}},
  \href{http://arxiv.org/abs/2103.08222}{{\tt arXiv:2103.08222}}.

\bibitem{Rizzo:2022qan}
T.~G. Rizzo, {\it {Portal Matter and Dark Sector Phenomenology at Colliders}},
  in {\em {2022 Snowmass Summer Study}}, 2, 2022.
\newblock \href{http://arxiv.org/abs/2202.02222}{{\tt arXiv:2202.02222}}.

\bibitem{Csaki:2008zd}
C.~Csaki, A.~Falkowski, and A.~Weiler, {\it {The Flavor of the Composite
  Pseudo-Goldstone Higgs}},  {\em JHEP} {\bf 09} (2008) 008,
  [\href{http://arxiv.org/abs/0804.1954}{{\tt arXiv:0804.1954}}].

\end{thebibliography}\endgroup

\end{document}